\documentclass[a4paper,11pt,letterpaper,final]{memoir}
\usepackage{natbib}
\usepackage{ulem}
\usepackage[french]{babel}
\usepackage{graphicx,amssymb,amsmath,color}
\usepackage[linkcolor={blue},citecolor={red},colorlinks=true]{hyperref}
\usepackage{relsize}
\usepackage{enumerate}
\usepackage{blindtext}
\usepackage{geometry}
\usepackage{cleveref}
\usepackage{cancel}

\def\be{\begin{equation}}
\def\ee{\end{equation}}
\usepackage{natbib}
\usepackage{graphicx}

\begin{document}

\setlength{\parindent}{4mm} \setlength{\intextsep}{20pt plus 6pt
minus 5pt}
\title{\Huge{Un avenir commun au sein de la société numérique}
\hspace{1mm}\\
{\large \textbf{Mémoire de Master M2 pour le cursus en actualité de la philosophie}}

\hspace{1mm}
{\large \textbf{Supervision et révision: Prof. Vincent Beaubois}}
}
\author{Miguel Vanvlasselaer }
\maketitle

\centerline{ \textbf{Résumé}}
\begin{quote}

Les données et les informations sont devenues aujourd'hui les denrées (surabondantes) d'un système mondial 
de machines en réseau et échangeant des signaux à une vitesse proche de la vitesse de la lumière. Dans cet écosystème surchargé, la communication est devenue le mode d'échange le plus central, laquelle est assurée par un complexe technique en mutation de plus en plus rapide. Alors que le téléphone fixe et la télévision ont dominé les échanges d'information pendant au moins un siècle, ils ont récemment cédé la main en quelques années aux ordinateurs portables connectés, aux téléphones portables, aux tablettes, qui peuvent suivre leur utilisateur partout. 

Ces nouvelles machines de la communication ont rapidement proposé une offre d'information tellement abondante qu'aucun humain n'aurait pu en suivre le mouvement. Par le biais de plateformes de mise en commun et de partage de contenu, comme Facebook, LinkedIn, Youtube, Airbnb, Uber et Google, la recherche d'information s'est redoublée d'un nouvel appareillage de suggestion, l'algorithme, offrant à l'utilisateur un contenu qui avait été calculé sur mesure pour plaire, pour coller à son utilisateur. Ainsi les algorithmes de suggestion se sont répandus sur toutes les plateformes en simplifiant tout le travail de traitement de l'information. Ne pas avoir à chercher, trier, soupeser, évaluer était le nouveau luxe que ces machines offraient au commun des mortels. 

Pourtant, en s'interposant entre les utilisateurs et les flux de données, les plateformes ont acquis la possibilité inédite de manipuler, d'orienter, de nourrir les goûts, les préférences, les choix  de consommation des utilisateurs. Au cours de ce processus, les techniques numériques captent les désirs de leurs utilisateurs, leur capacité à former un goût, à construire des volitions à partir d'une histoire personnelle, et elles parviennent à capter même les savoirs les plus techniques. De par ce fait, elles risquent d'induire une prolétarisation d'un type nouveau, par laquelle les individus perdent l'ensemble des capacités et compétences qui ont été captées. Dans ce mémoire, en nous appuyant sur les études de Bernard Stiegler, nous étudierons en détail les causes et les conséquences de ce phénomène de captation et l'inscrirons dans le contexte plus vaste d'une modernité qui encourage l'accélération sans bornes de l'innovation. 

Nous étudierons par la suite les prémisses de l'essor de ces nouvelles techniques algorithmiques et argumenterons, en utilisant les études de Fredric Jameson portant sur la postmodernité, que le contexte culturel de la postmodernité était déjà favorable à l'expansion de ces technologies. Par la dissolution des catégories modernes, la postmodernité a produit un contexte culturel propice à la captation et à la récupération des savoirs et des désirs humains par le système marchand. 

Finalement nous verrons comment les techniques numériques, en tant qu'elles sont un \textit{pharmakon}, peuvent être mobilisées dans un contexte collaboratif, sans produire la prolétarisation massive qu'elles induisent via les plateformes dérégulées. En prenant l'exemple des plateformes collaboratives telles que Wikipédia et arXiv, nous avancerons une ébauche de théorisation de ce à quoi une économie du commun numérique -- en prenant le parallèle des communs étudiés par Elinor Ostrom -- devrait ressembler.

\centerline{ \textbf{Abstract}}

Today, data and information have become overabundant resources within a global network of machines that exchange signals at speeds approaching that of light. In this highly saturated environment, communication has emerged as the most central form of interaction, supported by a rapidly evolving technical infrastructure. Traditional technologies like landline phones and television, which once dominated information exchange, have quickly been replaced by mobile devices such as laptops, smartphones, and tablets that stay in constant proximity to users.

These new communication tools have created an overwhelming surplus of information—so much so that no human could process it all. In response, platforms like Facebook, YouTube, and Google use algorithms to filter and suggest content. These algorithms act as sorting mechanisms, reducing the “informational noise” and presenting users with content tailored to their habits and preferences. This simplifies the cognitive work required to search and evaluate information—essentially offloading it to machines.

However, by placing themselves between users and data, these platforms gain control over what users see and, in doing so, shape their preferences and behaviors. In physical terms, we might say they function like selective filters or control gates in an information system—directing flows and creating feedback loops. Over time, this can lead to a kind of informational inertia, where users become increasingly shaped by algorithmic influence and lose the ability to form independent judgments.

This process reflects a broader trend that philosopher Bernard Stiegler describes as a new kind of proletarianization—where individuals lose the knowledge and skills that are absorbed and automated by digital systems. Borrowing from physics, we can see this as a shift toward higher entropy: a state where complexity is reduced and differentiation is lost.

This thesis explores how such algorithmic systems have emerged, and how the cultural context of postmodernity (as analyzed by Fredric Jameson) helped make this possible. With the breakdown of traditional structures and categories, postmodern culture paved the way for the appropriation of human knowledge and desire by commercial platforms.

Finally, we argue that digital technologies, like a pharmakon (both poison and remedy), can also be used in constructive ways. Platforms like Wikipedia and arXiv demonstrate how digital tools can support collective knowledge without leading to cognitive degradation. Inspired by Elinor Ostrom’s work on commons, we propose a model for a digital commons economy—where information is shared and governed collaboratively, helping to restore a balance between entropy and organization in our digital environments.
\end{quote}

\clearpage
\noindent\makebox[\linewidth]{\rule{\textwidth}{1pt}} 
\setcounter{tocdepth}{2}
\tableofcontents
\noindent\makebox[\linewidth]{\rule{\textwidth}{1pt}}

\chapter{La démocratisation du marché de l'information}

Une observation assez triviale sur les temps modernes, et plus spécifiquement sur les vingt cinq dernières années (pour prendre contre-appui sur le millénaire précédent) est que les moyens de transmission de l'information se sont colossalement démultipliés. Relativement, le temps par jour que chaque individu passe à absorber de l'information via ses nombreux moyens de transmission, à publier sa propre information ou opinion, à communiquer, échanger, converser a explosé. Alors que, dans le contexte d'une communication à grande échelle, certains craignaient une globalisation du monde, de la pensée et de l'action, c'est plutôt l'inaction généralisée, par la réaction constante, qui a émergé de l'hyper-connection. Alors que nous parlons et échangeons beaucoup, nous ne sommes que très rarement capables de construire une réflexion sur un sujet et ensuite imprimer une direction à un projet lié (malgré quelques notables exception comme les hashtags MeToo, BlackLivesMatter, ...). 

Bien au contraire, les exemples les plus marquants d'action démocratique et populaire contemporaine trouvent leurs racines dans une certaine forme de détachement, de déconnexion des grands canaux d'échange. Prenons l'exemple de la ``Convention Citoyenne pour le Climat" (CCC), l'initiative citoyenne mandatée ayant amené en France à la proposition de toute une série de projets de lois pour contrecarrer le réchauffement climatique. Faisant valoir l'idée qu'il n'y a pas de compétence spécifique liée à l'exercice de la  politique, Jacques Rancière a défendu que la manière la plus démocratique d'élire ses représentants est par la tirage au sort. Ainsi, les 150 membres de la conventions ont été tirés au sort parmi les citoyens français. Opérant via des canaux de communication et d'échange en présentiel, ainsi que des cours magistraux et des séances de questions et de réponses avec les experts notamment du GIEC, en six mois, cette Convention a produit 149 propositions de loi couvrant des domaines centraux comme le logement, l'alimentation ou les transports. L'expérience de cette démocratie directe a démontré que des citoyens ordinaires peuvent formuler des politiques solides lorsqu'ils sont bien accompagnés et présente donc une référence en matière de démocratie participative. A l'opposé, même si la démocratisation des plateformes d'échange a eu pour conséquence un accroissement spectaculaire des prises de position et des discussions, elles ne sont que très rarement le berceau de réelles propositions politiques.

 Cette observation paradoxale rend cruciale et centrale l'analyse de l'effet qu'une telle extension du domaine de la communication et de l'absorption d'information peut avoir sur nos sociétés, leurs structures et sur les individus qui les peuplent. Récemment, les plateformes ont commencé à se développer dans divers secteurs qui n'étaient plus liés à l'échange d'informations entre personnes mais qui concernaient toujours une certaine mise en commun d'individus. Quelques grands exemples sont Amazon, Airbnb et Uber. Chacun a détourné à son avantage un secteur marchand en mettant en relation les consommateurs et les marchands avec une grande efficacité via une plateforme unifiée et une interface pratique. Leurs plus grands atouts est qu'elles sont simples à utiliser, intuitives et qu'elles guident tout du long leur utilisateur. Elles classifient, évaluent, suggèrent, mettent en lumière ou invisibilisent, au gré de leur algorithme, pour le plus grand bonheur des utilisateurs. 

Il ne semble pourtant pas que ce processus de ``plateformisation" s'arrête à ces grands secteurs, hôtelier, de transport et d'échange de biens. Bien au contraire, il se répand aujourd'hui dans tous les secteurs où il faut mettre en contact deux catégories de personnes: il existe maintenant une plateforme pour prendre un rendez-vous chez le coiffeur, au guichet de sa banque, pour prendre des cours en ligne et même pour fixer des rendez-vous dans certaines administrations publiques. C'est l'idée même de la plateforme, en tant que strate de centralisation de l'information, de traitement et de suggestion, entre deux catégories d'individus qui s'est emparée de la société. 

C'est un objectif un peu trop vaste pour le contenu d'un court mémoire de master que de développer une théorie complète de ce phénomène, de ses conséquences et de ses origines. L'objectif de cette thèse est de marquer les tensions qui existent dans l'architecture actuelle d'internet et de les étudier avec les outils conceptuels notamment de deux philosophes du XXème siècle, Bernard Stiegler, philosophe français de la technique et Fredric Jameson, philosophe américain et spécialiste de la culture moderne et postmoderne. Nous verrons comment rapprocher leurs différents ensembles de concepts nous permettra de converger vers une voie prometteuse pour la constitution d'un internet communautaire, que nous étudierons à partir d'exemples de plateformes collaboratives telles que Wikipédia.

Au delà de ces conclusions, une autre raison me pousse à développer ce mémoire. Les moments les plus marquants de ma consommation culturelle sont ceux qui m'ont procuré un goût de déjà-vu, qui sont parvenus à poser une \textit{formulation}, souvent étonnamment simple, sur un problème, ou une solution que je ``connaissais" déjà, qui était déjà là en latence. Un exemple des plus marquants pour moi a été la lecture de ``l'insoutenable légéreté de l'être" de Milan Kundera, livre avec lequel je me suis senti tellement en accord que j'ai eu le sentiment que, présomputeusement, j'aurais pu l'écrire moi-même, que je le connaissais déjà avant de le lire. L'un des exercices les plus difficiles est l'exercice de la formulation de nos savoirs en latence, à tel point que l'on pourrait dire que le leitmotiv de la culture devrait être: ``le plus grand service que l'on peut rendre à quelqu'un, est de formuler pour lui ses idées."  Ce mémoire a donc pour but, au moins pour moi, de me forcer à formuler de façon plus ou moins continue un certain nombre de mes idées, notamment celles qui ont été pendant trop longtemps en latence, tout en espérant qu'elles ne se présentent pas de façon trop décousues. Un exercice de Maïeutique en quelque sorte.

\section{L'éveil du web}

Face au besoin de partager rapidement et efficacement les résultats de recherches entre scientifiques du monde entier, Tim Berners-Lee, ingénieur au CERN près de Genève, a développé au début des années 1990 le World Wide Web, que l’on appelle aujourd’hui simplement le Web. Cette plateforme permettait un échange simple d’informations et de documents entre les universités et les centres de recherche, facilitant ainsi la collaboration scientifique. Bien que le Web n’ait pas été créé spécifiquement pour le Grand collisionneur de hadrons (LHC), il constitue aujourd’hui un outil indispensable pour le traitement et le partage des énormes volumes de données générées par des expériences comme celles menées au LHC sous la ville de Genève. Cette innovation marque l'ouverture de la distribution internationale des données, à une vitesse défiant tout autre médium. Elle marque également le début d’une révolution dans la communication, l'échange d’information et la société. Le Web, à l’origine outil scientifique, devient rapidement un espace global transformant les usages culturels, économiques et sociaux, notamment suite au développement des technologies du Web commercial (apparition des navigateurs, des moteurs de recherche, et finalement des plateformes), développement qui trouve sa source dans l'initiative américaine centralisée géographiquement dans la fameuse Silicon Valley.
Le début du millénaire voit la démocratisation du Web, qui colonise en quelques années toutes les sphères de la vie privée. Les premiers réseaux sociaux voient le jour en 1997, avec SixDegrees.com, en 2002 (Friendster). Ceux-ci ne reçoivent tout d'abord qu'un succès modéré. L'essor des réseaux sociaux commence avec la création de LinkedIn, spécialisé dans les relations professionnelles, en 2003 et puis de Facebook, spécialisé dans les relations personnelles, en 2004. Ce dernier passe la barre du milliard d'utilisateurs dès 2012 et culmine aujourd'hui au nombre énorme de plus de trois milliards d'utilisateurs de par le monde. C'est l'ère du networking globalisé, à l'échelle mondiale. En parallèle, les plateformes de rencontre amoureuse se développent d'abord timidemment avec notamment Match.com en 1995 et Meetic en 2001 et puis plus franchement avec l'apparition du géant Tinder en 2012, suivi the Happn et Bumble en 2014, lesquelles briseront le tabou honteux de la rencontre en ligne. Selon les enquêtes Ifop reprise dans \cite{Plaire}, 25\% des internautes français déclarent avoir été inscrits sur un site de rencontre, aux US, 35\% des couples mariés entre 2007 et 2012 se sont rencontrés en ligne. Pour citer Gilles Lipovetsky à ce sujet \cite{Plaire}
\begin{quote}
   \textit{Le fait est là: on cherche de plus en plus à se rencontrer par écrans interposés.}
\end{quote}

Pour les jeunes nés au tournant du millénaire, les réseaux sociaux et de rencontre font partie intégrante du paysage culturel et s'imposent  comme des outils incontournables pour leur insertion sociale, amoureuse et professionnelle. L'omniprésence et l'incontournabilité de ces outils numériques creusent pourtant un fossé d'incompréhension entre la génération des jeunes adultes, nés dans les années 90 et 2000 et la génération antérieure, nées dans les années 70 et 80.   

Sur le plan de la consommation et de l'échange de contenu, l'essor spectaculaire de Youtube à partir de sa création en 2005 marque le début de la démocratisation de l'échange de vidéos. Youtube permet à tout individu de publier des vidéos et à tout autre individu de les consulter, sans que presque aucune vérification ne s'insère entre le créateur et le consommateur.   Cette révolution de la distribution et de la consommation audiovisuelle a permis notamment l'essor d'une nouvelle sorte de vulgarisation scientifique, culturelle, ainsi que d'une diversification presque à l'infini de l'offre vidéo-ludique, tant humoristique que du contenu d'opinion. Alors que l'offre télévisuelle était étroitement enserrée dans le carcan des contraintes télévisées, avec souvent une poignée de programmes disponibles dans chaque domaine (prenons l'exemple de la vulgarisation scientifique, laquelle se résumait presque, pour les chaînes francophones à \textit{C'est pas sorcier}), c'est aujourd'hui par le \textit{surplus} d'une offre diversifiée à l'infini et de qualité parfois très variable que se pose le problème de la consommation culturelle. De manière similaire, la propagation du contenu d'opinion, libéré de la médiation et de la régulation historique opérée par les médias, explose et permet un débat entre les internautes presque en temps réel, pour le meilleur et pour le pire.  

Après cette large captation du traffic internet, les gigantesques plateformes, gorgées d'un contenu beaucoup trop important pour être exploré dans sa totalité, ont déployé un outillage algorithmique. Cet outillage algorithmique avait pour objectif assumé de concentrer le traffic internet sur quelques informations et contenus clefs, notamment par la suggestion et la proposition constante à direction de l'utilisateur, et ce à tel point qu'on les a accusés de ``tuer la vérité" \footnote{https://www.youtube.com/watch?v=cooqSiRHaig}.  Dans les faits, cet outillage technique fut rapidement redirigé pour servir la promotion d'une offre culturelle déjà concentrée.

\section{La captation par les plateformes}
\label{sec:Simondon}

Si la philosophie de la technique est un sujet aussi vieux que la philosophie elle-même, la technique évoluant, la pensée de la technique se doit de suivre et de se réinventer, sous peine de devenir rapidement obsolète. De tout temps la conceptualisation, la théorisation a été lente et laborieuse. Aujourd'hui l'évolution de la technique est plus rapide que jamais, mais la conceptualisation reste presque aussi lente qu'auparavant. Pourtant, dans le contexte actuel, de concepts nous avons grandement besoins.

Effectivement, le propre de la technique est souvent de nous mettre devant le fait accompli. Lorsque la recherche en physique a découvert la bombe atomique, elle n'a pas pu revenir en arrière, et en réponse il a fallu développer de nouveaux programmes de démocratie, une nouvelle façon de faire de la diplomatie (une diplomatie qui prend toujours en compte la possibilité de l'entre-annihilation) pour encadrer la possibilité de la guerre nucléaire. De la même façon, sauf cataclysme majeur, la mondialisation de l'information, l'usage d'outils comme Youtube, Google, Facebook, et toutes la suite d'algorithmes qu'ils charrient ne peut sans doute pas être démantelées, ni vraisemblablement combattues. Et leurs effets, que nous explorerons dans ce mémoire, seront difficilement évitables. Ainsi il semble que la philosophie de la technique se résume souvent à la théorisation après coup de l'inarrêtable. Aujourd'hui, les évolutions techniques, dont le rythme de renouvellement est proche de la décennie, sont si rapides que les structures sociales sont bien souvent trop lentes à réagir: lorsque les législateurs commencent à imposer des restrictions, les lois sont souvent  déjà obsolètes. Le problème des temps modernes concernent donc autant l'impact des techniques sur les individus, que sur la société. 

Dans la théorisation de Bernard Stiegler, et notamment dans son livre ``Dans la disruption: Comment ne pas devenir fou?"\cite{Disrup}, le processus de \textit{disruption} se manifeste lorsque l'innovation technologique, agissant plus rapidement que l'adaptation sociale, dissout les corps sociaux, les groupes de rassemblement, de coopération et d'échange d'information. De nouvelles techniques se substituent alors aux groupes sociaux disruptés, avec comme exemple paradigmatique les réseaux sociaux, et se rendent indispensables. Ces groupes sociaux, absents ou affaiblis,  ne peuvent alors plus participer à l'individuation des personnes, concept de Simondon explicitant comment une matière encore informe se façonne au contact de l'altérité. 

Le principe d'individuation chez Gilbert Simondon  est au coeur de la pensée qui sera développée dans ce mémoire et qui est développée en détails dans son ouvrage ``L’individuation à la lumière des notions de forme et d’information".  Celui-ci repose sur l’idée que l’individu ne peut être compris indépendamment du processus qui le constitue. Contrairement aux approches de philosophie classiques qui considèrent l’individu comme un noyau d'être individuel, Simondon affirme que l’individuation est un devenir, une opération dynamique qui naît d’un état de tension qu'il nomme l'état ``préindividuel". D'une manière importante, ce champ préindividuel contient des potentiels non encore actualisés. Ainsi, l’individu se constitue par résolution partielle de ces tensions, cependant ce processus ne s’achève jamais totalement. Il reste toujours une part de préindividuel, qui peut par la suite devenir une source d’une possible transformation future. Comme nous le verrons, ce processus d'individuation s'articule alors naturellement avec le principe de croissance de l'entropie, qui est à la source de la résolution des tensions du système. 

En pratique, l’individuation selon Simondon peut prendre plusieurs formes.  Celle-ci peut être physique (dans le cas d'un système physique comme un cristal, exemple que Simondon emprunte souvent), vitale (dans le cas de l'émergence d'un être vivant, tel qu'étudié par la biologie), psychique (dans le cas de la formation de la personnalité d'une personne) et collective (dans le cas de la formation d'une association de personnes). Elle concerne non seulement les êtres vivants, mais aussi les objets techniques. 

Ainsi, on peut noter que l'approche de Simondon est largement transversale et analogique. Celui-ci construit sa conceptualisation et ses intuitions à partir de la modélisation physique des systèmes complexes et propose d'appliquer cette conceptualisation aux systèmes psychiques, collectifs, ect ... Cette approche offre une vaste applicabilité aux modèles physiques, sans pour autant tomber dans un réductionnisme matérialiste vieillot.

Ainsi, on peut comprendre l'individuation comme une éducation personnelle, un façonnement de l'énergie libidinale, qui devient alors le propre de chaque individu. 
Selon Stiegler,  cette énergie libidinale -- concept Freudien d'énergie latente en attente de se fixer sur n'importe quel objet de désir -- est, en l'absence de corps sociaux individuants, captée et détournée par les outils numériques, notamment par les plateformes. Ce détournement se fait au profit d'une individuation prédéfinie et d'un façonnement de l'énergie libidinale dictée, non par l'individu, mais par l'algorithmique. Comme exemple très connu de ce type de détournement, on peut citer les algorithmes de suggestion de contenu sur les plateformes qui, bien qu'ils semblent personnalisés, dirigent tous les utilisateurs vers un contenu standardisé\footnote{Un exemple assez frappant de ce phénomène est à l'oeuvre sur la plateforme Youtube, où un florilège de chaînes entièrement automatisées sont apparues, sur lesquelles toutes les vidéos sont réalisées pour une IA générativs et dont tous les commentaires sont des bots, qui répondent à des bots. Il s'agit donc d'un bruit de fond, un contenu sans contenu. }. Cela induit pour Stiegler une paradoxale isolation des individus qui, bien que similaires dans leur désir profond ne peuvent former de véritables liens.   Je pense qu'il faut remarquer ici que pour Stiegler, la culture semble jouer un rôle très crucial pour l'individuation et pour la formation des groupes sociaux, comme on peut par exemple le remarquer avec son retour incessant à la musique Jazz comme créateur de lien social. 

Néanmoins, cette relation au contenu culturel, non plus médiée par les corps sociaux traditionnels mais par l'algorithmique, constitue un nouveau rapport que nous entretenons avec la culture. L'un des penseurs les plus profonds de l'évolution culturelle récente est Fredric Jameson. Dans son livre sur le postmodernisme, ``le postmodernisme, ou la logique culturelle du capitalisme tardif"\cite{Post},  il énumère les caractéristiques qu'il reconnaît dans l'horizon culturel de la deuxième moitié du XXème siècle et fixe la définition du postmoderne. Parmi celles-ci, on peut voir poindre la prévalence pour l'effacement des frontières, une nouvelle forme de superficialité culturelle et la normativité d'un goût démocratique.  Selon Jameson, ce nouveau paradigme économique qui  donne forme à notre paysage culturel, entraînerait l'effacement des frontières par exemple entre la haute et la basse culture, et induit une consommation d'une culture standardisée. Dans ce contexte, les questions qui traverseront ce mémoire seront les suivantes: Peut-on analyser les effets de la culture algorithmique mis en exergue par Stiegler, dans les coordonnées capitalistiques de Jameson et de son approche du postmodernisme ? Est-ce que la disruption algorithmique diagnostiquée par Stiegler, que l'on peut voir comme un amour de vitesse et de la perte de contrôle au profit de la machine, peut être comprise comme un symptôme de la perte de l'authenticité et de l'originalité comme on peut l'entendre chez Jameson ?

Une autre caractéristique centrale que Jameson assigne au postmodernisme (en écho à Lyotard entre autres) est la chute et la disparition des grands récits politiques tels que le Marxisme et l'Humanisme de la modernité, lesquels offraient une cohérence et une unicité à l'action individuelle, au profit d'une multiplicité de discours fragmentés contemporains. Ces discours fragmentés, constitués d'une myriade de composantes politiques, philosophiques, spirituelles et idéologiques, deviennent alors des pièces d'un puzzle identitaire qui sont utilisables et mobilisables par l'individu dans sa vie de tous les jours. Peut-on voir dans ces discours fragmentés auquels Jameson fait référence la masse de discours disponibles sur les plateformes ? Chaque discours globalisant semble y avoir son contrepoint; au discours féministe répond le discours masculiniste, au discours anti-raciste répond répond un discours d'anti-racisme contre les blancs.  Cette grille de lecture est souvent simplifiante et globalisante et offre un prêt à penser algorithmique, une démocratisation de la pensée. 

La direction de ce mémoire sera de partir du contexte contemporain de l'algorithmique des réseaux sociaux et de les étudier avec le prisme de Bernard  Stiegler tout d'abord, et de voir par la suite si on peut inclure cette analyse dans le cadre d'une étude du postmodernisme capitaliste proche de celle de Jameson.

\section{De l'importance des plateformes}

    Je ne veux cependant pas refermer ce chapitre introductif sans modérer ce qui semble être une position unilatéralement critique des plateformes et de l'internet.  
    Je ne saurais surestimer l'importance que Youtube, qui s'inscrit dans la mouvance dite du web 2.0, ou web faiblement participatif,  et son contenu a eu sur ma trajectoire professionnelle, et qui sait, peut-être également personnelle. Je pense pouvoir dire sans me tromper que c'est grandement ma consommation de vidéos de vulgarisation scientifique qui orienta mon choix d'études universitaires vers la physique théorique. Je peux citer de mémoire Science étonnante, Veritasium, Science4All. Bien que maintenue et nourrie par des vidéastes souvent amateurs, ces chaînes de vulgarisation offrent un contenu d'une qualité spectaculaire au vu des moyens à la disposition d'une personne seule ou d'une petit groupe de personnes. 
    Cependant, la richesse du contenu de Youtube ne s'arrête pas à la science physique et à ses curiosités. 
    Dans beaucoup de domaines de recherches, quelques vidéastes ont construit un corpus de vidéos de la plus haute qualité. Pour ne citer que celles que j'ai le plus personnellement suivies: Heureka, portant sur l'économie, Hacking social, portant sur la sociologie et la psychologique sociale, et finalement les chaines de philosophie proprement dite, avec notamment Monsieur Phi. Mon suivi assidu de ce dernier a joué un rôle de premier plan dans la croissance de mon intérêt pour les questions philosophiques et mon envie d'y consacrer une partie de mon temps. 

    Il ne faut donc pas se méprendre, bien que critique dans sa forme, la réflexion menée dans ce mémoire est dans les faits avant tout une déclaration d'amour à l'incroyable ouverture des possibilités et des perspectives que des plateformes telles que Youtube ont pu apporter au monde. Et c'est à cause de ce même impact qui a jalonné non seulement ma vie mais également celle de bien des jeunes de ma génération que j'estime que ces nouvelles technologies de l'information méritent une réflexion en profondeur, laquelle n'a été qu'esquissée dans les études actuelles (me semble-t-il). 

    Le Web est encore et toujours porteur d'un rêve. Prenons pour exemple le passage d'une thèse sur les plateformes en tant que médias\footnote{Médias traditionnels et acteurs du Web 2.0 : vers la cohabitation ou la convergence des acteurs de l’information et du divertissement ? https://www.mediafire.com/?7ygmwwjz0cm}
\begin{quote}
    \textit{Le 2.0 se détache du seul Web et de ses fonctions principales, l’information et le divertissement. Le Web a permis aux citoyens de s’approprier des moyens d’interagir. Tous les canaux de communication sont bientôt à sa portée. De nombreuses sphères, comme l’économie et la politique, sont également concernées par ce besoin d’expression communautaire. Le Web a finalement permis la concrétisation d’un modèle en théorie appliqué depuis longtemps, mais dépourvu des moyens de s’appliquer concrètement: la démocratie.}
\end{quote}

Voici la promesse initiale portée par le Web. L'espoir d'un internet collaboratif n'est pas tout à fait dévoyé, dans le sens où internet, et en particulier Youtube offre une plateforme pour la mise en place de médias alternatifs au mode de financement participatif. On peut citer Elucid, Thinkerview, Livrenoir... Autant de plateformes qui sont toutes financées par les auditeurs.
 
\paragraph{Les plateformes comme espace de création de liens}
Je ne connais pas mes voisins, ni ceux d'en face, ni ceux d'à côté, ni ceux d'au dessus ou d'en dessous. Je ne suis même pas spécialement nouveau dans l'immeuble. Avant, j'aurais fait l'effort d'essayer d'en rencontrer quelques uns, mais aujourd'hui l'énergie me manque un peu. Pourtant je passe une quantité de temps assez impressionnante, plus souvent passivement qu'activement sur Youtube. Je m'égare souvent à lire les commentaires écrits sous telles ou telle vidéo, parfois je m'hasarde à en écrire un. On décrit souvent l'internet, et notamment les espaces commentaires, comme un lieu où chacun s'écharpe avec chacun. C'est sans doute souvent la cas, mais on rencontre ici et là des curiosités, en assez grand nombre si on sait où chercher. On y découvre un espace de débats passionnés et passionnants entre des inconnus qui ne savent ni où les autres vivent ni à quoi ils ressemblent. Ce qui lie ces individus, c'est un intérêt commun, ou une consommation culturelle en commun. Ainsi, on se sent facilement plus proche d'eux que de ses fantômes de voisins. Cela étant, je ne décris sans doute qu'une banalité qu'internet à rendu non-locale, et que beaucoup voient comme un théatre de la nouvelle solitude. 

Pourtant, on ne peut s'empêcher de vouloir rapprocher les deux observations ci-dessus, de chercher à leur donner un socle commun. On est très tenté de dire que l'on se gave des interactions en lignes car on souffre du manque d'interaction avec ses voisins, ou qu'à l'inverse on boude ses voisins car on est rassasié par les montagnes de contenus en ligne. Si la direction de la causalité entre ces deux observations est certainement difficile à établir, il semble évident qu'elles appartiennent toutes deux à un paysage commun, paysage que l'on se doit d'embrasser d'un seul regard, pour ne pas perdre quelqu'élément crucial.

\paragraph{Les plateformes comme espace d'organisation  de la contre-culture}

En mettant en lien des personnes séparées physiquement (dans l'espace) dans une situation commune, les plateformes ont permis de produire des types d'organisation que le bouche-à-oreille et les pancartes n'auraient pas pu créer si efficacement. De cette façon, les plateformes ont été un lieu d'organisation de certaines formes de résistance contemporaine au travail et aux contraintes patronales.

Ainsi, en guise d'exemple, TikTok et Reddit servent d'espace de formation de groupes d'employés et d'ouvriers qui ne voient pas de sens dans leur travail et qui rejettent la performance pour la performance. De la même façon, Facebook a beaucoup servi de plateforme d'organisation pour certaines manifestations, notamment les mouvements tels que les Gilets Jaunes.

\section{Plan et organisation}

Pour inclure le lecteur dans la trame de ce mémoire, je l'écrirai en utilisant le pronom ``nous". Cette utilisation générique du ``nous"  m'offrira une flexibilité qui me permettra également de revenir à l'occasion au pronom ``je", que j'utiliserai pour marquer une opinion moins consensuelle ou qui ne découle pas directement du raisonnement ou de l'opinion d'un autre auteur. 

Ce mémoire s'organise en cinq chapitres qui contiennent chacun une partie du raisonnemenent que nous comptons suivre. Dans un premier chapitre, \ref{chap1}, nous dressons un large panorama de l'impact de l'évolution des techniques informationnelles, vers une mise en place à grande échelle des algorithmes et l'étude du big-data, sur les institutions sociales. Nous mettons en évidence quelques effets potentiellement dangereux de cette évolution, notamment la formation de ``bulles de filtre", la possibilité de l'obsolescence de la modélisation et l'esthétisation de la politique.

Le panorama dressé dans ce premier chapitre, nous amène dans un second chapitre, \ref{chap2},  à étudier systématiquement la captation des désirs et des savoirs opérée par les algorithmes. En nous appuyant sur l'analyse opérée il y a plus de 70 ans par Adorno et Horkheimer, portant sur la captation de la libido par l'industrie culturelle, nous étudierons comment  les plateformes ont capté les flux informationnels pour influencer les utilisateurs. 

Dans un troisième chapitre, \ref{chap3}, nous abordons les effets de la marchandisation et de l'accroissement des flux informatiques sur les structures sociales et les interprétons sous le prisme de l'idée contemporaine de la \textit{disruption}. Nous présentons les mécanismes de la disruption et les conséquences qui émergent, telles que la désindividuation, la disparition de l'époque et l'émergence d'un présentisme généralisé, et les étudions au moyen d'une analogie avec les études physiques portant sur l'entropie. Cette étude amène à la définition d'une nouvelle quantité portant spécifiquement sur l'organisation des sociétés, à savoir l'\textit{anthropie}. 

Dans un quatrième chapitre, \ref{chap4}, nous recherchons les origines de l'essor très rapide de ces nouvelles techniques informationnelles, notamment sous leur forme actuelle de plateformes, dans une prédisposition culturelle que l'on appelle la postmodernité. En invoquant les analyses de Jameson, nous argumentons que cette prédisposition à la captation a une origine culturelle.   

Finalement, dans un dernier chapitre, \ref{chap5}, nous présentons un type de plateformes qui semblent résister à la captation informationnelle et à l'entropie décrite dans les chapitres précédents, en prenant l'exemple de Wikipédias et d'arXiv. Après en avoir décrit les caractéristiques, nous les rapprocherons des principes des communs édictés Ostrom. Nous conclurons en tâchant de mettre en évidence ce que pourrait être les principes du commun numérique. 

\textbf{Note finale:} Alors que je terminais l'écriture de ce mémoire, j'ai découvert le très récent ouvrage\cite{PostVeri} de Michaël Lainé, qui aborde des problématiques très similaires à celles de ce mémoire. J'ai eu le temps d'en faire une lecture sommaire et de le discuter à certains endroits importants de ce mémoire.

\chapter{Les mutations de la société par le numérique}
\label{chap1}

Avant leur algorithmisation, les réseaux d'échange numériques, basés sur l'internet naissant, ont fait l'objet d'un fantastique espoir d'émancipation basé sur la possibilité d'une économie collaborative à des échelles mondiales. C'est la naissance de ce que l'on appelle le  Web 2.0, également appelé Web participatif\cite{Myles}. Dans son article, ``Government as a Platform", Tim O'Reilly présente son concept du Web 2.0 comme l'émergence d'un gouvernement horizontal se servant de la nouvelle plateforme d'internet comme d'une base pour le débat entre des internautes égaux. De simple portail sur lequel l'internaute peut accéder à des données, le Web se transforme à ce moment en un médium d'un nouveau genre sur lequel n'importe quel internaute peut non seulement accéder à des données, mais aussi en publier, et les discuter avec n'importe quel autre internaute. Dans le Web 2.0, l'internaute devient acteur en alimentant les sites en contenu, comme les blogs, ou de manière collaborative avec les wikis. L'internet était devenu une vaste plateforme d'échange et de mise en commun, une économie presque horizontale. Cette ère de la participation donnera lieu à toutes sortes de plateformes (dans le cas présent ce sont des plateformes non-algorithmiques), allant de l'échange de données en peer to peer, contexte dans lequel chaque internaute est supposé donner autant qu'il reçoit, aux blogs et autres journaux personnels. Les espoirs étaient grands, comme l' expose magistralement Eli Pariser dans son ouvrage ``The filter bubble: what the internet is hiding from you"\cite{Pariser}
\begin{quote}
\textit{Quand je grandissais dans une zone rurale du Maine dans les années 1990, un nouveau numéro de Wired arrivait chaque mois à notre ferme, rempli d’histoires sur AOL, Apple, et sur la manière dont les hackers et les technologues étaient en train de changer le monde. Pour l’enfant préadolescent que j’étais, il était évident qu’Internet allait démocratiser le monde, en nous connectant à de meilleures informations et au pouvoir d’agir sur celles-ci. Les futurologues californiens et techno-optimistes dans ces pages parlaient avec une certitude inébranlable: une révolution inévitable et irrésistible était sur le point d’éclater, une révolution qui allait aplatir la société, renverser les élites et inaugurer une sorte d’utopie mondiale généralisée.}
\end{quote}

La nouvelle connectivité se répandait et allait changer le monde pour le mieux. Il était fini le temps des hiérarchies des sachants et des ignares, des puissants et des faibles. L'information allait se démocratiser, et les gouvernements ne pourraient plus garder aucun secret. Le monopoles des chaînes d'information, avec leurs lignes éditoriales biaisées allait exploser en une myriade de reporters privés et autant d'opinions qui s'entrechoqueraient sur le net. C'est sur cet espoir, nous dit Bernard Stiegler dans son ouvrage ``Dans la disruption"\cite{Disrup}, que se base le storytelling de la disruption. Dans la section 25 dudit ouvrage, ``néganthropologie de la disruption", le philosophe énonce 
\begin{quote}
\textit{Le storytelling qui accompagne la réticulation disruptive consiste à poser que celle-ci renverse les pouvoirs exercés de haut en bas, top down, en permettant aux gens d'exercer le leur de bas en haut, bottom up.} 
\end{quote}

Pour Bernard Stiegler, la disruption qui était donc avant tout l'espoir d'une disruption des hiérarchies institutionnelles et des inégalités, s'est muée en une excuse pour camoufler le détournement de cet espoir. Nous reviendrons dans les chapitres suivants sur ce que Bernard Stiegler diagnostique vraiment dans cette disruption par les réseaux. Insistons néanmoins sur le fait que de la possibilité d'un internet collaboratif, il ne reste aujourd'hui qu'un espoir partiellement déçu. Si il est vrai que l'internet reste une plateforme reliant des personnes séparées spatialement, ainsi permettant de créer des communautés qui étaient autrement physiquement impossible, Stiegler dénonce la machinerie algorithmique des réseaux qui s'est rapidement interposée entre les internautes, transformant l'interaction entre internautes en une interaction d'un nouveau type entre des individus individualisés et le tentaculaire réseau. Ce détournement de l'internet était déjà diagonistiquée par O'Reilly, dans les premières lignes de son article 

\begin{quote}
\textit{le secret du succès de leaders comme Google, Amazon, eBay, Craigslist, Wikipédia, Facebook et Twitter tient au fait que chacun de ces sites, à sa manière, a su exploiter la puissance de ses utilisateurs non seulement pour ajouter de la valeur à ses services, mais plus encore, pour co-créer ces services avec eux.}
\end{quote}
Citation qui a trouvé sa postérité dans le dicton; ``Si tu ne paies pas un produit, c'est que le produit, c'est toi". Bien sûr l'usage que chaque plateforme tire de ses utilisateurs varie largement  d'une d'une plateforme à l'autre, le résultat étant largement différent si celle-ci appartient au domaine public, collaboratif, ou privé. En particulier, les plateformes privatisées de partage et d'échange ont rapidement réalisé l'importance des données que la navigation des internautes laissait derrière elle, avec les interminables applications en étude du comportement auxquelles elles ouvraient les portes.

Dans ce premier chapitre, nous reprenons la tâche que nous avons esquissée dans le chapitre introductif précédent en approfondissant la description des rapides mutations que l'essor du numérique a induit dans les différents secteurs de notre société. Nous porterons un intérêt tout particulier à l'essor de l'algorithmique, dans tous les recoins d'internet et mettrons en exergue les dangers qui semblent particulièrement cruciaux à notre époque. Nous laisserons cependant une lecture plus globale et systématique de ces phénomènes pour le chapitre suivant.

\section{Automatisation, suggestions et bulles de filtres}
\label{sec:bulles}

La libération du carcan des médias traditionnels qu'a permis l'internet a rapidement produit un nouveau type de carcan, celui de l'algorithmique. Face au nombre colossal de publications et de contenus présents sur la plupart des grandes plateforme, l'internaute ne peut s'y retrouver seul.  En 2025, environ 500 heures de vidéos sont mises en ligne chaque minute sur YouTube. Cela implique que  la simple recherche par titre ou par thème semble devenir insuffisante. 

    La mise en place de l'architecture algorithmique des géants du Web commence lorsque Google annonce via un communiqué le 4 décembre 2009, l'activation de son nouvel algorithme de personnalisation des résultats de recherche. A partir de cette date, les résultats de recherche des utilisateurs du premier outil de recherche au monde allaient dépendre de leur zone géographique, de la fréquence de leurs clics, de leur historique de recherche. La même mécanique se met rapidement en place sur la plateforme Youtube, qui commence à suggérer les nouvelles vidéos que les internautes devraient consulter. Aujourd'hui, statistiquement, moins de 10 pourcents des vidéos regardées ont été recherchées\cite{Souzam}; le reste a été suggéré ou provient d'un média externe. On peut en conclure que la vidéo consommée n'est que très rarement la vidéo recherchée mais plutôt la vidéo suggérée. Suggérée le plus souvent, non par un ami ou un parent, mais plutôt par l'algorithme générateur de suggestions de Youtube. 

    Selon Stiegler, le Web, ses plateformes et la récente algorithmisation de celles-ci   contribuent  à ce que l'on peut appeler la \textit{troisième vague d'automatisation}. La première de ces vagues d'automatisation est celle décrite par Marx, que l'on appelle généralement l'industrialisation. La seconde vague d'automatisation, au sortir de la seconde guerre, à savoir la Taylorisation,  a amené les robots mécaniques dans les industries, et a déplacé une grande part de l'expertise des ouvriers vers ces tout nouveaux robots. Finalement, la troisième vague d'automatisation, celle qui se déroule en ce moment, prend appui sur les technologies du Web et sur les robots algorithmiques, notamment déployés sur les plateformes internet, comme nous allons le voir plus en détails, pour assister à la recherche, suggérer et finalement capter des données sur l'utilisateur. La dernière innovation de cette vague d'automatisation est le ``machine learning et le deep learning", qui ensemble forment ce que l'on appelle aujourd'hui l'intelligence artificielle. Cette nouvelle vague s'accompagne naturellement de toute une série de nouveaux acteurs, et donc de nouveaux ``milliardaires", comme Zuckerberg, Besos et consort.

    Les détails techniques de l'algorithme de Youtube constituent à ce jour encore un secret de polichinelle, dont on ne connait que les grandes lignes. L’algorithme est un système de recommandation qui sélectionne les vidéos les plus susceptibles d’intéresser chaque utilisateur. Il analyse des données comme l’historique de visionnage, les recherches, les likes, le temps passé sur chaque vidéo, et les abonnements\cite{Pariser}. Son objectif principal est de maximiser le temps de visionnage et l’engagement sur la plateforme. Il agit sur plusieurs zones de la plateforme: la page d’accueil, les vidéos suggérées, les résultats de recherche, et les Shorts. Les vidéos sont classées selon leur taux de clics, durée de visionnage, rétention, et feedbacks (positifs ou négatifs). L’algorithme favorise les contenus pertinents, engageants et regardés jusqu’au bout. Il ne privilégie pas uniquement les vidéos populaires, mais celles adaptées à chaque profil. Pourtant le seul objectif de l'algorithme est de maximiser le temps d'utilisation de la plateforme, d'absorber le plus de temps d'attention possible de la part de l'utilisateur. Le fait est là : Youtube suggère des contenus qui flattent nos croyances, qui s'insèrent parfaitement dans la vision du monde d'un individu\cite{Pariser}. 

Ainsi, le système se veut personnalisé et en constante étude de son propre utilisateur. Lorsque nous regardons Youtube, youtube nous regarde. Pour certains, cette information ne peut sembler qu'anecdotique, alors que pour d'autres elle sonne comme une catastrophe. Quoi qu'il en soit, l'algorithmique déployée sur la plateforme est à la racine d'un effet absolument nouveau engendré par les outils de communication numérique: celui des chambres d’écho et des ``bulles de filtres", un concept inventé par Eli Pariser, dans son ouvrage du même nom ``The filter bubble: what the internet is hiding from you"\cite{Pariser}. Celui-ci nous décrit l'effet que la distribution des informations par les algorithmes peut avoir sur les opinions des utilisateurs d'internet.  Effectivement, l'esprit humain tend à se tourner naturellement vers les sources d'information qui vont dans le sens de ses opinions. Ainsi pour le garder d'autant plus connecté sur ses plateformes, les algorithmes sont conçus pour déterminer les opinions de l'utilisateur et, connaissant le pouvoir attractif des opinions concordantes, lui recommander des contenus porteurs de ces opinions, encore et encore. Eli Pariser résume la situation dans les premières lignes de son ouvrage

\begin{quote}
\textit{Avec Google personnalisé pour chacun, la requête ``cellules souche" peut produire des résultats diamétralement opposés pour un scientifique favorable à la recherche sur les cellules souches et pour un militant qui s’y oppose. ``Preuves du changement climatique" pourrait donner des résultats différents pour un activiste environnemental et pour un cadre d’une compagnie pétrolière. Dans les sondages, une large majorité d’entre nous suppose que les moteurs de recherche sont impartiaux. Mais cela tient peut-être justement au fait qu’ils sont de plus en plus biaisés pour refléter nos propres opinions. De plus en plus, l’écran de votre ordinateur devient une sorte de miroir sans tain, reflétant vos intérêts, tandis que des observateurs algorithmiques scrutent vos clics.}
\end{quote}

L'effet immédiat des bulles de filtre est d'amplifier spectaculairement l'impact du biais de confirmation. Le biais de confirmation est un phénomène cognitif par lequel une personne accorde plus de valeur ou de crédit aux informations qui confortent ses croyances ou hypothèses initiales, tout en minimisant ou en écartant celles qui les contredisent. Ce mécanisme conduit souvent à une résistance au changement d’avis et à une illusion d'avis éclairé chez l'individu. Il se manifeste notamment lorsqu’un individu sélectionne ou interprète les faits de manière orientée, en se remémorant de façon partielle ou biaisée les éléments qui vont dans le sens de ses convictions. Ce biais peut fausser l’analyse d’une situation ou d’une recherche, en influençant inconsciemment la sélection des données prises en compte. L’esprit humain, par réflexe, tend à s’exposer davantage à des opinions similaires aux siennes, créant ainsi une impression trompeuse de consensus autour de ses idées\cite{Biais}.

Le ressort fondamental du biais de confirmation n'est pas totalement clair. Il pourrait reposer sur une inclination naturelle à l'économie mentale. Cette hypothèse voudrait que les informations conformes à des opinions déjà formées soient moins coûteuses à traiter cognitivement et donc positivement sélectionnées pour économiser l'énergie. Il pourrait également reposer sur un effet d'ancrage, qui nous pousse à préférer une information déjà acceptée à une opinion nouvelle. Quel que soit le mécanisme fondamental de ce biais de confirmation, Youtube, en installant son algorithme de suggestion, serait donc devenu le supermarché des opinions individuelles, tout en enfermant chaque individu dans son opinion préconçue.

Tocqueville, dans ``De la démocratie en amérique", nous met en garde contre ce qu'il pense être l'un des plus grands dangers de la république. Selon lui, il s'agirait de la constitution de petites sociétés qui s'isolent du monde et entretiennent un ``entre-soi" idéologique. Il pose la question: Dans un tel monde, comment maintenir une politesse de l'esprit ? Comment peut-on rester courtoi avec ceux qui ne partagent pas l'étroite opinion ressassée par le groupe ? Dans un tel monde, la vérité et lé recherche de la vérité seraient mises à mal. C'est une thèse qui est notamment défendue par Michaël Lainé, dans son ouvrage ``Dans l’ère de la post-vérité"\cite{PostVeri} et sur laquelle nous reviendrons longuement dans les pages qui suivent. En faisant de l'information une affaire de communautarisme  et en renforçant les biais portant sur la formation de l'opinion, les algorithmes qui gèrent la distribution des informations sur les réseaux facilitent la propagation de fausses informations.

Cela étant dit, on peut rapidement se poser la question suivante: en quoi est-ce que les bulles d'information de Youtube, qui dominent la distribution des opinions pour les plus jeunes générations, sont différents des journaux d'opinion d'antan ? En France, ceux qui entretiennent des opinions plus à gauche tendent à lire ``Libération" alors que ceux qui ont des opinions initiallement plus à droite n'achèteront que le ``Figaro". Alors que les individus ont toujours activement sélectionné leurs sources d'information, qu'est-ce qui rend les bulles d'information plus psychiquement dommageables? La différence tient dans l'échelle et la systématicité, ainsi que l'opacité qu'entretiennent les réseaux. Selon Michaël Lainé, les algorithmes brouillent les pistes, alors que des moteurs de recherche comme Google ou des hébergeurs comme Youtube en apparence se présentent comme apolitiques, impartiaux, neutres, ils contribuent à endormir l'esprit critique des usagers.

Lorsque l'on utilise Youtube, le biais de confirmation n'est plus seulement un biais psychologique de sélection des informations confirmant nos préjugés, il devient le harpon par lequel notre attention est captée et entrainée dans une boucle d'informations concordantes, donnant souvent l'illusion d'un consensus factice. La place de biais de confirmation, dans l'économie de notre consommation de contenu numérique, est donc l'outil, ou du moins l'un des outils, par lequel l'algorithme encourage la consommation de vidéos. L'algorithme de Youtube trie, renforce et radicalise les idéologies mais sans pour autant vouloir transmettre une quelconque idéologie en elle-même, ni ne semble vouloir en favoriser une, en tout cas de prime abord. Il semble idéologiquement neutre. Là où les journaux portent une ligne éditoriale, parfois politique claire, Youtube est neutre.  En cela, l'algorithme de Youtube a une place spéciale, il utilise, répand et renforce les idéologies à des fins commerciales.  La polarisation des opinions n'est alors qu'un effet pervers secondaire de l'organisation de la plateforme.  Cette organisation  est nouvelle et ne constitue que la partie la mieux comprise de l'effet des algorithmes sur la psyché des utilisateurs.

S'ensuit un étrange paradoxe de rapetissement de l'horizon perçu, où chacun ne peut voir que sa propre opinion multipliée à l'infini par les algorithmes. Comme le commente Clay Shirky, auteur de ``Here Comes Everybody and Cognitive Surplus": ``Les entreprises d’Internet nous montrent de moins en moins l’étendue du monde, nous enfermant de plus en plus dans le voisinage du familier". Le danger, argumente Eli Pariser, est que ``chacun d'entre nous finisse par habiter 
un ghetto intellectuel solitaire.”

Un autre exemple, sans doute plus grossier, de manipulation systématique de l'opinion politique est celui des dites ``fermes à trolls". Ici, le troll référait initialement à un individu qui, faisant irruption dans un débat et prenant avantage de l'anonymat d'internet, défend avec force et/ou vulgarité un point de vue outracier ou infondé. Aujourd'hui, le troll est plus souvent un bot, à savoir un robot postant des messages sur internet, qui publie en continu et sur différentes plateformes des contenus de désinformation, sous une forme souvent ultra-émotionnelle, dans le but de faire tourner l'opinion publique sur un point d'actualité controversé. La manipulation est souvent grossière mais mise sur un large nombre de publications et sur leur cadence effrénée. 

Concernant l'année 2024, Imperva a estimé que près de 50 \% du traffic internet était généré par ce type de robots \cite{Imperva}, dont une majorité vient des robots malveillants. D'une manière similaire la startup de détection de contenus Originality AI estime que plus de 54 pourcents des longs posts en anglais de LinkedIn seraient probablement générés par des robots. Cette tendance est en hausse rapide sur de nombreuses plateformes. Mêmes les plateformes de contenus vidéos, telles que Youtube, ne sont plus épargnées par ce phénomène et dénombrent un nombre croissant de chaînes aux contenus, vidéos et commentaires, totalement générés par IA. Il est aussi tout à fait notable que ces contenus reçoivent une attention croissante de la part des internautes.  

Ces différents phénomènes, et notamment les bulles de filtre et les fermes à troll, ainsi que l'augmentation des contenus produits par des robots tel que décrit plus haut,  contribuent tous à la polarisation des opinions, notamment politiques. 

\section{L'économie des données et l'espoir de la fin de la théorie}
\label{sec:Eco}
Après avoir esquissé l'impact que les algorithmes pouvaient avoir sur la formation des opinions, nous nous tournons maintenant vers la question de la relation du contemporain, plongé dans la nouvelle économie des données, au savoir. Comme nous l'avons déjà présenté dans le chapitre précédent, la captation toujours plus importante des données des utilisateurs ainsi que l'accroissement de la puissance computationnelle a commencé à révolutionner notre rapport au monde, par ce que nous pouvons appeler ``l'économie des données". Notre rapport au savoir ne saurait y échapper, et de fait, il se trouve lui aussi dans le viseur d'une tentative de révolution.

Dans son ouvrage, ``la condition postmoderne" \cite{Lyotard}, Jean François Lyotard, remarque que les sciences de pointe portent sur l'information et l'échange d'information, ainsi que sur le développement des langages informatiques. Dans ce virage de la recherche scientifique, il remarque la nature du savoir ne saurait rester intacte, 
\begin{quote}
\textit{Dans cette transformation générale, la nature du savoir ne reste pas intacte. Il ne peut passer que dans les nouveaux canaux et devenir opérationnel, que si la connaissance peut être traduite en quantité d'information. On peut donc en tirer la prévision que tout ce qui dans le savoir constitué n'est pas ainsi traduisible sera délaissé, et que l'orientation des recherches nouvelles se subordonnera à la condition de traduisibilité des résultats éventuels en langage machine. }
\end{quote}

On a abondamment documenté le poids que les mathématiques, après la révolution Newtonionne, ont eu sur la structure et la représentation du savoir. On parle de la mathématisation des sciences. Cette mathématisation du savoir a opéré un grand tri dans les disciplines pouvant se réclamer scientifiques et a partiellement exclu celles que l'on ne pouvait que difficilement mathématiser. Un tri similaire s'est naturellement opéré dans la définition des ``questions scientifiques", laquelle a dû amputer un grand nombre de voies de recherche pour payer le prix de la mathématisation. Par la voix du langage des mathématiques, le scientifique se fait exégète de la nature. 

Selon Lyotard, une nouvelle traduction des savoirs est cours. Ce que décrit ici Lyotard dans le passage ci-dessus, sans doute déjà avant l'heure, c'est le poids de l'informatique, de l'algorithmique, sur le savoir. Il exprime l'intuition selon laquelle ce que l'on pourrait appeler \textit{l'informatisation du savoir} va profondément remodeler notre rapport au savoir ainsi que la forme de ce dernier.

Un premier corollaire immédiat à cette traduction du savoir en langue informatique est sa mise en commun. Le savoir est mis  en dehors des personnes. En ce sens, ce que l'on appelle savoir doit être exprimable et exprimé dans une langue que les machines peuvent comprendre et transmettre, langage qui n'est fondamentalement pas celui des humains. L'humain ne pouvant dialoguer avec la machine qu'au travers d'une nouvelle traduction,  le savoir traduit en langage machine n'est alors plus compréhensible directement par l'humain, il ne lui appartient plus en propre (comme pourrait l'être un savoir pratique acquis par l'expérience par exemple), il existe en premier lieu et avant tout en dehors de lui, en tant qu'il est transmissible à la machine.

Cette extériorisation n'est pourtant pas fondamentalement nouvelle et a lieu depuis des millénaires au travers de l'écriture. On ne considère communément comme savoir que l'information factuelle que l'on peut transmettre sous forme de mots ou de symboles. Cet impératif de traductibilité disqualifie du savoir ce que l'on appellerait l'intuition et la sagesse, qui ne peuvent être transmises que par les expériences humaines et, pourrait-on dire, par l'expérience de la vie. La conséquence de ceci est que le savoir se renforce par la disqualification de non-savoirs. 

Ce qui est beaucoup moins trivial et ce qu'entrevoit en 1979 Lyotard, c'est que le savoir deviendrait à partir des années 2000 l'un des enjeux principaux de la course économique. Comme nous l'avons vu, l'architecture des algorithmes sert, non un but idéologique, mais un but commercial. En nous gardant plus longtemps sur leur plateforme, les distributeurs peuvent collecter plus d'informations sur les comportements des usagers, données qu'elles revendent pas la suite, essentiellement à des annonceurs. Ces données individuelles, bien que  le plus souvent anonymisées constituent la goutte d'eau d'un torrent de données statistiques portant sur la population, et donc sur les consommateurs. Données biométriques, historiques de navigation, de déplacement, d'activité, d'achat et de visionnage, en s'aggrégeant, elles forment ce que l'on appelle le ``big-data". 

Au premier abord, le big-data ne semble être qu'un amoncellement de données brutes, informes, lesquelles, même si elles peuvent présenter un intérêt commercial certain, ne saurait être appelée savoir, car elle manque drastiquement de la forme structurée sous laquelle se présente généralement le savoir scientifique. Par exemple, le savoir en sciences physiques a une forme synthétique, laquelle le rend capable de décrire beaucoup plus de phénomènes que ceux qui sont nécessaires pour le formuler. Ce savoir est théorique et synthétique à l'extrême, dans le sens où la dizaine de symboles nécessaires pour écrire l'équation d'Einstein permettent aussi de décrire \textit{tous} les phénomènes gravitationnels connus à ce jour. C'est sans doute l'exemple de synthèse la plus spectaculaire que l'on connaisse, et l'exemple paradigmatique du savoir scientifique contemporain, en tous cas dans le paradigme réductionniste. 

Le big-data est aux antipodes de ce savoir hautement structuré. Pourtant, malgré cet éloignement par rapport à la définition du savoir scientifique traditionnel, le nouveau savoir-big-data entend rivaliser avec son grand frère. Dans son fameux article ``la fin de la théorie", sous-titré ``Le déluge de data rend la méthode scientifique obsolète", qui a lui valu autant la reconnaissance des technocrates que l'ire de tous les autres, Chris Anderson déclare que le savoir-big-data va rendre la théorie obsolète.  Le sous-titre de l'article, ``Le déluge de data rend la méthode scientifique obsolète" doit se comprendre non comme la méthode scientifique, mais comme le savoir scientifique, car ce qui est en jeu dans ce court article du journaliste, c'est la valeur des modèles scientifiques par rapport à l'apport prédictif des grandes bases de données.

L'argument de Anderson est le suivant; le but ultime de la science dans sa construction traditionnelle est de construire des réseaux de causes et de conséquences, autrement dit, des réseaux de causalité. Comme candidats pour ces réseaux de causalité, le scientifique se penche sur les corrélations, dont seulement un petit sous-ensemble de corrélations peuvent être promues au rang de causalités. Cette causalité est ensuite actée lorsque qu'un modèle peut relier les deux termes de la relation de cause à effet.
Le modèle théorique validé permet ensuite de relier de façon systématique des données en input à la prédiction de conséquences futures. En guise d'exemple d'actualité, on peut considérer les modèles climatiques, qui à une certaine quantité de gaz à effet de serre font correspondre, entre autres, une température moyenne. 

 Anderson rétorque que le big-data -- et il disait cela en 2008, avant même l'avènement des LLM (Large language model) -- va permettre sous peu de relier les inputs aux conséquences sans nécessiter la médiation d'un modèle, et ce, promettait-il, dans toutes les branches de la science. Un réseau de pures corrélations, sans causalités attestées
\begin{quote}
 \textit{Il existe désormais une meilleure approche.
Les pétaoctets nous permettent de dire: La
corrélation suffit. Nous pouvons cesser de
rechercher des modèles. Nous pouvons analyser
les data en nous passant d’hypothèses
sur ce qu’ils pourraient montrer. Nous pouvons
charger les chiffres dans les plus grandes
grappes de serveurs que le monde ait jamais
vues et laisser les algorithmes statistiques trouver
des configurations
là où la science en est bien
incapable.}
\end{quote}

Les ordinateurs verront les corrélations qu'aucun scientifique, qu'aucun humain ne saurait percevoir. Traduisant cela dans le language des LLMs modernes, on peut dire qu'entraînés sur des quantités de données titanesques, les LLMs ont pu enregistrer des corrélations qui échappent à la théorie. Pire encore, ces corrélations ne pourront jamais être exprimées dans un langage compréhensible par les humains. A ce jour, on n'a pas pu décrypter les raisonnements sous-jacents aux coups gagnants de alpha-go, LLM qui est venu à bout du champion du monde de go, ou alpha-zero, la machine championne en titre d'échecs. Des coups purement \textit{inhumains}, car hors de la pensée analytique de l'homme. De cette observation, Anderson extrapole que la puissance de calcul des ordinateurs, secondés par les réseaux de neurones, permettront sous-peu de se passer de la formulation de modèle, et de se baser uniquement sur les corrélations infinitésimales détectées par le réseaux de neurones. 

Dans la section suivante nous défendrons que comprendre le modèle comme seul trait d'union entre une cause et un effet est une façon très réductrice et en réalité assez malavisée de décrire le statut du modèle. Nous offrirons une perspective plus complète et plus scientifique du modèle en expliquant en quoi il ne peut être facilement remplacé par des seules corrélations. Ce développement sera utile en ce qu'il illustre déjà les différences cruciales entre ce que nous appellerons, dans les chapitres suivants, une captation algorithmique du savoir et un savoir néguantropique.

\subsection{Rôle de la sciences, construction de l'heuristique et du modèle}

Nous avons présenté ci-dessus le récent espoir tout technocratique de l'éviction des modèles de la méthode scientifique, laquelle signifierait probablement l'éviction de la science en tant que telle. Le modèle, scientifique ou intellectuel y était présenté comme une médiation de moins en moins  nécessaire entre la relation de corrélation et celle de causalité. Nous allons maintenant argumenter qu'une telle conception de l'idée de modèle relève d'une épistémologie erronée autant que d'une définition de la science réductrice, laquelle se trouve toute résumée dans la phrase ressassée et tout aussi réductrice ``tous les modèles sont mauvais, mais certains sont utiles". A supposer même que la taille des banques de données nous permettent effectivement de nous passer des modèles pour produire des descriptions et des prédictions efficaces et précises sur le monde, ce postulat perd de vue le but du modèle scientifique, car il confond finalité et moyen. 

Dans la construction scientifique, loin d'être un moyen pour la description du monde, comme semble le supposer Anderson, le modèle est une finalité, la constitution de celui-ci est le terminus de la recherche scientifique, qui n'est actée que par la comparaison avec les données empiriques, lesquelles ne sont ici que le moyen de la confirmation ou de l'infirmation du modèle. Après cette constitution, ce modèle scientifique devient pour le scientifique le porteur d'une vision du monde, d'un horizon de ce qui est possible, probable, envisageable. Nommons ce résultat de la science, non plus un modèle, mais une heuristique. Ainsi l'activité scientifique ne cherche pas seulement à construire des modèles reliant des données en entrées à des données en sortie, mais construit un champ nettement plus large d'intuitions de ces interactions et de la façon dont elles communiquent. Ce champ d'intuitions s'inscrit lui-même en bordure de la science elle-même, puisqu'il ne peut pas vraiment s'exprimer sous la forme du langage mathématique dans lequel la science s'exprime habituellement. Il s'agit d'un savoir pratique construit sur un savoir théorique, en ce sens, celui-ci produit des potentialités.

Pour rendre ces considérations plus palpables, prenons l'exemple de la théorie de l'évolution par la sélection naturelle, que l'on pourrait taxer de modèle de l'évolution du vivant. Celle-ci établit des lois générales que nous connaissons bien et qui à une situation en entrée -- telle et telle espèce, pourvue de tels et tels caractères, plongée dans tel et tel environnement -- nous prédit, via le modèle, avec plus ou moins de succès, l'évolution des populations de chaque espèce.  A ceci Anderson rétorque que l'exploitation de large corpus de données pourrait prédire l'évolution des populations sans jamais édicter aucune loi générique de l'évolution, ni par la sélection, ni par aucun autre mécanisme. 

Pourtant il est clair que le but, la finalité de la recherche en biologie n'est pas, ou seulement très marginalement, de prédire l'évolution de populations d'espèces dans tel ou tel milieu. Le but de la démarche scientifique en biologie est construire une heuristique de la nature à partir des lignes directrices actées par la simplicité du modèle. Cette heuristique du monde naturel peut être fausse, malavisée, détournée, comme nous l'a montré l'interprétation de la sélection naturelle comme compétition de tous contre tous défendue par Galton et Spencer. Cette heuristique peut tout à fait être renversée par la suite, par exemple par Albert Jaquard dans son ``éloge de la différence", qui met en avant l'importance de la coopération dans les systèmes vivants ou par Olivier Hamant qui, dans son ouvrage ``de l'incohérence", met en évidence la robustesse et la résilience des espèces, par leur manque de performance.  

En tous les cas, il est tout à fait évident que l'heuristique du modèle dépasse de loin la simple corrélation entre entrées et sorties. Elle possède des possibilités d'ouverture, car elle peut être interprétée, et même déformée, ou mal-comprise. Pour l'exprimer dans les termes de Stiegler, ces ouvertures portent le noms de ``bifurcations", et elles sont ce qui portent la capacité à la néguanthropie, que nous expliciterons en détails plus bas. En effet, ces interprétations contiennent ensuite des implications politiques, pour le meilleur ou pour le pire. La théorie de la nature pouvant servir de modèle à la théorie des sociétés, Spencer appelait à appliquer la sélection naturelle plus durement sur les humains, alors que Hamant appelle maintenant les sociétés humaines à augmenter leur robustesse en créant de la redondance dans les chaînes d'approvisionnement.

Prenons un autre exemple sans doute plus constructif, celui des huit règles pour établir des communs du prix Nobel d'économie Elinor Ostrom. Le but de ces règles est de formaliser les différentes conditions qui doivent être réunies pour que l'exploitation \textit{en commun} d'une ressource épuisable -- les arbres d'une forêt, les poissons d'un ruisseau -- puisse atteindre un équilibre entre la ressource en commun et l'exploitant ou les multiples exploitants. On peut comprendre ces huit règles comme un modèle du commun. Les huit règles énoncées par Ostrom sont: 1. des limites nettement définies des ressources et des individus qui y ont accès, 2. des règles bien adaptées aux besoins et conditions locales et conformes aux objectifs des individus rassemblés, 3. un système permettant aux individus de participer régulièrement à la définition et à la modification des règles, 4. une gouvernance effective et redevable à la communauté vis-à-vis des appropriateurs, 5. un système gradué de sanction pour des appropriations de ressources qui violent les règles de la communauté, 6. un système peu coûteux de résolution des conflits, 7. une auto détermination reconnue des autorité extérieures, 8. S'il y a lieu, une organisation à plusieurs niveaux de projet qui prend toujours pour base ces bassins de ressources communes.

Ces huit règles sont le résultat d'une recherche portant sur la description du monde, recherche de laquelle elles émergent comme des résultats robustes. Ces principes, issus d’observations empiriques menées dans des contextes variés à travers le monde, constituent une alternative théorique solide à la thèse de la tragédie des communs (Thèse de Garrett Hardin), selon laquelle les individus en présence d'une ressource commune épuisable agiraient inévitablement de manière égoïste au détriment de l’intérêt collectif.

De plus, ces règles nous donnent un exemple de condition initiales dans lesquelles un commun pourra perdurer. A partir d'une série de conditions initiales,   il est très possible que l'algorithmique couplée à une grande quantité de données puisse prédire sa stabilité future; par le jeu des corrélations. Seulement il est encore peu probable que la machine soit capable de formaliser ces huit principle. Et, de plus, ces huit règles ne se présentent pas seulement comme une description de ce à quoi un commun peut ressembler dans le monde, mais inévitablement elles sont aussi un guide pour en créer de nouveaux. C'est ce que je voulais désigner par le terme d'heuristique.  Et c'est ce type d'heuristique que l'algorithmique, même couplée à un grand nombre de données, n'est pas encore capable de construire. En ce sens, les lois de communs ont un impact sur la politique et les institutions bien au delà de leur capacité à décrire le monde. Comment nous venons de le voir, l'heuristique construite sur un modèle n'est pas totalement déterminée par celui-ci. La formulation des lois de l'évolution, ou des communs, s'ouvre sur une heuristique du monde naturel, laquelle à un impact sur nos interactions effective avec lui, mais elle laisse aussi une large place à l'interprétation.

L'intérêt premier du modèle est qu'il puisse être formulé dans un language concis et intelligible tout en gardant une puissance prédictive. Pour pouvoir constituer une heuristique intelligible, le modèle doit étrangler les données en quelques formules concises écrites dans un langage intelligible. C'est cette formulation même qui est porteuse de l'intelligibilité. 
  L'exemple le plus frappant de cet étranglement du réel dans l'étroitesse des modèles est celui de la physique, avec comme exemple central la relativité générale.

  \textit{In fine}, l'observation la plus étonnante que l'on puisse tirer des théories de la physique, et sans doute la plus heuristiquement enrichissante, est le fait que le réel se laisse pour une grande part étrangler dans l'étroitesse des modèles. Ainsi, abandonner la recherche de modèle nous fait perdre la connaissance que ceux-ci existent et sont formulables. Elle constitue également ce que nous appelerons plus loin la \textit{captation} des savoirs.

\subsection{Fin des modèles, fin du progrès et accélération de l'innovation}
\label{fin_modeles}

J'aimerais maintenant argumenter que l'accélération du temps, qui s'accompagne de la chute des grands récits et du recul de l'idée de progrès, est liée à la désagrégation prédite des modèles.

Comme nous l'avons vu au dessus, le terme d'heuristique est préférable à celui de modèle d'abord parce que l'heuristique inclut également l'intuition et l'expérience qui accompagnent la connaissance des modèles. L'importance de ces deux composantes pour le scientifique se retrouve dans l'usage abondant de la croyance \textit{a priori}, de ce que l'on appelle un \textit{prior}, qui est la plausibilité que l'on attribue à un modèle avant même de le tester. Cette croyance à priori détermine les directions que prennent les recherches.  Nous voyons donc que ce que nous perdons en mettant à distance les modèles: la direction. Cette direction est ce que nous entendonc par l'idée de \textit{programme} politique ou de programme de recherche et se distingue strictement du \textit{projet} politique ou de recherche.  Ce programme est le levier par lequel nous pouvons espérer imposer une direction au cours des choses.

Un autre type de modèle qu'Anderson avait plus que certainement en tête lors de l'écriture de son article est le modèle sociologique. Basé souvent sur quelques principes sociologiques extrapolés et assisté par deux ou trois idées psychologiques, le modèle sociologique amène, comme les modèles d'évolution de l'univers, à des prédictions sur le futur, et contient souvent en germe une façon d'orienter cette évolution. Un exemple est le Marxisme, qui par les sombres rouages de la dialectique de l'histoire, a fameusement prédit l'effondrement du capitalisme sous son propre poids. Le recul de l'intérêt en la modélisation du monde, en sa simplification pour le rendre intelligible par l'intellect humain, se lie à un retrait des grandes idées totalisantes comme le Marxisme. On ne peut que difficilement ne pas voir dans cette éviction des modèles une tentative d'éviction des outils pour orienter l'évolution des sociétés humaines.

En même temps que la croyance en la fin des modèles, comme outil utile pour la compréhension du monde, se répand dans le monde des technocrates.  Celle-ci s'accompagne de l'effritement d'une autre idée, celle du progrès. Le progrès, en tant qu'il est dirigé \textit{vers un but}, requiert de se représenter le futur, de se former une représentation de ce que pourrait être le futur et de modéliser la trajectoire pour y aboutir. Il faut alors modéliser des trajectoires probables du monde. Quelques exemples paradigmatiques et de grande envergure de ces modélisations de trajectoire sont les modèles du climat développés par les nombreux climatologues ou, à une échelle plus modeste, les modèles d'évolution de l'univers théorisés par les cosmologistes. Dans le contexte de la société, le Marxisme a fait office de tel modèle d'évolution. Avec l'effritement de ces modèles d'évolution, l'idée de progrès, alors aveuglée, ne peut manquer de s'effriter également. Cet effacement s'accompagne d'une disparition du terme de progrès du langage public. Tagguief dans ``Faillite du progrès, éclipse de l’avenir", documente la lente disparition de l'idée de progrès, ainsi que la façon dont celui-ci s'efface du language public et médiatique, à tel point que celui-ci n'est presque plus utilisé aujourd'hui. Pour Etienne Klein, philosophe des sciences médiatique, l'idée de progrès est infiniment liée à l'idée de futur. Dans son article ``Progrès et innovation : quels liens ?"\cite{Klein}, il explique
\begin{quote}
\textit{Vivre implique d’accorder à l’avenir
un certain statut, ce qui suppose de l’investir avec des idées, des projets,
des représentations, des désirs. N’est-ce pas là l’essence même
de l’idée de progrès ?
}
\end{quote}

Le médium par lequel nous avons investi l'avenir, c'est le modèle. Le modèle, une fois chargé d'un poids normatif, devient une boussole, nous permettant de penser le progrès. Ce sont tous ces engrenages qui se sont aujourd'hui grippés, l'un après l'autre. Ils se sont grippés en silence, car ils ont été remplacés par une sorte de mécanique différente, celle de l'innovation.  

On peut voir l'actualité de cette position en citant la charte de l'union européenne sur l'innovation, nommée sobrement l'\textit{union de l'innovation}. Pour le goût de la mise en abîme, citons donc Etienne Klein citant cette charte\cite{Klein}, 
\begin{quote}
\textit{En 2010, la Commission européenne s’est fixé l’objectif
de développer une Union de l’innovation à l’horizon 2020. Le
document de référence commence par ces lignes :`` La compétitivité,
l’emploi et le niveau de vie du continent européen dépendent
essentiellement de sa capacité à promouvoir l’innovation, qui est
également le meilleur moyen dont nous disposions pour résoudre
les principaux problèmes auxquels nous sommes confrontés et qui,
chaque jour, se posent de manière plus aiguë, qu’il s’agisse du changement
climatique, de la pénurie d’énergie et de la raréfaction des
ressources, de la santé ou du vieillissement de la population." }
\end{quote}

Vu par la perspective du progrès, le temps est chargé d'un espoir en un avenir meilleur, aussi bien techniquement, que socialement et qu'éthiquement. Par contre, en relisant les phrases de la charte de l'union de l'innovation, il apparaît nettement que l'innovation se présente comme une force de maintien d'un ordre vacillant, un ordre que le temps ne cesse de mettre à l'épreuve, par une force\footnote{Le terme de force est ici un peu imprécis. Il n'existe pas de force entropique, cependant l'augmentation de l'entropie marque une tendance qui l'on peut mesurer au travers de l'entropie. Le terme de force se doit donc d'être compris comme une tendance, ou un pente. } de dissolution que nous nommerons plus bas entropie (et anthropie) et contre laquelle l'innovation doit constamment lutter. On voit donc que l'ontologie de l'innovation et celle du progrès ont un rapport au temps absolument contraire. Là où le progrès parie sur un temps constructeur, mis en mouvement par la force idéologique des hommes, l'innovation y voit une force entropique et destructrice, à laquelle on ne peut résister qu'au travers une innovation de plus en plus frénétique. Pour rester sur le même point, l'humanité doit continuer d'accélérer, de produire toujours plus d'innovations, non pour avancer, mais pour ne plus reculer. Reprenons une phrase particulièrement efficace de l'article de Klein, 

\begin{quote}
\textit{Le progrès s'appuie sur un temps qui fait et l'innovation sur un temps qui défait. }
\end{quote}

La fin de la théorie et du progrès, qu'ils soient espérés,  actés, ou en puissance portent en eux le retournement de notre rapport au temps que nous observons tous aujourd'hui. Notre temps a vu la disparition des utopies, qui ont été remplacées par une crainte vague du futur, l'extinction du progrès, remplacé par l'innovation, et finalement l'assaut actuel sur la modélisation du monde, remplacée par le big-data. C'est cette tendance de fond que Stiegler, comme nous allons le voir, appelle \textit{l'absence d'époque}.  

\subsection{La captation des savoirs techniques}
\label{sec:capt_sav}

La théorie de la fin des modèles relève d'un effet plus profond que celui que nous avons présenté jusqu'à maintenant. Pour décrire la notion de modèle et leur impact dans l'économie de la pensée nous avons introduit le terme d'heuristique, notamment pour souligner que ceux-ci constituent un savoir qui dépasse largement la simple connexion entre des inputs et des outputs. Ainsi, absorber les capacités des modèles dans des machines de traitements du big-data, et les laisser agir aveuglement est sans doute l'exemple le plus transparent de ce que Bernard Stieger entend par, premièrement, \textit{destruction} du savoir, dans le sens où le savoir, une fois prisonnier du big-data, n'est plus accessible aux communautés et aux humains en général, et deuxièmement,  \textit{captation} de ce savoir par l'appareil de production capitaliste. 

Dans la conférence à l'adresse\footnote{https://www.youtube.com/watch?v=pfJnDTPRhIk}, il présente le mécanisme de ce processus de la façon suivante: en dissolvant les modèles conçus par des humains, et en les remplaçant par des algorithmes qui repèrent les corrélations et qui lui sont incompréhensibles, les savoirs qui sont associés à ces modèles deviennent inaccessibles à l'humain. Ces nouveaux types de savoir sont alors mûrs pour être utilisés à grande échelle par des firmes privées ou des organismes publics. Ils ont été captés, et détournés.

Cette captation sera l'un des fil rouge guidant ce mémoire et se manifestera sous différentes formes, notamment celle des savoirs et des désirs. Nous verrons comment cette captation est à l'origine d'une forme nouvelle de prolétarisation, prolétarisation touchant même les métiers les plus techniques. Il sera donc crucial de ne pas perdre de vue que sous des atours souvent d'aide pratique, les robots algorithmiques répondent à un agenda de la captation et du détournement.

\section{Démocratie et politique numériques}
\label{sec:esth_pol}
On a coutume de dire que la politique est grandement, peut-être majoritairement, affaire de communication. Avec le tournant des technologies de la communication, nous voyons poindre un tournant dans la manière même de faire de la politique. Il est encore difficile de s'assurer que ces tendances se confirmeront dans le temps long, pourtant il reste crucial de les relever.

\subsection{L'esthétisation de la politique}

La communication autour de la politique n'a pas attendu le développement et l'extension de l'internet pour opérer une mutation, elle a en réalité commencé dès la télévision et a suivi le mouvement que l'on appelle avec Adorno et Horkeihmer l'esthétisation du capitalisme, jusqu'à devenir le théâtre que l'on connait aujourd'hui, théâtre qui semble recouvrir tout ce que l'on peut voir de la politique.
Avant de revenir dans le chapitre suivant sur Adorno et Horkeihmer et l'esthétisation du capitalisme, suivons tout d'abord les lignes de ``plaire et toucher; essai sur la société de la séduction" de Gilles Lipovetsky\cite{Plaire}. 

Gilles Lipovetsky analyse comment la politique contemporaine s’est transformée sous l’influence des logiques de séduction issues de la société de consommation. Il explique que la politique ne repose plus sur de grands récits idéologiques ou des engagements rationnels, mais sur l’image, l’émotion et la communication. Le mythe de la ``conquête" de l'électorat par le candidat s'est calqué sur le mythe de la ``conquête" de la femme par l'homme. Les candidats leaders doivent séduire l’opinion publique, adopter un style personnel, charismatique, voire ``marketing", et susciter l’adhésion en jouant sur l’affect plutôt que sur les idées. Dans ce contexte de l'omniprésence de spectacle, la communication politique devient spectacle, storytelling, mise en scène de soi, orientée vers la conquête de l’attention et de la sympathie. Mesurons le changement que cela implique; le politicien ne se réfère plus dans son discours à un ensemble plus grand que lui, plus vaste et plus important, dont il serait le porteur et l'émissaire, il se met lui-même au centre de son propre discours et devient une sorte de vedette dont le discours est un monologue télévisé.

Lipovetsky y voit une ``esthétisation" du politique, qui favorise la personnalisation du pouvoir au détriment des partis. L’électeur, devenu consommateur d’images et d’émotions, choisit davantage en fonction de l’attrait que du programme politique. Cette observation est en accord avec le recul observé et décrit plus haut, dans la section \ref{sec:Eco}, des programmes scientifiques et politiques. La politique se ``déréalise", elle s’apparente à un jeu de séduction permanent. Un exemple typique de cette théâtralisation se rencontre avec le président français Emmanuel Macron publiant des vidéos TikTok ``pour parler aux jeunes".  Cette évolution marque, selon lui, un tournant nouveau pour la démocratie, plus fluide, mais aussi plus instable et superficielle. Pour prendre un peu d'avance sur les analyses que nous adapterons de Fredric Jameson, on peut voir dans cette esthétisation de la politique l'effet général du tournant postmoderne de la culture globalisée. 

Pour l'exemple, considérons le rôle du débat dans l'économie politique moderne.

\paragraph{Le débat}
L'inflation de l'importance du débat politique est l'un des exemples les plus spectaculaires de l'effet de la démocratisation des médias sur la fonctionnement des sociétés démocratiques. Le débat, que l'on imagine, encadre et présente comme un pugilat à gagner plus par l'éloquence et la bagout que par l'argumentation et la justesse des prises de position, est l'un des moments les plus importants de la campagne des candidats. C'est par le débat que l'on appelle les candidats à se départager, par un duel en face à face à la télévision, duel modéré par un intermédiaire en la personne du journaliste censé diriger et orienter les questions. Le traditionnel débat de la présentielle française en est un exemple  marquant. C'est par le sondage ensuite des auditeurs que l'on décide du vainqueur du débat. Mais sur ce qui est de la grille d'évaluation, on reste toujours dans le flou. Sur quels critères l'auditeur se base-t-il dans son jugement ? On est bien en mal de remonter tous les ressorts. Les critères d'évaluations des duellistes sont aussi flous que les règles du jeu sont claires. Le duel se fait par échanges de slogans interposés où chaque intervention doit tenir dans un lapse de temps de l'ordre de la minute, le message doit être clair et plus percurtant qu'informant, les convinctions de l'intervenant doivent être clairement formulées, en évitant tout développement superflu. Seuls quelques sujets médiatiques, dûment listés au départ, sont abordés. Le débat en entier durera aux alentours d'une heure, la durée d'un film à rebondissements auquel l'auditeur est habitué. 

L'exercice du débat est l'examen du politicien, sa forme en est aussi fixe que l'exercice de la dissertation. Les similarités avec les jeux romains et ses gladiateurs dont l'on jugeait la performance par la voix des spectateurs,  sont confondantes. Ce type de formule est le terreau fertile du culte de la personnalité qui a envahi la politique.

\paragraph{Une ère post-débat par la révolution d'internet}

Pourtant le débat, avant d'avoir connu la critique qu'il méritait, s'est aussitôt métamorphosé suite à l'arrivée d'internet, et avec lui les façons d'approcher la politique et les échanges d'opinion. Là où la pratique du débat régnait sur les chaînes d'information et sous le contrôle de celles-ci, et sur lequel elles imposaient l'étroit étau qui les caractérisaient; contraintes de temps, communication par le slogan, opposition binaire des points de vue, l'internet aura en quelques années seulement fait exploser cette pratique. 

Qu'en est-il aujourd'hui? Quelle est la forme du débat en 2025 ? Il est devenu pluriel. Tout d'abord, répondant à l'étau souvent vécu comme oppressant et réducteur de l'exercice du débat, l'internet a multiplié les formes de l'interview, offrant souvent des supports d'interview plus long, avec des prises de paroles permettant un développement en profondeur des idées de chaque intervenant. Dans ce contexte, on peut citer à titre d'exemple les plateformes d'Elucid, Blast et Thinkerview. Cette nouvelle organisation de l'échange politique se présente souvent comme subversif, notamment en opposant une information basée sur la profondeur et la lenteur de la réflexion avec la pratique des slogans et la rapidité de traitement présentes dans les médias de plus grande écoute. Ces chaînes francophones, toutes essentiellement basées sur Youtube, totalisent pour certaines un million d'abonnés ou plus, 1.26 pour Thinkerview, 1.44 pour Blast, 0.33 pour Elucid. Elles donnent la parole à des personnalités moins en vue, proposent des vidéos plus longues, avec des interviews allant jusqu'à deux ou trois heures. L'importance de ces médias étant telle que des figures politiques comme que Mélenchon\footnote{https://www.youtube.com/watch?v=FzzQgJntbqQ} se sont livrées à l'exercice du petit fauteuil plongé dans l'obscurité de Thinkerview, tout en laissant la parole pour de longues heures à des personnalités beaucoup plus controversées ou marginales; l'ambassadeur d'Iran au lendemain de l'attaque Israélienne\footnote{https://www.youtube.com/watch?v=byA8gKMazs4}, les influenceurs de DATAGUEULE \footnote{https://www.youtube.com/watch?v=507HStKtw-I}, et même Bernard Stiegler \footnote{https://www.youtube.com/watch?v=YDT5f5sQSGA}. 

A côté de ce format nettement plus long, les plateformes telles que Twitter, aujourd'hui X, offrent aux politiciens un moyen de communiquer qui est direct, efficace, instantané et contournant le filtre traditionnel des grands médias. Ce type de communication correspondant aujourd'hui essentiellement à des textes relativement courts, ainsi que d'une communication faite le plus souvent de slogans. Là où le débat se voulait une lutte entre différentes positions idéologiques, théâtralisées et codifiées, ce nouveau moyen de communcation évince toute contradiction, tout contrepoids. Un exemple récent de ce type de communication est celle du président américain Donald Trump, dont la communication se base largement sur ses publications sur X et Truth Social et par des conférences de presse avec des journalistes triés sur le volet. Il refuse aussi systématiquement de répondre aux questions qu'il estime gênantes et les balaie.  

Avec le recul de l'importance du débat, on peut voir une diversification et une décentralisation de la communication politique. Avec cette décentralisation, les grands médias, ceux que l'on appelle le quatrième pouvoir, perdent une large partie de leur influence tout en se désintitutionnalisant. Avec la démocratisation de l'information, l'institution des médias s'effritent également, laissant place à un paysage beaucoup plus contrasté et conflictuel.

\subsection{Le rôle et le statut de la manipulation médiatique}

Cette révolution dans la production et la distribution des contenus d'opinions, autrefois le monopole de quelques médias, ne saurait laisser la structure des démocraties intactes. Quelques hommes politiques ont commencer à montrer comment ceux-ci pouvaient contourner les méthodes d'annonces traditionnelles, telle que la conférence de presse, en utilisant \textit{immédiatement}, sans la médiation des médias, des annonces via Twitter, X, TruthSocial ... Ceux-ci ont alors capitalisé sur l'effet d'annonce nouveau engendré par ce nouveau format court.

Evidemment, la manipulation de l'opinion publique ne date pas de l'essor des outils de communication numérique et est aussi ancienne que la communication et la politique elle-même. Laissons volontairement de côté les manipulations qui peuvent avoir lieu dans les régimes autoritaires pour nous concentrer sur celles qui ont eu lieu au sein même des démocraties. L'exemple paradigmatique de telle manipulation médiatique est celui de l'entrée en guerre des US, dans un contexte où la population américaine était pourtant largement opposée à une entrée en guerre. Après une propagande à grande échelle menée par Edward Louis Bernays, et racontée dans son ouvrage ``Propagande", il parvient avec brio à retourner l'opinion publique en faveur de l'entrée en guerre. Cet épisode est par la suite systématisé et théorisé par Walter Lippmann dans son ouvrage ``Public Opinion", que l'on pourrait presque voir comme le ``Prince" de Machiavel des temps modernes. Son argumentaire se base sur l'observation que la structure des sociétés implique que les individus, pour des raisons aussi bien techniques que cognitives, ne peuvent pas être aussi bien informés que les experts travaillant pour les gouvernements. Par conséquent, seuls ces derniers sont à même de produire des décisions éclairées et sont en droit de le faire. Lippmann propose donc que de manière systématique, les gouvernements exercent dans chaque domaine leur expertise et trouvent pour chaque problématique une solution. Cette solution choisie au préalable par le gouvernement sera ensuite \textit{suggérée} à l'opinion publique via la propagande des médias, d'une manière assez subtile pour que l'apparence du choix populaire demeure. Ce processus est ce que Lippmann baptise ``la manufacture du consentement", processus que Chomsky repère et retrace (voir son ouvrage``Manufactring consent") dans toutes le strates des médias et dans leur manière de transmettre l'information.  Ainsi, pour Lippmann, la démocratie ne peut donc être qu'une apparence -- même si c'est une \textit{apparence de la plus haute importance pour la stabilité de la société}. Cette manufacture du consentement est une manipulation à grande échelle et largement structurée par les décisions de quelques uns.

Cependant la manipulation de l'opinion publique a pris une toute nouvelle forme lors de l'essor des moyens de communication numériques, en permettant une propagande ciblée et personnalisée pour chaque individu. Un exemple moderne est celui de Cambridge Analytica. Reprenons rapidement les événements qui ont entouré l'incident. Cambridge Analytica a illégalement recueilli les données de plus de 87 millions d'utilisateurs Facebook via une application de quiz, "This Is Your Digital Life". Ces données ont été utilisées pour établir des profils psychologiques sans le consentement des utilisateurs. L’entreprise s’en est servie pour cibler des électeurs avec des publicités politiques personnalisées, notamment lors de la campagne de Trump en 2016 et du Brexit. Le scandale a éclaté en 2018 grâce au témoignage du lanceur d’alerte Christopher Wylie. Dans ce cas-ci, par la collection de données, la manipulation, qui repose sur les mêmes mécanismes de surexposition à certaines informations déformées, devient personnalisée.

L'une des conditions à la constitution et à la survie d'une démocratie réside dans un armement des individus à la manipulation à grande échelle. C'est par ces arguments, d'éducation et d'enseignement, que l'on a répondu au postulat de Lippmann selon lequel les populations ne sont pas capables de former des opinions éclairées. Cependant, avec la nouvelle méthode d'influence portée par les outils numérique, on peut se demander comment on peut encore espérer que le citoyen décide en son âme et conscience.

\chapter{Captation de la libido par le marché}
\label{chap2}

La longévité et l'efficacité du système capitaliste à survivre à ses concurrents est une question hautement discutée par la sociologie et la philosophie. D'une part, l'intuition et le sentiment d'une inadéquation du système capitaliste au monde et à son épanouissement, aujourd'hui exemplarisé par son impuissance face à la crise climatique, est un sentiment largement partagé. D'autre part, le système capitaliste se maintient et a survécu à toutes les formes de critique qui lui ont été adressées. Plus encore, il semble toujours s'être nourri de cette critique et en être ressorti plus fort.

Une réponse partielle à ce dilemme a été proposée par les sociologues Boltanski et Chiapello, notamment dans leur ouvrage ``Le Nouvel Esprit du capitalisme"\cite{Boltan}. Pour ces auteurs, c'est la capacité du système capitalistique à absorber sa critique et à l'incorporer, et par là à changer partiellement de forme, qui a fait toute la force de ce système sur plusieurs siècles. Dans ``Le Nouvel Esprit du capitalisme", Boltanski et Chiapello \cite{Boltan} identifient trois grands stades du capitalisme, chacun associé à un esprit particulier, ainsi qu'à une forme de la production et une forme de la consommation. Le premier, le capitalisme bourgeois, domine jusqu’au début du XXème siècle: il repose sur la figure du patron propriétaire et la discipline des travailleurs. Il glorifie aussi l’épargne des petites gens et la morale du travail, ainsi que la notion de mérite. Ainsi, il ressemble à ce que l'on pourrait appeler aujourd'hui un capitalisme responsable. La précarité du travail, dont beaucoup de postes sont des journaliers ou des saisonniers est vigoureusement critiquée. Le deuxième, le capitalisme managérial ou fordiste, s’accompagne de l’industrialisation de masse. Dans ce contexte, l’entreprise devient hiérarchisée, l'emploi des travailleurs est stabilisé et sécurisé, en échange d’une soumission des travailleurs à l’organisation managériale. Ce modèle est hautement critiqué par l'exploitation qu'il fait des travailleurs en leur donnant un travail pénible et répétitif, ennuyeux à souhait. Cette critique opposait notamment la figure de l'ouvrier spécialisé sur un des rouages de la vaste chaîne de montage et étranger au fruit global de son travail, à la figure de l'artisan, capable de construire un produit par ses multiples aptitudes, et attaché à des valeurs de créativité, d'authenticité et de flexibilité. Le troisième stade, le capitalisme connexionniste, émerge dans les années 1980, et s'étend jusqu'à aujourd'hui. Il voit l'émergence des cadres et des cols blancs et correspond à l'extension énorme du secteur tertaire. Du point de vue moral, il valorise la flexibilité, l’autonomie, la mobilité et la créativité. L’entreprise devient un espace de projets temporaires, sans garantie ni attachement durable. Ce nouveau modèle récupère les critiques artistiques du fordisme (contre l’ennui et la hiérarchie) pour les transformer en outils de performance.
    
    Tout au cours de cette évolution, le système du capitalisme change ainsi de visage pour mieux se maintenir, en adaptant ses justifications aux attentes sociales de chaque époque et en introduisant tout un répertoire d'expressions nouvelles, une novlangue, pour soutenir le processus de transformation. Pourtant à chaque étape, il étend sa sphère d'influence en incorporant en son sein des pans entiers de la critique. Par exemple, lors du tournant connexionniste du capitalisme, le système absorbe la critique opposant l'ouvrier à l'artisan pour forger l'image de l'auto-entrepreneur, porteur et défenseur de son projet créatif et capable d'apprendre toutes sortes de tâches. Figure emblématique de l'individualisme triomphant, l'auto-entrepreneur et la start-up disruptrice forment aujourd'hui le symbôle principal du néolibéralisme. Au sein de cette nouvelle forme de la production, les communautés humaines ne sont que des groupes momentanément homogènes entre lesquels l'auto-entrepreneur doit sauter, au gré de ses besoins.  Il fait beaucoup de choses, mais jamais rien en profondeur, il appartient à beaucoup de groupes, mais n'a pas de liens forts. Il ne jure que par son ``réseau de connaissances". 
    
    L'émergence de cette figure peut également se lire sous le prisme culturel, par ce qu'on appelle la mutation postmoderne de la culture, sur laquelle nous reviendrons dans un chapitre prochain. L'auto-entrepreuneur est un atome individuel, une illusion de maneouvrabilité insérée dans un système hiérarchique devenu invisible mais médié plus que jamais par le capital. Car tout l'effort de l'auto-entrepreneur est tourné vers l'obtention et le maintien d'un capital, relationnel et culturel évidemment, mais avant tout financier. Gardons également à l'esprit qu'ici déjà nous avons en germe l'interaction entre l'individu solitaire et le réseau tentaculaire, lequel n'est encore ici que son ``réseau de connaissances", mais celui-ci ne tend qu'à s'élargir. 

Face à cette figure de l'auto-entrepreneur start-upeur individualiste et réseauté, certains ont pu voir en l'internet naissant une contre-culture. On percevait alors l'internet comme un nouveau terrain fertile à l'expérimentation de l'économie participative, et donc comme plateforme de discussions et d'échange d'information au sein de communautés en croissance rapide. L'analyse du capitalisme moderne est compliquée notablement, car, dans les mots de Adorno et Horkheimer, ``la civilisation actuelle confère à tout un air de ressemblance".  Cependant, nous défendons ici que la captation, au cours des années 2000, de l'internet par le marché a alors constitué un démantèlement tout aussi rapide de cet espoir initial. Dans cette section, nous entendons montrer comment l'essor de l'algorithmique a pu tirer profit des mécanismes de l'esprit pour ``disrupter" l'écosystème de l'internet et transformer l'interaction entre individus au sein d'une communauté en une nouvelle sorte d'interaction, à savoir l'interaction entre un individu solitaire et le réseau tentaculaire. Ce détournement de l'interaction entre humains en une interaction entre humain et machine s'est appuyée sur ce que l'on appelera la ``captation de la libido par le marché", et que nous allons maintenant développer.

\section{L'élargissement de l'industrie de la culture}

Pour mieux appréhender l'évolution que nous allons décrire, il faut commencer par remonter dans le temps et reprendre les analyses d'Adorno et Horkheimer sur l'industrie de la culture. 
Historiquement, selon Michel Foucault, la peur primordiale de l'homme était de devenir fou, de perdre la raison. Pour lui, le fait d'être performativement rationnel participait à l'inclusion de l'individu dans la société, alors qu'au contraire, la perte de la raison faisait immédiatement de l'individu un paria, l'excluant de la société.

Pourtant, dans le fameux chapitre de la \textit{dialectique de la raison} publié en 1944 \cite{DiaRai}, \textit{Kulturindustrie: raison et mystification des masses},  Adorno et Horkheimer, tous deux traumatisés par les ravages et les barbaries de la seconde guerre mondiale, avancent que c'est le monde globalement rationnalisé qui perd la raison. Ils défendent en effet que la rationalité devient elle-même irrationnelle quand elle est subordonnée aux logiques de domination, de rendement et de marché. Ainsi, pour les deux philosophes, cette perte de la raison globalisée est la conséquence de la rationnalisation, de l'uniformalisation du monde et de la captation et du détournement par le marché du beau et de l'esthétisme. Elle est aussi la conséquence directe du retournement dialectique de la ``raison des Lumières".  L'analyse de ces auteurs est de la plus haute importance pour deux raisons. Premièrement, parce que l'analyse d'Adorno et de Horkheimer contenait déjà en germe une analyse des internautes actuels qui utilisent les algorithmes et qui sont regardés par ceux-ci dans le but recolter des données\cite{DiaRai}
\begin{quote}
\textit{
Les consommateurs réduits à du matériel statistique sont répartis sur la carte géographique des services d'enquêtes en catégories de revenus, signalés par des zones rouges, vertes et bleues.
 }
\end{quote}
et, ensuite parce qu'elle jette les bases d'une relation entre la consommation culturelle et la santé psychique des consommateurs, que nous permettra d'appuyer l'analyse de Stiegler et son analyse psychique de l'internaute moderne.

Pour eux, la culture moderne des sociétés occidentales, présentée sous la forme de cinéma, musique, télévision, est produite en série selon des logiques commerciales et  vise désormais le profit plutôt que l’émancipation. Ils assènent\cite{DiaRai}
\begin{quote}
\textit{ Le film et la radio n'ont plus besoin de se faire passer pour de l'art. Ils ne sont plus que business : c’est là leur vérité et leur idéologie qu'ils utilisent pour légitimer la camelote qu'ils produisent délibérément. }
\end{quote}

Cette standardisation appauvrit les oeuvres, les rend uniformes et prive le public de la réflexion critique inhérente à l'ancienne production artistique, au nom d'une forme de démocratisme qui les rend tous égaux en médiocrité\cite{DiaRai}
\begin{quote}
\textit{
Démocratique, la radio transforme tous les participants en auditeurs et les soumet autoritairement aux programmes des différentes stations, qui se ressemblent tous.
 }
\end{quote}

Le divertissement devient un outil de contrôle social, qui diffuse les valeurs du capitalisme et empêche toute remise en question. Les individus, exposés à des contenus répétitifs, deviennent passifs, conformistes, et mis en face d'une série d'options toutes standardisées, croient faussement à leur liberté de choix et d'action. L’industrie culturelle crée ainsi une fausse diversité et renforce les mécanismes de domination. La culture n’élève plus : elle endort. Pour nos générations qui sont nées et ont été bercées dans cet environnement, nous ne voyons même pas que l'art s'est éclipsé entièrement de la vie quotidienne. 

L'un des points cruciaux de l'argumentaire d'Adorno et d'Horkheimer qu'il nous faudra garder en tête pour le reste de ce mémoire est le renversement de la justification apportée par les producteurs de cette culture\cite{DiaRai} 
\begin{quote}
\textit{ 
Les parties intéressées expliquent volontiers l’industrie culturelle en termes de technologie. Le fait qu'elle s'adresse à des millions de personnes impose des méthodes de reproduction qui, à leur tour, fournissent en tous lieux des biens standardisés pour satisfaire aux nombreuses demandes identiques.  Les standards de la production sont prétendument basés sur les  besoins des consommateurs: ainsi s'expliquerait la facilité avec laquelle on les accepte.  
 }
\end{quote}

La conséquence de ce processus est que ce n'est plus l'artiste qui décide de l'oeuvre, mais ce sont les contraintes du marché qui la déterminent entièrement. La ``sélection économique"  des biens culturels enferme alors les consommateurs dans sa consommation de l'identique. Là où les producteurs (disons de musique pour fixer les idées), défendent qu'ils produisent le produit que le consommateur leur demande, ils refusent d'admettre que c'est la production même qui façonne la demande et qui enferme le consommateur dans un ``tout identique", qu'ils revendiquent par la suite comme ``ce qui est pourtant voulu par les masses": La qualité d'une oeuvre se réduit alors à sa rentabilité, enfermant de fait à la fois l'artiste créateur et le consommateur et ainsi\cite{DiaRai}
\begin{quote}
\textit{ 
  le cercle de la manipulation et des besoins qui en résultent reserrent de plus en plus les mailles du système. 
 }
\end{quote}

Cette rhétorique \textit{a posteriori} du rejet de la responsabilité sur le consommateur, l'argument usuel ``s'ils l'utilisent c'est qu'ils doivent y trouver leur compte", sera réutilisé, comme nous allons le voir, dans le contexte des réseaux algorithmiques. 

L'une des conséquences de cette standardisation grandissante des contenus culturels réside dans l'effacement des différences culturelles\cite{DiaRai}
\begin{quote}
\textit{ 
Pour le moment, la technologie de l’industrie culturelle n'a abouti qu'à la standardisation et à la production en série, sacrifiant tout ce qui faisait la différence entre la logique de l'œuvre et celle du système social. 
 }
\end{quote}
 
Cet effacement entre le la spécificité du travail artistique et les spécificités du monde marchand est ce que l'on nomme par l'expression d'\textit{esthétisation du capitalisme} et constitue un autre exemple de la capacité du capitalisme à englober et à phagociter les sphères hétérogènes, ici la sphère artistique, dans la sphère marchande, tout en modifiant en profondeur ses réseaux de relation. 

De cette situation, nous pouvons tirer une conclusion, que sans doute Adorno et Horkheimer n'auraient pas adoubée. En standardisant l'offre culturelle, les producteurs ont disrupté les interactions entre l'individu amateur d'art et les communautés créatrices, tout en organisant et favorisant l'interaction entre l'individu-consommateur et la vaste machine productrice. Il faut voir ici une réorganisation des modes d'interaction, qui après une longue phase d'interactions de type individu-individu pour entre dans une phase dominée par l'interaction individu-machine de production. 

Ce nouveau type d'interaction est précurseur de l'interaction que nous voyons aujourd'hui entre l'internaute et le réseau internet tentaculaire, le Web, lequel interpose maintenant globalement la vaste \textit{plateforme} et organise l'interaction individu-plateforme. Creuser cette observation, ainsi que ses conséquences, est le thème principal de ce mémoire. On remarquera que cette évolution dans les domaines professionnel et culturel d'une interaction exclusive entre un individu et un réseau tentaculaire élargit l'importance du marché et  offre la possibilité d'une captation. Cette captation s'exemplifie d'une part avec l'essor des plateformes de mise en liens par exemples des travailleurs et des employeurs, avec des plateformes telles que LinkedIn, de mise en lien des producteurs et des consommateurs, avec une pléthores de domaines, notamment culturels telles que Netflix, Amazon,  ... 
Cette évolution est également similaire à celle du travailleur dont le mode d'interaction auparavant était une relation immédiate entre le travailleur, l'employeur, et le groupe syndical et s'est muée en une relation avec un réseau de compagnies susceptibles d'embaucher le travailleur flexible.

\section{Impact sur la santé psychique}
Nous nous tournons vers le lien qu'Adorno et Horkheimer tissent entre la consommation culturelle et la santé psychique.  Reprenons les analyses d'Adorno et de Horkheimer\cite{DiaRai}
\begin{quote}
\textit{
Ce que la production culturelle impose, c'est de suivre un chemin tout tracé par l'industrie. [...]  Pour le consommateur, il n’y a plus rien à classer : les producteurs ont déjà tout fait pour lui. L'art prosaïque sans rêve destiné au peuple réalise un idéalisme de rêve que l’idéalisme critique a condamné. }
\end{quote}

 Le premier service que l’industrie apporte au client est de tout schématiser pour lui. Ce que l'industrie offre au consommateur, c'est la satisfaction d'un format qu'il reconnaît, qu'il peut facilement classifier et anticiper. Même la quantité d'inattendu est codifiée, et cette codification a pour effet d'atrophier les facultés du consommateur\cite{DiaRai}
\begin{quote}
\textit{
Aujourd'hui, l'imagination et la spontanéité atrophiées des consommateurs de cette culture n'ont plus besoin d’être ramenées d'abord à des mécanismes psychologiques. [...]
Tous les autres films et produits culturels qu'il doit obligatoirement connaître l'ont tellement entraîné à fournir l'effort d'attention requis qu'il le fait automatiquement
.}
\end{quote}

 L'industrie culturelle, après avoir façonné le style  façonne le consommateur\begin{quote}
\textit{
 C'est ainsi que l’industrie culturelle, qui est le plus rigide de tous les styles, apparaît comme l'objectif même du libéralisme auquel on reproche l'absence de style.
 }
\end{quote}

 A cette observation, Adorno et Horkheimer redoublent une conséquence psychologique.  En croisant Freud et Marx, ils observent que la culture de masse, produite par l’industrie culturelle, favorise un retour à une forme de narcissisme régressif chez les individus. L'individu, toujours confronté par l'industrie culturelle à la répétition du même, toujours conforté dans ses croyances, se replie alors sur lui-même, sur ses affects et sur ses propres besoins immédiats. L'industrie culturelle fonctionne par la fausse satisfaction personnelle.  Ainsi, finalement, l'impact de l'industrie culturelle induit une \textit{régression narcissique}. Dans la théorie Freudienne de la formation de la psyché, la régression narcissique est un concept qui désigne un retour à un état psychique centré sur soi-même. Cette régression peut avoir lieu, par exemple, en réaction à une blessure affective, un échec ou un manque de reconnaissance. L’individu se replie sur lui-même pour restaurer une image idéalisée de son moi, souvent au détriment de la relation aux autres et aux parents. Selon Freud, ce mécanisme défensif de l'individu peut entraîner une rupture avec la réalité extérieure et un désinvestissement affectif du monde. Donc, il s’agit d’un retour à une phase plus infantile où l’amour de soi prédomine sur l’investissement des objets extérieurs. A ce stade gardons en tête, le fil rouge du motif du repli sur soi. 
 
Dans la théorisation d'Adorno et d'Horkheimer, il n'y a pas vraiment de blessure qui induit la régression, seulement l'impact répété de la consommation. Cette régression narcissique retourne l'impact du narcissisme primordial freudien, pourtant étape nécessaire du développement,  en outil du marché en enfermant l'individu dans cette étape. Une fois l'individu replié sur lui-même, il se retrouve isolé et le narcissisme peut devenir un outil d’adhésion au système. 

Après avoir enfermé les individus dans leur phase de narcissisme primordial, l'industrie peut librement exacerber les désirs de ceux-ci. Cependant, elle n'entend pas les satisfaire, ni les sublimer, mais seulement en faire usage pour que plus de consommation soit par la suite nécessaire 
 \begin{quote}
\textit{
 C'est là le secret de la sublimation dans l’art: représenter l'accomplissement comme une promesse brisée: L'industrie culturelle ne sublime pas, elle réprime. 
 Les œuvres d’art sont ascétiques et sans
pudeur, l'industrie culturelle est pornographique et prude. }
\end{quote}

Le désir est invoqué sur la mode de l'exacerbation, de l'excitation pure et simple de celui-ci. Pour le consommateur, la torture est ensuite la répression qui suit cette exacerbation. L'industrie propose plusieurs fuites devant cette répression, la première étant plus de consommation. La seconde réside dans le rire, la fuite par l'arrêt de la pensée. L'amusement comme détournement, 
\begin{quote}
\textit{
 La libération promise par l'amusement est la libération du penser en tant que négation. 
  }
\end{quote}
 L'industrie culturelle rappelle encore et toujours au consommateur combien il est vain et coûteux de résister, jusqu'à ce qu'elle soit hégémonique.

\begin{quote}
\textit{
Plus les positions de l’industrie culturelle se renforcent, plus elle peut agir brutalement envers les besoins des consommateurs, les susciter, les orienter, les discipliner, et aller jusqu’à abolir l'amusement: aucune limite n’est plus imposée à un progrès culturel de ce genre.
 }
\end{quote}

Comme nous l'avons vu, pour Adorno et Horkheimer, ce que cette culture standardisée empêche, c'est l'individuation, parce qu'elle offre à l'individu la liberté de se conformer à un modèle commun. Elle repousse tous les individus dans une phase pré-individuelle, le narcissisme primordial, et puis informe les désirs de tous par le même modèle d'individuation. Ils concluent  
\begin{quote}
\textit{
Le principe de l’individualité fut lourd de contradictions dès le départ. L'individuation n'a jamais été réellement réalisée. 
Dans les visages des héros de cinéma ou des personnes privées, qui sont tous confectionnés sur le modèle des couvertures de magazines, une apparence à laquelle nul ne croyait d'ailleurs plus disparaît et la popularité dont jouissent ces modèles se nourrit de la secrète satisfaction éprouvée à l’idée que l'on est enfin dispensé de l'effort à accomplir en vue de l’individuation, parce que l'on n’a plus qu'à imiter, ce qui est beaucoup moins fatigant.
 }
\end{quote}

On peut par exemple penser aux modèles de la virilité et de la féminité de Hollywood, aux idéaux de beauté. Il faut bien comprendre pourtant que ce n'est pas l'instinct d'imitation, celui que certains appellent avec beaucoup de mépris, avec Nietzsche, l'instinct du troupeau, qui est critiqué ici par les deux auteurs. Autrement, on pourrait penser que la critique d'Adorno et Horkheimer s'applique également à la formation des standard de beauté par les statues pendant, disons, l'époque Hélléniste. Ce n'est pas le cas. Ce que ceux-ci critiquent, c'est le fait récent que l'imitation induite par l'offre culturelle oriente les consommateurs vers des modèles standardisés dont la construction même poursuit un but d'optimisation commerciale. Le résultat est une détournement et une captation des désirs individuels, et qui entrelacent ceux-ci avec la consommation.

Cette tendance se comprend facilement lorsque l'on prend quelques exemples particulièrement grossier de manipulations. L'omniprésence de la cigarette chez les héros du cinéma, accompagné de son slogan ``La cigarette, pas pour les enfants" associe l'aura du héros au fait de fumer, culminant dans un ``fétichisme de la cigarette" \cite{articleCiga}. Dans ces cas extrême, l'``héroïcité" du héro s'entremêle étroitement avec la consommation, dans le cas présent de cigarettes.

Nous approfondirons plus bas les conséquences sociales d'une telle captation et essayerons de comprendre dans quel sens cette captation du rôle du modèle par les industries culturelles vient court-circuiter la transmission intergénérationnelle.

Il faut aussi insister sur l'absence, dans les analyses d'Adorno et d'Horkeimer, du rôle de l'appareil technique de captation et de distribution des contenus culturels, lesquels étaient de leurs temps essentiellement la télévision et la radio. Avec la démultiplication spectaculaire des média de partage de contenus culturels et leur nouvel appareil algorithmique, le support de la transmission lui-même devient actif dans cette captation.  L'analyse de Stiegler reprend le thème de la santé mentale dans le contexte de la normalisation algorithmique de la culture, ainsi il reprend la critique de ce qu'il appelle les \textit{industries de l'esprit}.

\subsection{La mondialisation et les captations des rétentions}

Il est clair que cette standardisation de la culture ne peut avoir lieu que par l'intermédiaire d'une révolution de grande ampleur, une révolution de la communication. Cette révolution de la communication permet un échange d'information et de biens efficace partout dans le monde entier, c'est la mondialisation. Dans cette section, nous allons donner une définition plus fine de ce concept de mondialisation de la communication.

Il nous faut commencer par définir le concept de rétention. Les rétentions sont les traces que produisent les événements sur les individus, lesquelles se déclinent en trois types différents de traces\cite{Disrup}: 
\begin{quote}
    \textit{
Les rétentions primaires et secondaires sont des réalités psychiques - les primaires appartenant au temps présent de la perception, les secondaires au temps passé de la mémoire. Les rétention tertiaires sont des rétentions artificielles, non pas psychiques, mais techniques, telles que les  archives, les enregistrements, les reproductions techniques en général.
}
\end{quote}

Ainsi, le troisième type de trace forme un type particulier de traces, qui ne sont pas propres à la psyché des individus, mais appartiennent à la société en entier, car inscrites dans son système technique. L'un des exemples paradigmatiques de telles rétentions tertaires est l'écriture alphabétique, laquelle a formé un système mnémotechnique qui a stabilisé toute la société pendant plus de deux millénaires.

En contrepoint des rétentions, les protentions sont les attentes, sous toutes ses formes, d'un individu dans une société, à savoir les désirs, les attentes, les volontés et les volitions. Ces volitions peuvent être individuelles ou collectives, elles peuvent être stimulées par l'entourage et les parents, par les récits que raconte et se raconte la société, et finalement par l'industrie culturelle elle-même.

Dans le chapitre 4 de ``la technique et le temps; le temps du cinéma et la question du mal-être" \cite{CinemaS}, Stiegler nous propose une explication de l'évolution du système technique, 

\begin{quote}
    \textit{Un système technique ainsi entendu a une aire de diffusion et une durée. L'analyse montre que, avec le temps, son extension est tendanciellement de plus en plus vaste, tandis que sa durée est de plus en plus courte. Il est traversé par des tendances évolutives et entre régulièrement en crise, ce qui induit des ruptures de système. }
\end{quote}

Chez Bernard Stiegler, le système mnémotechnique  désigne l’ensemble des techniques et dispositifs qui permettent l’enregistrement, la conservation et la transmission de la mémoire humaine, ce sont les appareils permettant de créer une rétention tertiaire. Depuis les premières inscriptions écrites jusqu’aux larges machines numériques, ces dispositifs extériorisent la mémoire individuelle, lesquels sont appelés par la suite à devenir une mémoire collective. Cette externalisation est constitutive de l’humain, qui se différencie largement des autres espèces animales par sa dépendance à la technique. Comme nous pouvons l'observer dans la vie de tous les jours, chaque nouveau support mnémotechnique (écriture, imprimerie, audiovisuel, numérique) transforme notre rapport au temps, au savoir et à la culture. Un exemple très récent est l'essor, en une petite vingtaine d'années, du téléphone portable, dont tout le monde possède dans nos pays occidentaux un exemplaire. Celui-ci nous connecte en temps réel à l'ensemble des flux mondiaux courant dans les larges artères du Web. Avec lui, nous suivons les news, suivons nos amis, nousorientons, choisissons nos restaurants, communiquons. Grâce à lui, nous ne pouvons presque plus ne pas être connectés. Ces techniques  sont à la fois bénéfiques (par exemple pour la transmission du savoir, comme dans le cas des vidéos de vulgarisation sur les médias comme Youtube) et dangereuses (à cause de la perte d’attention induite par ces techniques, de la captation des mémoires, des savoirs, des désirs, que nous décriront tout au long de ce mémoire). Comme toutes ces techniques, le numérique déplace la mémoire hors du sujet humain, au risque de l’aliéner et de le prolétariser, de lui dérober sa mémoire. Pour ces raisons, les mnémotechniques ne sont pas neutres : elles façonnent les conditions de la pensée. Ce que nous défendrons dans ce mémoire, c'est que ces mnémotechniques jouent un rôle central dans l’évolution de l’humain et de ses institutions.

Une particularité de ce monde nouvellement mondialisé soulevée par Stiegler est l'érosion de l'indépendance entre le système mnémotechnique, que nous venons de décrire, et le système technique, 
\begin{quote}
    \textit{le système technique devenu planétaire est aussi et en premier lieu un système mnémotechnique mondial et il y a en quelque sorte une fusion du système technique et mnémotechnique, et du même coup mondialisation. }
\end{quote}

Cette fusion peut donc se comprendre comme une définition du processus de mondialisation. 
Historiquement, Stiegler voit l'origine de ce mouvement de fusion dans le système de télécommunication naissant. Cette fusion effaça l'hégémonie du pouvoir théologique et politique sur les rétentions tertaires humaines. Les histoires, les contes, et l'imagination collective n'étaient plus filtrées par les pouvoirs théologico-politiques, mais elles avaient été remises entre les mains mécaniques des systèmes de rétentions techniques, sous formes d'abord d'écriture, puis d'enregistrements vidéos et radios, puis aujourd'hui de façon encore plus massive avec l'internet et le big-data. Le résultat de ce processus de captation généralisé est résumé par Stiegler dans la formule

\begin{quote}
    \textit{Le système technique mondial est devenu essentiellement un système mnémotechnique de production industrielle de rétentions tertaires,  }
\end{quote}

Le système technique n'a plus pour but que la captation de la mémoire humaine, et, après le recul de toutes les autres institutions théologiques et politiques, seul le systèmetechnique est détentaire de cette mémoire. 
Ainsi, le système technique est devenu  

\begin{quote}
    \textit{le critère de sélection rétentionnelle pour des flux de conscience inscrites dans des processus d'adoption.  }
\end{quote}

Ainsi, l'aboutissement de cette fusion des systèmes technique et mnémotechnique est que le choix de ce qui est inscrit comme rétention tertaire n'est plus entre les mains des humains, au travers de ses institutions, mais se retrouve aujourd'hui entre les mains du système technique.

\subsection{Captation et détournement du désir}

L'explosion des approches qu'a offert l'arrivée d'internet et du Web a induit une nouvelle éclosion du pluralisme, lequel a germé dans tous les secteurs. Soudain, ``il n'y avait pas -- plus -- de vérités, que des interprétations". Pourtant par l'algorithmisation de ces moyens, les individus sont devenus des ressources pour les algorithmes eux-mêmes. Dans cette écologie algorithmique, le bien le plus précieux est la donnée, et celui qui est le plus à même de la fournir est le consommateur, qui se trouve donc transformé en fournisseur de data pour le nombre toujours croissant de capteurs que nous nous sommes implantés

\begin{quote}
    \textit{puces, capteurs, balises GPS, automobiles, téléviseurs, montres, vêtements.}
\end{quote}

Il s'agit de la société réticulaire, organisée en réseau et qui charrie un flux de données de l'individu vers les data center. Tout cela a presque un seul objectif, celui de comprendre l'individu consommateur pour le faire acheter. On pourrait s'arrêter à ce niveau et se demander: ``et alors?". 
Pour Stiegler, cela ne s'arrête pourtant pas là. La vitesse d'échange de l'information permet de faire circuler l'information entre les individus plus vite que jamais, et remplacer, à la force foudroyante de la vitesse de la fibre, l'homme, le parent, l'ami, par la machine. Le système technique, en tant qu'il a fusionné avec le système mnémotechnique de formation des rétentions, se substitue aux systèmes sociaux. Les impacts sur la psyché humaine rejoignent ceux évoqués par Adorno et Horkheimer ci-dessus, 

\begin{quote}
    \textit{Transformés en fournisseurs de datas, ceux-ci (Les individus et les groupes que les réseaux dits sociaux déforment et reforment selon des nouveaux protocoles d'association) s'en trouvent désindividués par le fait même: leurs propres données qui constituent aussi ce que l'on appelle des rétentions permettent de les déposséder de leurs propres protentions.  }
\end{quote}

Pour les psychanalistes la libre expression de la libido ainsi que sa stimulation par son environnement, et par les parents, constitue une étape indispensable dans l'évolution de l'enfant, et dans la formation de son ``moi". Elle est notamment au centre de la transmission intergénérationnelle.  

En se substituant à la génération précédente pour la formation de la pulsion libidinale, et donc du désir, les algorithmes forment la structure du désir des nouvelles générations, en court-circuitant l'apport des parents et des générations précédentes. En s'adressant directement aux désirs des nouvelles générations, les réseaux algorithmiques s'interposent entre les générations et la transmission de protentions, tout en les dévaluant. Ce détournement est d'autant plus sournois qu'il emprunte souvent le vocabulaire de la prise de liberté, de l'émancipation des carcans des générations précédentes.  Nous avons entrevu une esquisse de ce phénomène dans l'analyse d'Adorno et de Horkheimer de la captation du désir par l'industrie de la culture. La nouvelle spécificité de la captation moderne réside dans son intensité. Alors que les appareils connectés se sont rendus totalement indispensables, le système mnémotechnique et technique est devenu omniprésent, rendant la captation inévitable. Nous verrons néanmoins qu'une différence majeure avec la situation qu'Adorno et Horkheimer ont étudiée est la vitesse des changements du système technique, vitesse qui induit la disruption des systèmes sociaux.

Chez Freud, le désir est au cœur de la vie psychique. Celui-ci exprime le manque, la tension, et la quête de satisfaction. Ce désir est porté par une énergie psychique fondamentale qu’il nomme \textit{libido}. La libido au sens large est l’énergie du désir sexuel, elle ne se limite pas à la sexualité génitale, elle concerne tout ce qui relève de l’attachement, du plaisir et de la pulsion de vie. A partir de cette théorisation, Freud développe l’idée que le psychisme humain est structuré par des pulsions, lesquelles sont des forces inconscientes du psychisme qui visent la satisfaction et qui gouvernent les comportements des individus. Ces pulsions naissent dans le corps et cherchent à décharger leur tension par des objets. La libido est l’énergie de ces pulsions sexuelles.

En partant de cette théorisation, Freud distingue plusieurs phases du développement libidinal. Il introduit le stade oral, anal, phallique, période de latence, et génital. Chacune correspond à un type et à un mode d’expression du désir ainsi qu'à un rapport spécifique au corps. Les désirs refoulés (jugés inacceptables par le moi ou le surmoi) deviennent inconscients, mais continuent d’agir sous forme de rêves et de fantasmes. De cette façon, Freud insiste sur le fait que le désir est toujours conflictuel, celui-ci est soumis à la censure morale, notamment la censure personnelle due au \textit{surmoi}, au principe de réalité, via le \textit{moi}, et à la pression des pulsions, le \textit{ça}. Freud voit dans la libido, et dans les différents schèmes de résolution des conflits entre le ça, le moi et le surmoi, le moteur de la créativité, de la névrose, mais aussi de la civilisation. 

Dans ce processus de civilisation et de créativité, les tensions créées par le conflit constant entre le ça, le moi et le surmoi sont redirigées vers des buts non sexuels comme l’art, la science ou la religion, par le processus de sublimation. Pour lui, il appartient à la transmission intergénérationnelle de façonner et d'apprendre à sublimer la libido, même lorsque celle-ci est frustrée. Insérée dans un milieu social, cette libido individuelle devient collective et donne lieu à une économie: une économie libidinale\cite{Disrup},

\begin{quote}
    \textit{
Une économie libidinale est une économie du désir en tant qu'il est à la fois individuel et collectif. Le désir est structuré par un champs de protentions héritées par la désirant et projetées par lui singulièrement à partir des rétentions collectives transmises par le jeu intergénérationnel que régulent les modèles d'éducation aux différents âges de la vie.  
}
\end{quote}

Pour Stiegler, le rôle de la mécanique générationnelle est claire: celle-ci transmet par l'éducation les rétentions collectives inscrites dans la société au jeune désirant, en permettant ainsi de former ses protentions à partir de l'éducation et des rétentions. L'économie lidibinale est donc le jeu de la transmission des protentions d'une génération à une autre.
Cependant, l'effet de la captation du désir par les algorithmes est de façonner et de modeler les protentions du désirant, ainsi que la structure de la libido, en lieu et place des parents et sur les besoins marchands. On peut illustrer le détournement commercial de la libido à partir de l'histoire de la publicité, que l'on peut découvrir dans l'article de revue de Myriam Tsikounas, ``La publicité, une histoire, des pratiques"\cite{articleTsikounas}. Aujourd'hui, il nous semble difficile de concevoir que les premières publicités, à partir de 1830 et plus tard, étaient avant tout informationnelles et vantaient les mérites de leur produit d'un point de vue factuel, tellement ce modèle serait inefficace aujourd'hui. Elle énonce\cite{articleTsikounas}
\begin{quote}
    \textit{
Au départ, le journal et l’affiche sont les seuls supports publicitaires. Dans
la presse, l’annonce est rudimentaire, enfermée dans un cadre rectangulaire de
petites dimensions, insérée horizontalement. Elle se limite à du texte argumentatif,
souvent long, en noir et blanc, sans artifice typographique. 
 }
\end{quote}

Dans ces premiers temps de la publicité, le péché capital dont beaucoup de publicistes se rendaient coupables était celui de publicité mensongère. Factuellement mensongère. Cependant, à partir du XXème siècle, moment à partir duquel les étudiants ont eu accès à un curriculum dans l'art de la publicité, et avec l'arrivée de nombreux publicistes des états-unis, la publicité prend un tournant plus systématiquement pulsionnel et lidinal. Les publicistes ont alors recours à l'illustration par le dessin et par la photographie pour attacher leur produit à une image. 

Encore une fois, un exemple très grossier aide à appréhender le raisonnement. Dans la publicité pour une voiture audi accessible au lien\footnote{https://www.youtube.com/watch?v=GCQhoYrNxdI}, accompagnée de son slogan ``il a une audi, il aura la femme", la libido, d'origine protéiforme, est très explicitement détournée pour l'achat d'un produit confectionné à la chaîne. La publicité ne vante aucun mérite réel -- On ne penserait pas à attaquer audi en justice si, après avoir acheté une audi, on n'avait pas eu la femme -- elle tend seulement à exacerber les désirs.  D'une manière équivalente, l'allusion sexuelle, ou viriliste est rapidement devenue légion dans la promotion des marques de cigarettes. 

Suivant en ligne directe les algorithmes de suggestion ciblant spécifiquement les consommateurs, les publicités personnalisées, tirant avantages des énormes quantités de données amassées sur chaque individu, ont fait leur apparition et ont rapidement dominé les logiques du marketing computationnel. Entrainés à reconnaître les volitions, les intérêts et les pulsions des internautes, et sans doute à décrypter les désirs refoulés des utilisateurs, les algorithmes de suggestion des annonceurs se spécialisent dans la captation de la libido des internautes. Le but de toute publicité étant de faire correspondre à un objet de consommation, un affect ressenti par le consommateur, la forme de celle-ci s'organise comme une gigantesque machine à générer et en reformer les pulsions. 

Aujourd'hui, l'appel à la pulsion, pulsion de consommation, pulsion d'achat, que Stiegler associe à la pulsion de mort Freudienne, est ce qui structure l'entièreté de l'appareil du marketing mondial. Comme une captation est toujours aussi une refonte, en remodelant l'ensemble des rétentions, le marketing computationnel s'est octroyé la capacité d'influer sur la formation même des protentions des individus et des sociétés, et a orienté cette puissance vers des finalités marchandes. Cette tendance favorise à l'outrance les protentions de consommation et d'achat, favorables au marché, sur les protentions de créations culturelles, de production de savoir sans application commercialisable et d'apprentissage. En rédéfinissant les pulsions refoulées, la publicité pulsionnelle réprime toute forme de sublimation et rend l'individu incapable de former des protentions individuelles. C'est ce que Bernard Stiegler appelle la \textit{prolétarisation psychique}.

Ayant perdu la capacité de générer des protentions personnelles, l'individu est mûr pour l'objectification. C'est exactement ce que proposent les plateformes lorsqu'elles offrent alors au consommateur la possibilité de devenir le publicitaire de sa propre image qu'il exhibe sur ses divers réseaux sociaux amicaux, comme facebook, professionnels, comme LinkedIn, amoureux, comme Tinder. Le réseau-internaute mondial a appelé l'internaute à se faire produit et à s'afficher, et l'internaute a répondu. Comme un bon publicitaire, il vante alors son lifestyle, lequel n'est enviable qui si il est entièrement tourné vers le grandiose de la consommation. Par ce tour de maître, le capitalisme de consommation peut même se passer de la publicité, laquelle est fournie quotidiennement et gratuitement par des nouvelles petites stars des réseaux. Cette publicité spontanée des internautes est le dernier tournant opéré par la publicité moderne. 

Face à la détresse, inéluctable et bien visible, qu'induit chez l'individu la perte de la capacité fondamentale de la génération de protentions, le système pulsionnel répond par l'offre d'un choix encore plus large d'objets de consommation. Le système génère toute une série de solutions automatiques. Un exemple que nous allons explorer est la consommation de biens spirituels. D'autres exemples sont le développement personnel doublé de sa pensée positive, critiqué par exemple dans le livre de Gérard Neyrand, ``critique de la pensée positive". Ainsi, même si je ne peux développer ce point dans ce mémoire, je crois que le phénomène récent de la pensée positive, et de développement personnel, doit se comprendre en large partie dans ce contexte de détresse psychique et comme offre spontanée que le système marchand aménage automatiquement.

\section{L'évolution de l'offre culturelle}

Ici, nous devons mettre l'analyse en pause pour dénouer un noeud personnel pour le moins gênant. Nous avons argumenté que la standardisation de la culture amène à un appauvrissement personnel du consommateur, notamment par la consommation de biens culturels uniformisés, supposément destructeur d'individualité. A titre personnel, je dois pourtant remarquer que c'est ma consommation culturelle, sous forme de séries, de livres, de jeux vidéo contemporains, qui a souvent marqué le point de départ de l'exploration de quelques questions philosophiques majeures qui m'ont inquiété.

Après la ``consommation" du jeux vidéo français ``Clair Obscur" et la série allemande ``Dark", ou du film ``Matrix", je me suis intéressé à la question de la valeur de l'illusion, que l'on peut par exemple explorer avec Nietzsche. Après avoir visionné le film de Sean Penn, ``Into the wild", je me suis penché sur la question de l'authenticité de nos modes de vie, que l'on peut explorer avec le philosophe américain Charles Taylor, et sur ce qui constitue un mode d'existence en accord avec ses principes. Je pourrais continuer les exemples. Ce sont des questions certainement aussi profondes que celles soulevées et discutées par les grands classiques, par exemple de la litérature, comme ``crimes et châtiments". De plus, on ne peut que concéder à une grande profondeur aussi bien psychologique que poétique à un jeu vidéo comme ``Clair Obscur". 

Comment puis-je réconcilier cette observation avec la critique de la platitude du contenu culturel proposée plus haut ? Laissez-moi avancer une proposition de réponse, laquelle n'est sans doute pas tout à fait une solution.

\subsection{Les questions métaphysiques et le marché de la petite spiritualité}
\label{sec:petite_ques}
 Il faut commencer par remarquer que dans les oeuvres culturelles mentionnées ci-dissus les questions métaphysiques ne sont généralement que survolées, elles servent parfois de ressort à l'intrigue, mais ne sont que rarement explorées en profondeur -- je me dois néanmoins de marquer une exception pour ``Clair Obscur". Elles servent d'ornement à l'intrigue. Il semble que le fait de les aborder n'ait pour but que de laisser le consommateur pensif, avec un sentiment de (fausse ?) profondeur. Par ce mouvement c'est la métaphysique toute entière qui est mobilisée au service l'intrigue et du suspense. Dans sa critique artistique, on fait souvent valoir comme un point positif de telle ou telle série qu'elle \textit{soulève} un point profond de métaphysique, et on lui met à son crédit.

 On peut rapprocher cette tendance de la marchandisation de la spiritualité et de ce que l'on appelle souvent les philosophies de vie, qui a taillé une place importante à toute une offre spirituelle dans les rayons des libraires, sur le marché des vidéos Youtube, ect... Cela permet à tout un chacun de faire son libre choix au marché des offres spirituelles, d'en prendre un petit bout et de s'en parer. Aujourd'hui, on peut donc s'habiller d'un slogan stoïque, le joindre à une référence à la Bhagavad Gita, saupoudrée de spiritisme amérindien. Il suffit de se balader dans une libraire pour trouver tout un rayon dédié à ces petites spiritualités, et de discuter un peu avec les vendeurs pour observer que c'est là qu'ils font leurs meilleures ventes. Une caractérisque importante de l'adhésion à ces petites croyances spirituelles est qu'elle se fait à peu de frais. L'individu ne s'en pare que pour goûter au goût que cela fait, pour soi-même et pour les autres. Il est généralement aussi peu coûteux pour l'individu d'adhérer à une philosophie l'espace de quelques jours, semaines, mois, années, que de s'en défaire, lorsque le goût s'est estompé ou que l'originalité s'est fânée. Le propre de ces spiritualités est qu'on y va aussi vite qu'on en revient. Pour emprunter le vocabulaire de Stiegler, ce processus n'est qu'un exemple de la perte de l’attention durable et du soin apporté aux savoirs. 

Ce marché des petites spiritualités se redouble d'un marché des petites questions métaphysiques, qui ne sont abordées que pour l'effet de vertige, le \textit{vertige métaphysique}, que ces questions procurent au consommateur. Lorsque ces questions sont abordées, elles le sont pour produire l'effet de vertige métaphysique, presque jamais pour creuser la question métaphysique en tant que telle, ni pour pousser une réflexion qu delà des limites qu'on lui connaît déjà. Par ce mécanisme, de même que les petites croyances spirituelles, les petites idées métaphysiques deviennent des objets de consommation. Une fois que le vertige métaphysique d'une petite idée s'est estompé, on en redemande. Moi aussi. 

Cependant, cette consommation de la toute nouvelle et vaste offre spirituelle, toute parée de légéreté, ne saurait être aussi innocente qu'il y paraît. Les spiritualités, sans doute autant que les questionnements métaphysiques d'une époque appartiennent aux rétentions de cette époque, sous la forme d'écrits, de discours oraux, d'histoires racontées dans une société existant à un moment donné et dans un environnement donné. Ces rétentions appartenant à une société sont génératrices de tout un réseau de protentions, qui sont des impulsions, des désirs, des espoirs et des visions qui mettent les individus de cette société en mouvement. La marchandisation de ces rétentions importées peut, à mon avis, s'interpréter dans le cadre d'une disruption des protentions locales, induites par la mondialisation, comme nous les verrons plus bas. 

Cette marchandisation s'inscrit dans la tendance beaucoup plus globale à l'individualisation de la conviction religieuse, telle que décrite par exemple dans l'ouvrage de Michel Gauchet, ``La religion dans la démocratie : parcours de la laïcité"\cite{Gauchet}. Cet ouvrage nous décrit les mécanismes par lesquels les convictions religieuses se sont retirées peu à peu du domaine social et collectif pour devenir une affaire individuelle ou familiale. Suite à cette évolution, la parole religieuse a perdu une grande partie de son influence sociale et s'est effritée.

\chapter{La mécanique de la disruption,  et l'anthropos}
\label{chap3}

Dans les chapitres précédents, nous avons vu comment la numérisation des moyens de communication et des médias, bien qu'elle ait ouvert le champs des possibles et offre une opportunité inégalée pour l'économie horizontale, amène pourtant une série de problèmes rédoutables et donne un nouvel éclairage à d'anciens problèmes. Dans ce chapitre, avec l'apport des philosophes de la technique contemporains, nous tâchons d'apporter un peu de systématicité aux symptômes diagnostiqués dans les chapitres précédents. Cette systématicité nous permettra d'inclure la théorie de la \textit{disruption}, que l'on décrira comme crise constante du système technique et comme captation par le système marchand des protentions individuelles, dans une théorie du devenir et de l'avenir.

\section{L'idée de la disruption}

\paragraph{La disruption dans le monde de l'entreprise}

En tant qu'il est un mot profondément changeant et actif dans la société aujourd'hui, commençons par revenir aux sources du terme de ``disruption". Cette expression trouve son origine dans le monde du business. Dans son acceptation moderne, le terme est introduit par Clayton Christensen  dans son ouvrage de 1997, ``The Innovator’s Dilemma" où il l'introduit sous la forme de l'idée d'innovation disruptive, icône du temps nouveau, et qu'il oppose à l'amélioration incrémentale d'un produit déjà bien inséré sur le marché.  Comme exemple de ce phénomène on peut prendre le cas  de la marque de téléphone Nokia, exemple typique de firme ayant misé sur l'innovation incrémentale. Entreprise pourtant leader de la téléphonie, elle a pourtant échoué face à des innovations radicales et disruptives, alors même qu'elle était menée de la manière la plus rationnelle et la plus efficace\footnote{https://hbr.org/2015/12/what-is-disruptive-innovation.}
\begin{quote}
\textit{La disruption désigne un processus par lequel une entreprise plus petite, disposant de moins de ressources, parvient à concurrencer avec succès des entreprises établies et dominantes.}
\end{quote}

Les innovations disruptives s'adressent à un nouveau public, dans des situations nouvelles, ou portent des fonctionnalités novatrices.  La disruption c'est donc d'abord une théorie descriptive, qui nous présente l'image d'un danger permanent pour les entreprises établies et d'une opportunité de tous les instants pour les outsiders. Elle est à l'image d'une époque de la toute puissance d'un instant et puis de la chute vertigineuse. L'innovation disruptive des nouveaux arrivants est ce qui punit le trop grand sérieux avec lequel les entreprises établies ont suivi la règle incrémentale, ou encore pire, la stagnation. Quelques exemples de succès disruptifs dans des milieux pourtant compétitions sont Netflix, Airbnb, Uber.

Dans cette acception, la disruption rappelle la destruction créatrice de Joseph Schumpeter, qui est un concept économique selon lequel l’innovation détruit les anciens modèles économiques pour en créer de nouveaux, plus efficaces et plus performants. Si on veut être compétitif dans un marché toujours plus violent, il faut être disruptif. Sans doute parce qu'elle s'insérait si bien dans la description de notre nouvelle époque, la disruption a été appelée la théorie, ou l'idée, la plus influente du 21ème siècle \footnote{https://www.wsj.com/articles/clayton-christensen-has-a-new-theory-1475265067 et https://www.economist.com/britain/2017/06/15/jeremy-corbyn-entrepreneur}. De description, elle s'est alors muée en une injonction généralisée à la compétition.

\paragraph{La disruption dans le monde des idées}
Un autre exemple concerne les travaux de recherche, notamment scientifiques, milieu dans lequel le terme de disruption a su se répandre sans grande peine. Dans Nature, l'article         ``Papers and patents are becoming less disruptive over time"\cite{park2023papers} se désole de la prétendue observation que la ``disruptivité" des articles scientifiques semble décroître au cours des années. Dans ce travail de recherche, Park, Leahey et Funk définissent un nouvel index de mesure, le CD index, qui entend quantifier après coup la disruptivité d'un travail de recherche. Dans les grandes lignes, cet index est un nombre assigné à chaque article scientifique étudié en fonction de la manière dont celui-ci a été cité dans la litérature qui l'a suivi. L'index CD est d'autant plus grand que l'article est cité comme ``premier" article référence d'un domaine ou d'un sous-domaine. En pratique, Park, Leahey et Funk regardent si l'article étudié est cité en commun avec d'autres articles que l'article étudié lui-même cite. Si c'est le cas, on peut en déduire que l'article est moins ``disruptif", qu'il est plus ``cumulatif", incrémental, qu'il n'ouvre pas une nouvelle voie mais consolide ou continue une voie déjà ouverte. 

 Dans ce sens, le CD index prétend quantifier à quel point un article ouvre des voies nouvelles, qui seront exploitées par les autres chercheurs comme point de départ. Le point central de l'article est que cet indice global, lorsqu'on le moyenne sur un grand nombre d'articles publiés pendant une année donnée, semble tendantiellement décroitre au cours du temps. Les auteurs ne rechignent alors pas à conclure que la recherche, globalement, tend à devenir moins disruptive, qu'elle ouvre moins de voies nouvelles, 
\begin{quote}
\textit{We find that papers and patents are increasingly less likely to break with the past in ways that push science and technology in new directions.}
\end{quote}

L'analyse de Park, Leahey et Funk est hautement criticable -- dans une réponse à leur article, ``The disruption index is biased by citation inflation"\cite{petersen2024disruption}, Peteso, Arroyave et Pommeli rétorquent que la décroissance au cours du temps de l'index construit par Park, Leahey et Funk est un artefact de l'inflation des citations, ou simplement du fait que les chercheurs sont aujourd'hui plus complets et plus précautionneux dans leur recherche bibliographique et plus enclin à citer un grand nombre d'articles. 

Néanmoins, l'article Nature de Park, Leahey et Funk est aujourd'hui cité plus de 800 fois selon Google Scholar et a eu un large retentissement dans les médias. La question traitée dans les médias était généralement de savoir pourquoi la créativité humaine était en pleine décroissance. Etait-ce une maladie de notre époque? Funk dans un autre artice propose des méthodes pour réactiver la créativité des chercheurs\footnote{https://iap.unido.org/articles/papers-and-patents-are-becoming-less-disruptive-over-time}. 

    Ainsi, il semble qu'à l'appel pour les techniques disruptives, se double une peur de la perte de la créativité, créativité mesurée selon un indice aussi vulgaire que les réseaux de citation. En tirant les ficelles du raisonemment, une bonne recherche serait celle qui ouvre des voies, mais qui ne les explore pas, ou du moins qui laisse aux autres le déplaisir, ou le déshonneur d'explorer ces nouvelles voies. 

\paragraph{La disruption et la plateforme}

Quelques exemples de succès disruptifs dans des milieux pourtant compétitifs sont Netflix, Airbnb, Uber, Youtube, ect. Ces plateformes sont tellement tentaculaires qu'elles ont transmis leur nom à nos actions quotidiennes: on prend un airbnb, on appelle un uber, on se fait un netflix.  Au delà de leur spectaculaire succès, ces compagnies ont toutes en commun de n'offrir en soi aucun bien, ni à vrai dire aucun service. De fait, un gîte ou un hôtel n'a pas besoin d'Airbnb pour offrir une chambre à son client, Uber ne possède aucune voiture, Youtube ne produit presque aucun contenu. Les plateformes ont pourtant compris que dans un monde de moins en moins régulé, de plus en plus gorgé d'informations, de stimulus, de contenus, l'un des services les plus précieux résidait dans la suggestion. Dans un monde occupé à glorifier la liberté de choix, le luxe est maintenant dans le plaisir de se laisser guider, vers le airbnb le plus proche, d'une vidéo à une autre, d'une série à une autre, choisie selon nos préférences, lesquelles ont été étudiées par la plateforme. Ainsi, le produit qu'offre les plateformes, c'est la suggestion.

Jusqu'à maintenant, la disruption a été décrite comme un processus économique proche de la destruction créatrice. Cependant, comme nous l'avons également vu dans le chapitre précédent \ref{chap2}, les plateformes et leur réseaux algorithmiques captent avec une grande efficacité la libido en remodelant les protentions des utilisateurs. En faisant cela, elles se creusent une place dans l'économie libidinale collective et imposent un changement social qui désintègre les structures sociales existantes, ce que Stiegler explique par\cite{Disrup}
\begin{quote}
\textit{Ce pouvoir automatique de désintégration réticulaire s'étend sur la Terre à travers ce que l'on appelle depuis quelques années la disruption. }
\end{quote}

Ainsi, à la disruption technique décrite par Christensen, Stiegler adjoint une disruption sociale, inévitable selon sa théorie de la connexion entre l'appareil technique et social.   

Selon Stiegler, la disruption se rencontre en concomitance avec l'innovation, mais jamais avec le progrès. Comme le souligne à maintes reprises Etienne Klein dans ``Progrès et innovation : quels liens ?"\cite{Klein} que nous avons déjà repris plus haut, l'innovation n'est pas le progrès. Celle-ci se distingue de celui-ci par l'absence, acclamée, d'un récit englobant. Le progrès et les grandes idéologies sont devenues impensables, inatteignables.  Jean-François Lyotard voyait la fin de la notion d'époque dans l'absence de grands récits unificateurs. Dans ``La Condition postmoderne", il affirme que les grands récits qui donnaient un sens global à l’Histoire (comme le progrès, l’émancipation ou la raison) ont perdu leur crédibilité. Il parle d’une incrédulité envers les métarécits, caractéristique de la postmodernité. Ces récits ne parviennent plus à légitimer le savoir ni à guider les actions collectives. À leur place émergent des petits récits, locaux et fragmentés, qui reflètent une diversité de points de vue sans prétendre à l’universalité et l’histoire n’est plus perçue comme un chemin linéaire vers un but ultime, comme cela pouvait être le cas pour la pensée des Lumières ou la pensée Marxiste. Cela affecte la science, la politique et la culture, qui ne peuvent plus s’appuyer sur une vision unifiée du monde pour leur légitimation propre. Le savoir devient performatif, c’est-à-dire jugé à son efficacité plus qu’à sa vérité, et les sciences deviennent instrumentales, leurs usages se relativisent.

Là où le progrès s'incrit dans une histoire de l'avant et de l'après, qui appelle à faire quelques sacrifices aujourd'hui pour trouver des lendemains meilleurs, l'innovation entend faire table rase du passé, et ne l'incorpore dans son narratif que pour pouvoir mieux s'en distinguer, ou l'éliminer, comme nous l'avons vu pour l'indice de disruption pour les articles scientifiques. En d'autres mots, contrairement au progrès, l'innovation ne s'inscrit pas dans un récit totalisant et téléologique. L'innovation n'a pas de direction, car elle se définit uniquement par le fait de briser les liens qui l'unissait au passé, par la rupture avec celui-ci. Elle ne cherche ni histoire, ni origine, point que nous avons déjà illustré plus haut lorsque nous avons argumenté que la fin de la théorie se rattachait à une perte de direction pour la pensée. Cette perte du passé dans le processus de l'innovation se répand alors au futur lui-même. 
Dans les mots de Klein, bien qu'elle soit entièrement tournée vers le futur, l'innovation ne vise aucun devenir
\begin{quote}
\textit{Lorsque nous lisons les journaux, les pages web, ou que nous regardons
la télévision, nous avons le sentiment que le futur s’est absenté. [...] Cette situation fait de nous des hommes et des femmes bloqués
dans le présent.}
\end{quote}

 Ce type de situation se rencontre de façon assez fréquente dans les systèmes physiques. Les seuls systèmes capables d'oublier les conditions initiales en physique sont les systèmes dissipatifs, ceux qui créent de l'entropie, concept sur lequel nous reviendrons largement. Dans ce sens, et en termes de thermodynamique, on peut voir la disruption généralisée comme un oubli des conditions initiales, comme un effacement que la mécanique inévitable du temps induit. Un système qui a atteint un maximum d'entropie n'évolue plus, son temps est devenu un présent permanent et il n'a plus aucune marque du passé et ne porte aucun futur. 

Voici donc maintenant une caractérisation plus globale de la disruption, qui incorpore néanmoins les éléments précédemment cités de rupture avec le passé et d'innovation foudroyante et destructrice: la disruption est un processus de déstabilisation soudaine des repères culturels, sociaux, scientifiques et économiques par l’accélération de l'innovation technologique, notamment numérique. Cette innovation détruit les structures existantes et court-circuite les structures de transmission intergénérationnelles. Dans le cas particulier des plateformes, que l'on considère maintenant comme outils de captation de l'attention et du désir,  elle détruit les capacités de projection individuelle (protentions) et rend obsolètes les institutions sans leur laisser le temps d’évoluer. Parce qu'elle cherche essentiellement la rupture et tend à faire table rase du passé, la disruption est un processus entropique d'oubli. Elle impose un présentisme souverain.

    De ce point de vue, on peut se demander quel est le moteur de la disruption, qu'est-ce qui peut la rendre enviable ? La réponse réside partiellement dans la mécanique de l'entropie que nous allons maintenant développer.

\section{La mécanique de la disruption}

Comme nous le voyons, la disruption est un processus qui implique un changement culturel et social, autrement dit un changement d'époque. 
Pour Stiegler, le changement d'époque s'accompagne toujours d'un changement technique

\begin{quote}
    \textit{
Lorsque se produit un changement de système technique, l'époque au sein de laquelle il prend sa source s'achève: une nouvelle époque émerge, généralement au prix de conflits guerriers, religieux, sociaux et politiques en tout genre.
}
\end{quote}

Le changement technique est donc à la source du changement d'époque qu'est le changement culturel et social. Pour faire avancer notre réflexion, nous devons maintenant approfondir un concept important de la pensée de Bernard Stiegler, à savoir le double redoublement épokhal.

\subsection{Le changement d'époque et le double redoublement épokhal}

Après un changement technique, l'émergence de la nouvelle époque n'est pas automatique. Celle-ci n'émerge que lorsque les rétentions héritées des époques précédentes se reconfigurent en une série de protentions nouvelles qui reformeront le socle de la nouvelle époque, ou dans les termes de Stiegler

\begin{quote}
    \textit{
La nouvelle époque n'émerge que lorsque, à l'occasion de ces conflits, et du fait de la perte de prégnance des savoirs et pouvoirs de vivre, faire et concevoir de l'époque précédente, de nouvelles façons de penser, de nouvelles façons de faire et de nouvelles façons de vivre prennent forme. 
}
\end{quote}

La nouvelle époque apparaît suite à la reconfiguration de l'économie libidinale collective, à savoir l'économie des  protentions des individus dans la société. Cette économie libidinale se transmet ensuite de génération en génération via une structure bien définie, que l'on appelle la \textit{transmission intergénérationnelle} et qui est organisée par les rétentions tertiaires collectives. Cette mécanique du changement d'époque a été formalisée par Stiegler sous le terme de \textit{double redoublement épokhal}, qu'il formalise notamment dans son ouvrage ``La technique et le temps, vol. 1 : La faute d’Épiméthée", et qui  désigne ce changement au cours duquel la technique transforme radicalement notre rapport au temps et à la mémoire. 
Commençons par noter que pour Stiegler, la technique est un support de mémoire (mnémotechnique), qui extériorise et structure le rapport humain au passé, donc à l’épokhè, c’est-à-dire à la suspension du temps ou des repères temporels.

Le premier redoublement correspond à l’externalisation de la mémoire humaine dans des supports techniques (comme l’écriture, l’imprimerie ou le numérique). Cette période d'inscription marque un redoublement de la mémoire, qui est tout d'abord une mémoire humaine et qui devient par la suite une rétention technique tertiaire. Ce premier redoublement marque une rupture avec la temporalité précédente. Par exemple, lors de  l'essor de l'écriture, nouveau réceptacle pour la mémoire, la relation de l'humain s'en est trouvée radicalement modifiée par rapport à la temporalité biologique de la mémoire purement humaine.

Le second redoublement survient lorsque ces techniques évoluent au point de transformer elles-mêmes les conditions de ce premier redoublement, par exemple par l’automatisation (algorithmes, intelligence artificielle). En passant d'une technique à une autre, la forme des rétentions se redouble de nouveau. Ce double processus redéfinit l’humain, en déstabilisant la conscience, la mémoire, et les structures collectives du savoir. Ce double redoublement épokhal se produit donc en deux temps que l'on peut résumer par l'étape d'externalisation, et ensuite par l'étape de désajustement, suivi de réajustement qui achève le processus.

Depuis ce point de vue, la technique n’est pas seulement un support qui enregistre l'époque, mais devient un acteur autonome dans la production de la temporalité de l'humain. L’époque numérique illustre parfaitement ce phénomène et nous en avons cité plusieurs exemples. En accélérant les flux d’information, avec des appareils comme les ordinateurs et les téléphones portables, le monde bat au rythme des news internationnales que tout un chacun peut suivre à toute heure de la journée, au point de court-circuiter la réflexion. Le double redoublement épokhal marque donc une mutation anthropologique profonde.

Pourtant, ``dans l'état d'urgence entropique qui est la réalité concrète de l'anthropocène", l'un des dangers majeurs est celui de l'essouflement de la notion d'époque\cite{Disrup},
\begin{quote}
  \textit{Nous sommes aujourd'hui tout près de cette explosion. A présent, chacun de nous le sait, le craint, mais aussi le refoule et le dénie, comme pour tenter de continuer à vivre dignement. Or il n'est plus possible de le refouler; au stade où nous en sommes, cela devient précisément indigne, et litéralement lâche. }
\end{quote}

Pour Bernard Stiegler nous sommes entrés dans l'ère de l'absence d'époque. Cet essouflement de l'époque vient d'un dysfonctionnement de ce mécanisme du double redoublement épokhal.

Les systèmes techniques, exosomatisés par les sociétés humaines, évoluent en concomitance avec les autres systèmes sociaux humains et ne cessent de se développer, développement qui ne manque de les faire entrer en crise. Pendant cette crise, les systèmes techniques évoluent plus rapidement que les systèmes sociaux, avec lesquels ils se désynchronisent, se désajustent. Comme proposé plus haut, ce désajustement offre l'opportunité de l'émergence et de la stabilisation d'une nouvelle époque, si le double redoublement épokhal peut arriver à son terme. En effet,
\begin{quote}
    \textit{La stabilité revient lorsque ces autres systèmes ont adopté le nouveau système technique. }
\end{quote}
Par autres systèmes, il faut ici entendre l'ensemble des systèmes sociaux. Cependant, la mondialisation et la technisation du monde, particulièrement spectaculaire au cours des 20 dernières années, ont rendu le monde et le système technique encore plus instables et prompts aux crises, comme nous avons pu l'observer dans le contexte des crises économiques diverses et variées qui ont bercé ces deux décennies.

Ainsi, l'une des caractéristiques principales de la technique moderne est sa vitesse, qui va plus vite que la volonté humaine. Un exemple particulièrement clair de cet état de fait est la retard que prend la législation sur les nouvelles technologies. Alors qu'une nouvelle série de mesures est prise pour encadrer législativement un milieu numérique, ce milieu est souvent déjà dépassé et a été délaissé pour un autre. Pour reprendre les mots de Stiegler, ce stratagème relève de la théorie du \textit{choc} et est typiquement celui utilisé par les barbares\footnote{Cette idée de nouveaux barbares a été habilement décrite par Baricco dans son ouvrage ``Les Barbares", où il explore les bouleversements culturels contemporains provoqués par la numérisation, la vitesse et la mondialisation. Il nomme "barbares" les acteurs de cette transformation — jeunes, technophiles, amateurs de surface plutôt que de profondeur. Dans son livre, il cherche à comprendre leur logique, qui est celle du mouvement, de la connexion et du plaisir immédiat. Contrairement à ce que l'on pourrait penser, ce nouveau rapport au monde ne détruit pas la culture, mais la reconfigure selon d'autres valeurs. Nous approfondirons cette mutation culturelle dans le prochain chapitre avec Fredric Jameson.}.

Cette accélération du renouvellement technologique a pour conséquence, selon Bernard Stiegler, de nous faire entrer dans l'ère où le système technique est \textit{constamment} en crise. Cette crise constante empêche le réajustement entre les systèmes techniques et sociaux, qui permettraient d'achever le ``double redoublement épokhal" caractéristique du changement d'époque. Ainsi, dans l'ère de la disruption, le second temps du double redoublement épokhal ne peut plus avoir lieu, et la nouvelle époque n'émerge pas, 
\begin{quote}
    \textit{Cette épokhé est disruptive en cela qu'elle ne donne absolument pas lieu au second temps, ni donc à aucune pensée: elle ne donne lieu qu'au vide absolue de la pensée. }
\end{quote}

Cette chaîne de conséquence amène Stiegler à la conclusion que notre époque, dans laquelle la nouvelle époque ne parvient pas à émerger, est celle de l'absence d'époque. Nous en arrivons donc à une seconde caractérisation de la disruption comme échec du double redoublement épokhal. L'échec de la transmission intergénérationnelle, due en grande partie à la captation discutée ci-dessus par les réseaux algorithmiques, est à la fois conséquence et accélérateur de cette mécanique. La disruption  est une information qui se répand dans les systèmes sociaux plus rapidement que le temps d'adaptation typique de ces systèmes, trop vite pour que l'on puisse réagir\cite{Disrup}
\begin{quote}
  \textit{ La disruption est ce qui va plus vite que toute volonté, individuelle, aussi bien que collective des consommateurs aux dirigeants, politiques aussi bien qu'économiques. }
\end{quote}

Au creux de cette théorie de la pensée moderne, une question semble cependant rester en suspens, à savoir la question première de la possibilité d'émergence de l'absence d'époque. Qu'est-ce qui a rendu possible la disruption ? Qu'est-ce qui l'a rendu efficace ? Il semble difficile d'assigner cette absence d'époque à une conséquence mécanique de l'accélération de l'amélioration de la technique. 

Par une interpération plus personnelle de la théorisation de Stiegler, je peux commencer par avancer -- mais je dédierai tout le chapitre prochain à cette question -- que cette mise en crise constante du système a été rendue possible par l'assaut généralisé sur les structures et les savoirs autrefois stabilisants, que nous avons passés en revue ci-dessus. Les spiritualités, autrefois le ciment de cultures non-mondialisées ont été marchandisées, trivialisées, individualisées et ridiculisées, à tel point que l'adhésion profonde à une quelconque foi a perdu toute valeur collective. La pénible théorisation et modélisation scientifiques sont taxés d'intellectualisme superflu dans un monde de la donnée abondante. La parole politique, après de longues années de théâtralisation, a été grandement décrédibilisée.

Ces éléments appartenant à des mondes sociaux pourtant différents ont tous en commun d'être soumis à une forte pression déstabilisatrice, et leur lent effritement est la possibilité de la mise en place de la disruption généralisée. En effet, comme nous l'avons vu, la disruption est médiée par l'essor de l'algorithmique computationnelle, laquelle est une ennemie directe de l'organisation intellectualisante portée par les idéologies, les théories scientifiques, et les dogmes religieux et spirituels. Nous reviendrons dans le chapitre suivant sur les impératifs culturels qu'une telle structure implique. 

Bien sûr, dans chacun de ces domaines, des bastions de résistance demeurent. Quiconque d'un peu sensé a bien conscience qu'il ne peut pas s'improviser stoïque et commencer à se comporter comme tel sans avoir l'air ridicule. Tous ne sont pas dupes face à la théâtralisation réductrice dont sont  victimes la politique et la spiritualité. L'aura d'une gloire ancestrale demeure dans la mémoire collective, ne demandant qu'à être ravivée.

Parmi les domaines qui peuvent encore se défendre par l'aura de l'expertise, tellement glorifiée de nos jours, le modèle scientifique  garde vraisemblablement de beau jour devant lui. Le monde scientifique résiste à sa marchandisation, par la structure fermée de l'accès à la profession. A toutes les étapes de la montée dans la profession, ce sont les pairs, et non les impératifs marchands qui président à la sélection des candidats. Néanmoins, la marchandisation de l'entreprise scientifique gagne du terrain, notamment par les nouvelles mentalités du ``publish or perish" et les mécanismes de plus en plus répandus d'octroit de financement par projet. Ainsi, l'octroit de financement est aujourd'hui souvent soumis à la rédaction d'un projet argumenté et rédigé de façon codifiée, lequel exige une organisation coordonnée du temps et une anticipation, pourtant impossible, des résultats obtenus.  L'union européenne, au travers de ses appels à projets, présente aujourd'hui le plus pur modèle de ce type de marchandisation de la recherche.

\subsection{La disruption comme production d'entropie et captation des savoirs}
\label{sec:anthro}
L'une des questions scientifiques les plus épineuses qui aient traversé le XXème siècle concerne la possibilité même de la vie, et sa réconciliation avec les avancées théoriques de la physique de la matière. La physique, en tant que science des interactions de la matière, avait érigé nombre de lois d'évolution temporelle, notamment en mécanique, décrites par les équations parametrisées par un paramètre temporel, souvent dénoté t. Cependant, ces équations se comportent de la même façon que le temps avance ou recule.  Cette invariance est en tension avec l'intuition voulant que certains processus, par exemple briser un vase, soient irréversibles.

Pour résoudre ce dilemme de l'invariance  dans le temps des équations, pour introduire ce que l'on appelle une ``flèche du temps", les physiciens ont défini une nouvelle quantité, l'entropie. Cette quantité mesure l'état d'ordre d'un système, une plus petite valeur indiquant toujours un plus haut niveau d'ordre. Associée à cette nouvelle quantité, ils ont postulé également une loi d'évolution de cette quantité, sous la forme de la ``seconde loi de la thermodynamique", que l'on peut comprendre par l'observation intuitive que les situations de désordre sont plus probables. Techniquement, une situation de désordre présentant plus de réalisations possibles, son entropie, qui dénombre ces réalisations, est plus élevée, et le système tend vers ces configurations désordonnées au cours du temps. La configuration la plus désordonnées, la plus amorphe, la plus stable et stérile est la configuration à l'équilibre. C'est la configuration finale vers laquelle tend le système thermodynamique.   

La seconde loi de la thermodynamique énonce que dans tout système fermé, l'entropie doit augmenter au cours du temps, et que le désordre doit donc, d'une façon concomitante également augmenter. Notons à ce stade que la notion d'entropie n'a de robustesse scientifique que grâce à deux corollaires. Premièrement, l'entropie est une quantité qui peut être calculée, en principe, dans chaque situation donnée, et deuxièmement l'évolution de cette quantité est dirigée dans le temps par une loi, la loi de l'accroissement de l'entropie pour tout système fermé. Rétrospectivement, la loi de l'accroissement de l'entropie est le principe qui permet de fixer une succession définie des événements, de définir un avant et un après. L'invariance des lois de la physique est alors brisée par les phénomènes macroscopiques d'augmentation du désordre et le temps acquiert une direction. 

Cette direction du temps est exactement ce qui ouvre les potentialités du sytème. Dans un système physique à bas degré d'entropie, et donc loin de l'équilibre, de larges gradients de force existent et le système est prêt à traverser de nombreuses transformations. Dans le vocabulaire de Simondon, le système contient des potentialités de bifurcation. Ainsi des étoiles se forment et brillent car elles commencent à produire de l'entropie par la transformation de l'hydrogène en hélium. Ces systèmes recèlent des tensions internes qui permettent l’émergence de nouvelles formes ou organisations. L’individuation (formation d’un individu ou d’un objet) naît de cette dynamique. Ainsi, l’instabilité devient source de création et au contraire l'équilibre final, l'état du maximum d'entropie est un système vidé de ces potentialités. 

Néanmoins, alors que la physique a érigé, à raison, en loi immuable la croissance de l'entropie, et donc du désordre, l'apparition et le maintien de la vie, devient un problème. Un organisme vivant semble être un accroissement local de l'ordre, puisque le vivant est plus ordonné que son environnement, ce qui dont décroit localement l'entropie. Ainsi, par sa seule formation, le vivant semble entrer en collision avec la seconde loi de la thermodynamique. 

Pour contourner le problème, dans son ouvrage ``Qu'est-ce que la vie ?"\cite{Schrodinger} Erwin Schrodinger propose l'existence d'un principe connexe à l'entropie, qui en serait son envers, la \textit{néguentropie} et que seules certaines entités dites vivantes seraient capables de produire pour lutter contre l'accroissement constant de l'entropie. Il entend par là réconcilier les lois physiques avec les observations biologiques. Au final, la notion de néguentropie s'est révélée inutile et erronée\footnote{Le problème de l'apparente diminution de l'entropie due au vivant s'est résolu en observant que la terre n'est pas un système thermodynamique fermé, mais qu'il faut y inclure le soleil, lequel est créateur d'une quantité phénoménale d'entropie. En faisant le bilan du système soleil, terre, et vie sur terre, on obtient bien que l'entropie augmente au cours du temps. On conclut aussi que la vie sur terre ne peut thermodynamiquement pas exister sur une planète sans astre. La situation s'est aujourd'hui renversée en l'hypothèse que la vie permette en réalité au système terre entier de produire de l'entropie plus rapidement. Le malentendu de Schrodinger sur la négentropie s'est retourné en une nouvelle interprétation, même si celle-ci demeure très spéculative, de la vie comme mécanique permettant de créer encore plus efficacement de l'entropie au sein du système terre-soleil. Pour parler métaphoriquement, la vie serait une ruse de l'univers pour augmenter son rythme de création de l'entropie. }, cependant on peut en reprendre l'idée en tant que métaphore appliquée à la vie en société\cite{Disrup}
\begin{quote}
  \textit{Lorsque, voyageant, nous nous ressourçons par la diversité des modes de vies et de la singularité des cultures qui constituent ce que nous appelons le monde en le cultivant, nous avons une perception sensible et immédiate de ce en quoi consiste la néguentropie.}
\end{quote}

En suivant Bernard Stiegler, nous remarquons donc que dans un premier temps, la diversité des cultures bourgeonnent et demeurent, et que dans un second temps, cette diversité permet de nous ressourcer. Côtoyer l'extraordinaire, avant que celui-ci ne deviennent ordinaire, est pour lui un \textit{pharmakon}, à savoir une expérience qui, correctement dosée, permet le bien-être. Cette approche donne au tourisme un aspect beaucoup plus profond que celui que l'on admet généralement, comme une simple fuite de la quotidienneté. Pourtant, comme nous l'avons vu, l'homogénéisation des expériences et la standardisation de la culture mettent à mal la possibilité future de cet extraordinaire.

En analogie directe à la notion physique d'entropie, Stiegler introduit la notion d'\textit{anthropie}, qui est son pendant pour l'organisation des systèmes \textit{sociaux}. 
Avec cette nouvelle notion, la  \textit{standardisation} acquiert une signification bien précise, qui est celle de l'augmentation des taux d'anthropie. En effet, les savoirs captés par les machines sont rapidement soumis à la calculabilité, qui impose une large restriction sur ces savoirs, et qui les inscrit dans un système fermé, à savoir algorithmique. Ces systèmes clos stérilisent les savoirs, en les rendant calculables et systématiques.  Ces savoirs, que l'on voudrait appeler inutiles, sont devenus par cela non plus des savoirs, mais des informations, qui se distinguent des savoirs justement par le fait qu'elles peuvent être dupliquées à l'identique et calculées. Cette information s'oppose à l'heuristique du modèle que nous avions présentée dans la section \ref{sec:Eco}, et dont nous avions vu qu'elle permettait des nouvelles interprétations et donc des bifurcations.

Au contraire, à la diversité  culturelle humaine, Stiegler donne le nom de \textit{neg-anthropie}, appliquant ainsi le concept de Schrodinger à la société humaine. On peut proposer une mesure de cette néguanthropie en tant que marqueur de diversité culturelle. Au contraire, l'anthropie, ou l'anthropisation, est le processus par lequel l'homme impose une marque standardisée à un zone en friche, en en faisant une zone commerciale, un champs en monoculture, ou encore en standardisant la musique, ect ... Stiegler prend appui sur le sens commun pour nous rappeler que de telles situations sont déplaisantes;
\begin{quote}
  \textit{Lorsque nous nous sentons mal à l'aise devant un terrain vague, une chambre en désordre, une affligeante zone commerciale, c'est l'anthropie qui nous étreint.}
\end{quote}

L'anthropie, pour Stiegler, est ce qui ferme un système. Un algorithme, en tant qu'il est une suite de commandes qui ne peut que créer des copies de l'information, est un tel exemple de système fermé. Cette répétition, par la copie, est ce qui détruit la connaissance par l'augmentation des taux d'entropie. Au contraire, la connaissance, dans un système néguanthropique, se forme par l'ouverture, que seul un organisme non-algorithmique peut performer. 

Plus haut, dans le chapitre \ref{chap2}, nous avons présenté la captation de la libido par les machines. La fermeture du savoir fonctionne par un processus similaire de captation du savoir par les machines, que nous avons introduit dans la section \ref{sec:capt_sav}. De la même façon que les protensions humaines sont captées par le système algorithmiques, les savoirs sont également captés par la machine, après avoir été traduits en langage machine. Un exemple très actuel est la captation des langues par les LLMs. La calculabilité des langues, qui mène à une réduction des connaissances de l'orthographe et de la grammaire, relègue la connaissance des langues au statut de connaissances inutiles. Les GPS soumettent la connaissance géographique à un processus similaire. Ce sont également les savoirs des ingénieurs et des techniciens qui sont captés pour le moment par l'appareil marchand. Cette captation est aussi l'une des tendances profondes du capitalisme. 
Dans ce contexte, l'anthropocène, ère dominée par la race humaine et son impact sur l'environnement, est un entropocène, ère dominée par l'augmentation des taux d'entropie et d'anthropie.

Cependant, à ce stade, aucune loi d'évolution n'a été assignée à l'anthropie, et donc la diversité sociale et culturelle pourrait aussi bien croître que décroître au cours du temps. Cependant, par des arguments statistiques analogues à ceux produits dans le cas de l'augmentation de l'entropie, on peut postuler comme tendance naturelle l'augmentation de l'anthropie. De la même façon que l'évolution thermodynamique fait tendre le système physique vers la configuration la plus désordonnées, la plus amorphe, la plus stable et stérile, l'évolution algorithmique fait tendre le système réseau-internautes vers l'homogénéisation des désirs, vers une configuration ne contenant que des copies du même, en apparence différent.

Prenons l'exemple du système humain-plateforme, où la plateforme peut être n'importe laquelle des grandes plateformes internet. Dans le système humain-plateforme, qui a été le sujet de ce mémoire depuis ses premières pages, l'algorithme pourrait être traité comme un être vivant. C'est d'ailleurs une appellation qui est à peine une métaphore si l'on considère les LLMs, large langage machine, que l'on nomme de nos jours ``intelligence artificielle". Comme l'être vivant sur terre, l'algorithme sur le réseau a atteint un stade d'autonomie. Pour bien des algorithmes, le code informatique est bien trop vaste et compliqué pour être connu et compris par un seul homme. 
Les algorithmes autonomes, suivant leur savante programmation se sont partout infiltrés dans le réseaux, et entre les utilisateurs du réseaux, pour devenir des organes à part entière de l'immense réseau, organes avec lesquels nous ne cessons de communiquer, à qui nous offrons des données et recevons en échange des suggestions. La formation de ces suggestions est infiniment plus rapide que le traitement par nos systèmes nerveux.

Cependant, ces machines ne sont capables que d'imposer le même, partout où elles s'installent. Par ces répétitions, elles homogénéisent, elles standardisent. L'algorithmique computationnelle contemporaine, appliquée à grande échelle via la politique des plateformes mondiales est l'autoroute qui amène le système humain vers ce conglomérat humain-réseau mondialisé et vidé de ses potentialité, et donc maximalement anthropique. Voici la mécanique de la disruption qui est à l'oeuvre sur les réseaux actuellement. Cette perspective permet d'entrevoir le moteur de la disruption, ce qui la meut, malgré son absence évidente d'attrait intrinsèque. La disruption est une mécanique naturelle qui s'enclenche dès qu'on ne lui résiste pas activement par la création de néguanthropie.

Si nous postulons comme loi l'augmentation de l'anthropie, qu'est-ce qui permet de créer de la néguanthropie, de contrer la tendance ? Il nous faut postuler que c'est l'effort conscient et commun des population humaines.  Nous verrons plus bas, lorsque nous considérerons les plateformes comme Wikipédia, que ce que nous entendons par la création de néguanthropie est la collaboration \textit{non-médiée par les algorithmes}.

\subsection{Injection de la calculabilité et la calculabilité des savoirs}
\label{sec:calc}

Les algorithmes sont avant tout des outils de calcul et de comparaison numérique. Tout le traitement qu'ils peuvent faire d'un sujet doit le présenter sous cette forme calculable et cette forme seulement. Plus haut dans la section \ref{sec:Eco}, nous avons rattaché la standardisation à la captation des savoirs par la machine, savoirs qui se trouvent donc soumis à l'impératif de calculabilité.  Cette calculabilité se manifeste sous des formes diverses et nous commencerons donc par sa facette la plus évidente et palpable: la calculabilité est la possibilité de produire des tris, des classements, des arbitrages purement numériques.

\paragraph{L'évaluation:}
La troisième vague du capitalisme, comme l'ont montré Chiapello et Boltanski, a fait un large usage de l'évaluation généralisée dans son discours légitimateur. Le système a donc commencé à recommander des évaluations à grande échelle, d'abord des employés par les manageurs, et ensuite des professeurs par leurs étudiants, et aujourd'hui une large part des secteurs, public et privé, proposent une certaine forme d'évaluation, sous forme de ``êtes-vous satisfait du service de ... ?". Cependant, l'évaluation a atteint son stade le plus systématique et multilatéral avec l'essor des plateformes, qui demandent au consommateurs d'évaluer son marchand, au marchand d'évaluer son acheteur, au deux d'évaluer l'environnement, la plateforme ect ... Le mot d'ordre est la satisfaction. Ces évaluations servent de base pour classifier et suggérer, en d'autre mots, de calculer. Ce mot d'ordre tend par exemple, pour Airbnb, à rendre la qualité et l'hospitalité d'un accueil sanctionnable par une note, sur cinq étoiles, pour Amazon, la qualité d'un livre, aussi sur cinq étoiles. Dans le domaine scientifique, le h-index et la plateforme ``Google scholar" offre ce genre de calculabilité et d'évaluation pour le chercheur.  

\paragraph{L'informationnalisation:}
En parallèle de cette calculabilité par l'évaluation et la note, un second mode de calculabilité, sans doute moins évident et visible, touche la forme du savoir contemporain. Il s'agit de ce que l'on pourrait appeler ``l'informationnalisation du savoir" et celui-ci concerne la formation même du savoir et le critère de sélection de ce que l'on nomme savoir. La tendance générale a été de n'appeler ``savoir" que ce que l'on peut traduire en information.  

Nous avons déjà vu plus haut, dans la section \ref{sec:Eco}, que ce processus est un parallèle de la mathématisation de la physique, qui a éliminé la physique aristotélicienne et l'a transformée, non tellement en savoir faux, mais plutôt en non-savoir. De la physique aristotélicienne, on a dit, pour de très bonnes raisons, qu'elle empruntait le mauvais langage, un langage non-mathématique. Aujourd'hui, une théorie physique n'est recevable \textit{que} si on peut exprimer celle-ci en termes mathématiques. Cette contrainte restreint drastiquement les possibilités, et c'est sans doute ce qui en fait sa valeur, en plus du fait qu'elle offre des possibilités de vérification et de prédiction inégalées. Cette contrainte à même rejaillit comme une puissance d'explication nouvelle sur le monde, par les mathématiques. On a commencé à dire que le monde était mathématique, ou au moins mathématisable. 

Cet impératif de mathématisation ne s'est pourtant pas arrêté à la physique, mais s'est répandu jusqu'aux sciences sociales, en sociologie, en économie, en anthropologie. L'effet de la mathématisation a été une restriction thématique massive de ces sciences: pour répondre à l'impératif de mathématisation, ces sciences ont dû se délester de beaucoup de questions qu'elles n'ont pas su exprimer en termes mathématiques. Ces questions sont donc partiellement sorties du domaine du savoir. Les questions restantes étant typiquement de l'ordre du processus, des chaînes de causalité, du ``comment", le savoir s'est réduit à un catalogue, à une hiérarchie, et à des lois de processus. Ces lois de processus, sont justement ce qui est systématisable et calculables dans un algorithme. Exprimé en tant que lois de processus, ce savoir peut alors tout naturellement devenir information pure. 

La question qui émerge naturellement de cette discussion est: quelle est la forme que le savoir prendra lorsque le processus de ``l'informationnalisation du savoir" sera achevé ? Cette forme, il me semble, reste largement mystérieuse, même si elle prendra certains des traits que nous avons déjà abordé: effacement de la théorie et du modèle, perte de la direction de la recherche en science. Quoi qu'il en soi, former des conjectures sur cette évolution est bien au delà des objectifs de ce mémoire.

\paragraph{La disruption comme augmentation de la calculabilité:} Cette observation nous offre de la disruption une définition très précise et singularisant presque une mesure de l'avancement de celle-ci. Ainsi, la disruption est le processus qui transforme systématiquement les savoirs incalculables, en informations calculables. Nous pouvons donc mesurer l'avancement de la disruption dans un secteur en mesurant la quantité de savoirs réellement en circulation dans ce secteur, ainsi en l'opposant à la standardisation. Dans un MacDonald, quelles connaissances sur la cuisine sont en circulation ? Si la réponse est ``presque aucune", alors ce domaine est largement standardisé et disrupté, comme nous pouvons l'observer directement. 

\subsection{Un détour par l'acteur-réseau et le réseau-interface}

Dans les lignes précédentes, nous avons tenté d'aborder les mécaniques de la disruption au travers d'une relation entre l'internaute et le réseau-interface, que nous approfondirons également plus tard. Cependant, à ce stade, il est bon de voir en quoi cette interaction est crucialement différente de l'interaction acteur-réseau proposée par Michel Callon, Bruno Latour, Madeleine Akrich, notamment. 

Pour l'une des premières fois, cette théorie accorde aux objets techniques (comme les machines, logiciels, lois ou infrastructures) une forme d’agentivité qui en fait des acteurs et donc des actants dans le système. Ils réunissent donc dans la même approche l'humain et le non-humain. Ainsi, depuis cette perspective, un acteur existe aussi et surtout par ses relations: son pouvoir vient du réseau d’interactions dans lequel il est inséré. Ainsi, une porte automatique ou un algorithme peut imposer un comportement, au même titre qu’une norme sociale. Cette théorie de l'acteur-réseau considère donc les réseaux comme des assemblages concrets et instables d’éléments hétérogènes contenant des humains, des objets, des institutions, des discours, ... qui interagissent et qui forment ensemble le social. La traduction opérée entre les différentes strates du discours est centrale dans cette théorie et permet le voyage des connaissances dans les différents milieux sociaux, notamment entre les milieux scientifiques et non-scientifiques, c'est ce que nous appelons la vulgarisation. 

Précisons tout d'abord la notion de réseau dans la théorie de l'acteur-réseau: un réseau est un agencement qui est à la fois matériel et symbolique et est composé d’éléments hétérogènes. Ces éléments sont reliés par des interactions effectives, le tout formant le réseau. Le réseau est donc le produit d'interactions sous-jacentes, qui se forme par elles tout en les permettant. 

Dans le cadre de l'interaction entre l'internaute et le réseau, nous avons eu en réalité deux types de réseau présents dans notre discussion, mais n'en avons réellement discuté qu'un seul. Dans un premier temps, il y a le réseau internet matériel et l'ensemble de son infrastructure technique et dans un second temps le réseau-interface, que nous avons aussi appelé plateformes, qui relient un ensemble d'acteurs, parfois avec des fonction différentes, tout en s'interposant entre ces acteurs. Ce second type de réseau à la fois médie et s'interpose. En cela, il se distingue par de nombreux points de la notion du réseau-acteur développé par Michel Callon, Bruno Latour, Madeleine Akrich. 

Là où l'acteur-réseau est un produit des interactions et se co-construit avec celles-ci, le réseau-interface se présente comme le seul moyen permettant les interactions dans un certain but. Par exemple, l'interface Airbnb est ce qui rend possible l'interaction entre le locataire de courte durée et les hôtes. Ce type de réseau, avec pour exemple type Airbnb, Uber, UberEats, ect ... a pour but de coloniser un marché et d'en faire un monopole ou un quasi-monopole. En ce sens, il n'est pas un réseau informationnel qui permet la traduction des informations entre différentes strates sociales, mais il est un réseau de classification, par exemple de la qualité des hôtes et de la suggestion. 

Ceci s'oppose à la l'approche théorique de Michel Callon, Bruno Latour, Madeleine Akrich qui essayaient avec leur théorie de briser l'asymmétrie originelle entre la machine et l'humain pour les mettre sur un pied d'égalité dans le réseau, et par la suite les mettre en interaction, par exemple pour l'interaction humain-algorithme. Si on peut voir en effet en creux une interaction de co-apprentissage entre les algorithmes et les utilisateurs, par exemple dans le cas de chatGPT et Google, qui donne des informations aux utilisateurs tout en collectant des données statistiques sur ceux-ci, cette interaction semble construite sur le modèle de la prédation qui crée une asymmétrie de fait entre l'algorithme et l'utilisateur. Le réseau-interface, ou plateforme, se répand en s'insérant dans une relation commerciale entre un consommateur et un marchand, tout en évaluant les deux. En simplifiant cette interaction, il tend à se rendre indispensable, et par la suite modèle l'offre marchande par la suggestion. 

On voit donc aussi en quoi la relation le réseau-interface, ou plateforme, qui a émergé à la fin du Web 2.0, implique un type de socialité drastiquement différente de celle des réseaux interactifs, y compris sur internet, tels que les Webblogs, les e-mails, les listes de serveurs, qui mettaient en contact les internautes à des grandes distances et à une vitesse extrême mais ne s'interposaient pas entre les usagers.  En ce sens, par le passé, les réseaux interactifs sur internet pouvaient \textit{être utilisés pour construire des groupes}, mais ne constituaient pas par eux-mêmes une plateforme.

\section{Les conséquences de la disruption}

Alors qu'elle s'est répandu de façon fulgurante sur les réseaux algorithmés, la disruption nous apparait de façon la plus claire par ses conséquences sur l'atmosphère globale. 
Nous explorons maintenant les nombreuses conséquences palpables de la disruption. 

\subsection{La constitution d'un horizon du devenir}

Pour reprendre une méthode de Stiegler et de Simondon, un bon moyen de former de l'intuition  sur les processus humains statistiques est de s'inspirer des lois régissant les systèmes physiques complexes, comme dans le cas des transitions de phase chez Simondon, ou de l'entropie chez Stiegler. En physique, si on considère un milieu caractérisé par un temps de réponse donné, souvent nommé la vitesse du son dans le milieu, et que l'on y injecte un objet ou une information qui se répand dans ce milieu plus rapidement que la vitesse du son, alors se crée une onde de choc, et un horizon. 

    Un exemple bien connu est celui de l'avion supersonique qui forme une onde de choc, donc un horizon, se répandant plus vite que le son, ou celui du trou noir qui crée un horizon des événements. En filant cette analogie, si au sein des systèmes sociaux on injecte une perturbation, ici les nombreuses ruptures qui viennent de la disruption, qui évolue et change constamment sur un rythme plus élevé que la vitesse d'adaptation  des systèmes sociaux, alors nous nous attendons à voir apparaître un horizon lié à cette perturbation. 
    
    Le mécanisme lié à cette apparition d'un horizon se comprend aisément par le concept d'emballement. Alors que nous sommes capable de penser le futur à partir de l'état de nos systèmes sociaux actuels et avons une certaine intuition de la façon dont ceux-ci s'adaptent, par exemple en édictant des lois, si les techniques que sont censées structurer ces systèmes sociaux évoluent plus vite que ceux-ci et si nous ne savons prédire comment les techniques évolueront, car aucune rationnalité ne préside à leur apparition, alors on ne saurait même essayer de prédire l'évolution de nos systèmes sociaux et techniques.  Un exemple devenu aujourd'hui anxiogène est celui du remplacement des emplois humains par des intelligences artificielles. La forme du marché de l'emploi a un horizon relativement proche est aujourd'hui un impensable. 

La disruption a donc pour conséquence la constitution d'un horizon du devenir, une limite au delà de laquelle le futur est impensable et imprévisible. Une série de discours concordent à cette apparition d'un horizon du devenir, nous en analyserons deux en particuliers; la singularité par l'effondrement, et la singularité par la technologie. Bien qu'en tout point opposées par les milieux et par origines sociales de ceux qui les défendent, ces deux visions  du devenir concordent sur l'imprédictibilité du futur. 

\paragraph{La singularité par l'effondrement du système terre}

Cette vision repose sur le fait que notre monde actuel, ainsi que notre mode de vie, commence à agir sur l'environnement de manière significative. Elle se base sur l'observation factuelle que chaque année une fraction significative des ressources fossiles, gaz, charbon et pétrole présentes dans la croûte terrestre est extraite pour satisfaire aux besoins de la machine économique et que les déchets que le l'humanité a rejeté au cours de ces deux siècles post-industrialisation ont commencé à modifier significativement le système terrestre, aussi bien le système climatique, que les chaînes animales et végétales qui forment le biotope.  

Le premier exemple marquant d'un tel événement concerne le trou dans la couche d'ozone, événement au cours duquel l'humanité a pu constater que son action pouvait agir sur les processus physiques à l'oeuvre dans l'atmosphère et en modifier largement la composition. A cette occasion, la réponse institutionnelle s'est étalée sur une dizaine d'années et a permis l'interdiction de la vente et de l'utilisation des composants chimiques dissociant l'ozone. Ce n'était pourtant qu'une première secousse et rapidement, à partir des années 90, les sciences du  climat ont alerté sur un autre changement en cours dû aux émissions de la transformation entropique des combustibles: le changement du climat.  En même temps, les biologistes alertaient sur l'effondrement de la population de nombreuses espèces sur terre et en mer et sur la disparition même d'un nombre croissant d'espèces.  

Ces quatres modifications du système terre (raréfaction des ressources, modification du climat, perte du biotope, accumulation des éléments polluants dans l'environnement), à cause de la non-linéarité des réponses du système terre, rendent les prédictions à long terme très difficiles. Par exemple, en ce qui concerne la modification du climat, d'après les modèles physiques qui tentent de suivrent ces changements, les boucles de rétroaction (autrement appelées non-linéarités) rendent le système en son entièreté chaotique, dans le sens où une perturbation assez grande des  conditions initiales peut amener le système arbitrairement loin de son point initial.

Pour comprendre ce que cela soustend, il faut rappeler que la modélisation des systèmes repose sur l'analyse de ceux-ci proches de leur position d'équilibre, et sous des perturbations faibles. La description des systèmes fortement perturbés est beaucoup plus délicate et amène à des états finaux que la théorie souvent ne peut pas anticiper, ni décrire. Ce sont les fameux points de bascule, que l'étude linéaire des systèmes ne peut pas anticiper et au delà desquels aucune prédiction n'est fiable.  En d'autres mots le modèle du climat actuel est bien compris tant qu'on lui impose des perturbations relativement faibles.
Une perturbation suffisamment grande amènerait le monde dans un état que nous ne pouvons pas anticiper, ni imaginer, c'est ce que beaucoup appelle l'effondrement, car nous savons seulement que nos sociétés seront très différentes de ce qu'elles sont maintenant, mais ne n'avons aucune idée de ce à quoi elles ressembleront exactement.

Prédire l'évolution à long terme du système-terre, simplifié à l'extrême et soumis à ces quatre perturbations durant une longue période fut l'un des objectifs des études du Club de Rome, dont les conclusions furent publiées dans le fameux rapport Meadows.  En utilisant un modèle informatique, celui-ci montre qu'une croissance exponentielle conduit inévitablement à un effondrement des conditions de vie et des populations humaines et animales. Il met en évidence les liens entre population, industrialisation, pollution, production alimentaire et ressources naturelles. Le rapport préconise une stabilisation volontaire de la croissance pour atteindre un équilibre durable. Les prédictions sur l'avenir des sociétés humaines notamment par les rapport du GIEC font également valoir une grande incertitude et une possibilité d'un effondrement. 

Ainsi, pour conclure, les larges perturbations auxquelles nous soumettons le système terre, couplée à la non-linéarité intrinsèque de celui-ci rend la prédiction des conditions futures presque impossible.

\paragraph{La singularité par la technologie
}

Cette singularité par la technologie, souvent juste nommée \textit{singularité}, prédit que l'amélioration de l'intelligence artificielle, notamment renforcée par la multiplication des données et le big-data, pourra s'automatiser dans un futur assez proche. Cette autonomisation de l'amélioration de la machine par la machine, découplée de la nécessité de l'apport humain, créera une spectaculaire accélération de l'amélioration de la puissance de la machine. L'innovation se faisant par la machine elle-même, ses créations sont par définition inimaginables par le cerveau humain. Tout comme la singularité par les changements du système terre, la singularité par la technologie créée un horizon au delà duquel l'imagination humaine est impuissante à se représenter le futur. 

\paragraph{L'horizon de la disruption
}

Comme on peut le voir, ces deux positions que tout semble opposer,  amènent toutes deux à des avenir inimaginables, irreprésentables, et au final impensables. Aucune des deux n'est proprement positive, les deux nous racontent qu'au delà de cette singularité, le monde sera entièrement différent. Plongé dans le contexte actuel, l'homme se retrouve tout proche d'un futur pour lequel il doit se préparer sans avoir aucune idée de ce à quoi il peut ressembler. Cette omniprésence de l'horizon du futur est sans doute une première dans l'histoire humaine. 

Cette tendance, toute actuelle, se marie avec l'attrait que beaucoup semblent porter à la fin des modèles, telle que présentée plus haut dans la section \ref{fin_modeles}. Or le modèle scientifique est ce qui nous permet, comme nous l'avons vu, de former et développer une heuristique sur les systèmes physiques, de créer de l'intuition par rapport à ceux-ci. La fin du modèle scientifique, évincé par le big-data nous rendrait incapables de prédire le futur, et nous priverait même de l'envie de le prédire, menant ainsi à l'inutilité du modèle. En conclusion, on peut voir ces deux tendances comme une pente psychologique pour une paresse intellectuelle, un plaisir du laisser-faire et l'oubli de son futur. Cette pente amène pourtant un effet des plus pervers, celui de l'effacement de l'époque, tel que décrit par Stiegler, comme nous l'avons déjà abordé également dans la section \ref{fin_modeles}. Cette présence de l'horizon du futur travaille de concert avec la disparition de l'idée de progrès pour nous enfoncer dans un constant présentisme psychique, lequel ne saurait être que consumériste et a perdu tous les aspects possiblement sacrificiels qu'avaient une philosophie du futur, ou de la construction du futur. 

En guise de résumé, nous pouvons conclure ici que, alors que le futur a longtemps été pensé à travers des régimes religieux, humanistes, scientifiques ou politiques — du salut messianique à l’émancipation par la raison, du progrès technologique au dépassement dialectique des sociétés de classes —, ces matrices d’anticipation collective, aujourd’hui en crise, laissent place à un \textit{présentisme algorithmique} où l’horizon du devenir impose une limite imprédictible à notre projection collective. Ainsi, l’effondrement des philosophies du projet entraîne une perte de la capacité à se projeter, privant les sociétés d’un sens partagé du temps.

\subsection{La désindividuation collective et psychique par le réseau}

Alors qu'il semble s'être noyé dans le réseau, on peut se demander où est l'individu, ce qui le caractérise, et ce qui reste de la substance sur laquelle la philosophie a longuement disserté.  

Rappelons de nouveau, après les rappels déjà prodigués dans la section \ref{sec:Simondon}, les mécanismes d'individuation tels que présentés chez Simondon. Chez  Simondon, l’individuation est un processus ontologique par lequel un individu se constitue à partir d’un champ de tensions préindividuel, plutôt que d’être donné d’emblée comme un être fixe. Ainsi, au contraire des approches classiques qui partent de l’individu comme d'une substance ou d'un mode, Simondon place en premier le devenir de l’individu. Ce processus d'individuation se déroule dans un milieu associé, c’est-à-dire un environnement qui participe activement à la genèse de l’individu. L'individuation a lieu lorsque la matière préindividuelle métastable, riche en tensions se transforme à sa surface, par réaction avec un réactif, terme inspiré de la cristallisation en physique (Simondon part souvent d’exemples comme la goutte de solution saturée où un cristal se forme autour d’une impureté ou d’un grain). En ce sens, l’individuation, qui est l'effet de l'interaction entre le subtrat, encore largement préindividuel, et le réactif, est toujours incomplète, laissant subsister un reste préindividuel qui peut engendrer de nouvelles transformations. Dans l'approche de Simondon, l’individu et son milieu se co-constituent, et l’être est donc fondamentalement relationnel.

A partir de cela, on peut étudier quel modèle théorique d'individuation correspond le mieux à l'individu de notre époque. En dressant le portrait avec des traits grossiers, l'évolution des trois siècles passés a vu le déclin  de l'individu souverain, citoyen, et porteur de la raison des lumières en faveur de l'individu consommateur, non pas un homo œconomicus qui n'a jamais réellement existé, mais un individu inséré dans un réseau d'interactions ayant pour but d'abord de le pousser à la consommation. Le réseau était ici le réseau marchand, auquel l'individu se connecte périodiquement et qui le transforme et le modèle, d'abord en lui adjoignant une série d'objets connectés à ce réseaux marchand, comme la télévision, pour ensuite capter ses protentions, comme expliqués plus haut dans le chapitre \ref{chap2}. Cette absorption par la réseau est devenue complète lorsque les plateformes se sont implantées sur internet et que tous les aspects de la vie des individus se sont réseautisés via Linkedin, Facebook et Tinder.

Ainsi l'``individu  plongé dans le réseau"  est substitué à l'``individu entendu comme substance". Cet état de fait est d'autant plus marquant qu'il est en contradiction permanente avec les modèles promus publiquement par le système. Là où la captation des protentions travaille à la destruction de l'intériorité, les livres de développement personnel appellent tous et chacun à trouver leur vrai ``soi" en eux-mêmes, à interroger leur intériorité et à s'extraire des attentes sociales.  On ne peut s'empêcher de remarquer la contradiction de cette injonction, ``Pour ne pas être un mouton, il faut se conformer au modèle général", lorsque l'on se rappelle que ladite intériorité, qui n'est autre dans ce cas-ci que les protentions individuelles, a elle-même été activement détournée pour la réseau pour répondre aux besoins marchands. C'est ainsi qu'en interrogeant leur intériorité tous et chacun répondent en coeur que leur désir est d'avoir un travail libre d'auto-entrepreneur flexible leur permettant de profiter d'un lifestyle de voyageur tout en n'ayant aucune attache autre que les réseaux, car les réseaux lui permettent en tout lieu de subvenir à ses besoins.

\paragraph{La nature du réseau-interface}
On peut alors se demander la fonction que remplirait le réseau-interface dans la théorie de l'individuation de Simondon. Le réseau ne peut être ni le milieu préindividuel, ni un réactif aidant à l'individuation. Dans l'interaction entre le milieu préindividuel et le réactif, il est ce qui capte, ce qui se met entre les deux et modèle le milieu préindividuel à son image. Pour comprendre le danger, il faut élargir l'idée du système en interaction au delà du simple individu, et aller vers la multitude des milieux préindividuels. 

Là où les systèmes sociaux sont un réactif qui en entrant en contact avec un certain milieu préindividuel produisent un individu qui est le fruit des tensions inhérentes et idiosyncratiques à ce milieu préindividuel, les réseaux tendent à dissoudre ces tensions et à informer le milieu préindividuel, en lui imposant sa propre forme. La réseau en ce sens semble appartenir au schème hylémorphique, dans lequel il jouerait le rôle de moule. Notons que la façon dont le réseau dissout les tensions du milieu préindividuel originel ne s'apparente pas à une prise des protentions par la violence ou à une imposition d'une volonté externe par dessus les tensions du milieu préindividuel. Bien plutôt elle est une suspension, une désactivation \textit{momentanée} de ces tensions. Par son intervention, le réseau empêche la réaction entre le milieu et le réactif. L'interface du réseau devient une interface entre le milieu préindividuel et le réactif. On peut se convaincre de ce fait en expérimentant une consommation quelconque (je joue par exemple aux jeux vidéos). Par la consommation, les tensions internes se soulagent pour un temps (ici le temps du jeu) sans pour autant disparaître. En détournant les protentions psychiques, le réseau enferme les tensions irrésolues et les isole tout en offrant périodiquement un moyen de les soulager.

En résumé, l'interface plateforme empêche la formation de protentions psychiques et se rend indispensable tout en empêchant l'individuation. Elle isole l'individu du collectif en échange d'une suspension momentanée de ses tensions internes et d'un remodelage de celles-ci. Elle agit comme un moule.

Dans les mots de Bernard Stieger \cite{Disrup}, cela mène à une vaste désindividuation collective

\begin{quote}
  \textit{Avec le social engineering, ou le social networking, les groupes sociaux sont frappés comme jamais auparavant par la désindividuation collective, le social étant désintégré à sa racine même, c'est-à-dire en partant des rétentions secondaires psychiques, qui sont elles-mêmes à l'origine des rétentions secondaires collectives, et en les privant par avance de toute possibilité de protentions psychiques et singulières, c'est-à-dire, aussi bien, de capacité de projection dans des processus d'identification ouvrant à des processus d'idéalisation.}
\end{quote}

Les protentions collectives que sont les religions, les spiritualités sont dissolues par l'interposition du réseau et sont ensuite reprises sous la forme de ce que nous avons appelé ci-dessus, dans la section \ref{sec:petite_ques}, des petites spiritualités, au côté des petites questions métaphysiques. La religion comme la spiritualité,  perd son sens collectif, elle devient une affaire de conviction où chacun a sa façon de croire, de pratiquer, de spiritualiser. Ainsi, une autre conséquence de la disruption par la digitalisation est la désindividuation collective, une tendance d'autant plus étonnante et paradoxale que l'on condamne gaiement l'individualisme forcené de notre temps.

Ce processus, de perte des protentions secondaires collectives, suite à la captation des protentions psychiques par le réseau, et un soulagement en surface des tensions du préindividu, mène à la perte de la raison de vivre, en tant que convergence commune de protentions psychiques: 
\begin{quote}
  \textit{Les groupes sociaux frappés par la désindividuation collective sont et seront de plus en plus nombreux à perdre toute raison de vivre, à perdre ainsi la notion même de la raison comme convergence des protentions.}
\end{quote}

Pour Bernard Stiegler, cette désindividuation collective, et la perte de raison de vivre associée est la cause première du nombre croissant de suicides dans les sociétés industrialisées.

\subsection{Perte des facultés humaines: la critique platonicienne des plateformes}

Les nouvelles avancées techniques ont toujours été l'objet de la critique philosophique la plus serrée, notamment parce qu'elles raviraient les humains de certaines de leurs facultés fondamentales. L'un des exemples les plus célèbres de telle critique est la critique de l'écriture de Platon. Dans le Phèdre, Platon critique l’écriture en la présentant comme un substitut trompeur de la mémoire et du savoir vivant. Par la voix de Socrate, il raconte le mythe du dieu Theuth, inventeur de l’écriture, et du roi Thamous, qui rejette ce don en affirmant qu’il affaiblira la mémoire des hommes. Le premier point de la critique est donc celui de l'affaiblissement d'une faculté humain fondamentale, celle de la mémoire, à cause de l'utilisation abusive et malavisée de l'écriture. Cette critique est légion par exemple aujourd'hui en ce qui concerne le développement et l'utilisation des LLMs, tels que chatGPT et autres IA génératrices de texte. Par exemple dans cette étude du MIT\cite{2025arXiv250608872K} largement relayée par les médias (et aussi largement déformée par ceux-ci), les auteurs cherchent à montrer que les utilisateurs assidus de chatGPT pour l'écriture d'essai ont des capacités cognitives diminuées lors de l'écriture d'essai sans le support de LLM. Ils argumentent donc que malgré ses nombreux avantages pratiques, l'usage abusif de LLM a un coût cognitif qui correspond au déficit d'entraînement que son usage implique. Cette critique est largement protéiforme et a été adaptée à la vaste majorité des grandes inventions techniques, surtout juste après leur mise en place. Dans le cas des plateformes, et des réseaux sociaux associés, la critique a porté sur la perte de la capacité à nouer des liens spontanés et à former des groupes. L'usage des plateformes nous désapprend à faire la fonction qu'ils rendent facile pour nous, à savoir la socialisation, et donc, en plus de porter une socialité marquée par la superficialité, la plateforme se rend indispensable pour l'initier. 


Le second flanc de la critique de l'écriture concerne la fausseté de la reproduction écrite par rapport à la production orale. L’écriture, selon Platon, ne transmet pas une véritable connaissance, mais seulement l’apparence du savoir, car elle ne permet ni dialogue, ni questionnement. Un texte écrit ne peut ni se défendre, ni s’adapter à son lecteur, contrairement à la parole vivante qui s’ajuste à l’interlocuteur. Ici, Platon valorise donc la dialectique, l’échange oral où la pensée s’aiguise par la confrontation. Il voit dans l’écriture un outil figé, incapable de produire une pensée vivante et en mouvement. 

Cette critique se retrouve exactement dans la même forme en ce qui concerne l'usage des LLMs, qui amolirait les capacité cognitives, par son manque de réalité et d'authenticité. Cette même critique est adressée aux plateformes, sur lesquelles la fonction qui leur est attribuée, à savoir différentes facettes de la socialisation, est déformée, superficielle et inauthentique en comparaison de l'interaction réelle.

En creux de cette critique de la déformation, se déroule la critique platonicienne de la superficialité et de l'illusion contre l'essence. L'écrit ne peut porter l'essence de la vérité comme la plateforme ne peut porter l'essence de la sociabilisation. De la même façon que l'apparence est trompeuse chez Platon, car elle déforme, les plateformes nous montrent une vision déformée et excessivement glorifiée de nos amis et connaissances. On peut les interprêter comme les cavernes de Platon modernes, lesquelles nous enferment dans l'immobilité et la contemplation du faux, nous éloignant de la vérité. 

Cette critique est bien connue et s'applique presque sans heurt aux plateformes. Le prolongement naturel de cette critique est l'analyse sanitaire d'un monde gorgé de plateformes et d'écrans. L'étude de l'impact des plateformes reste largement à faire, cependant l'impact sanitaire des \textit{écrans} a fait l'objet de nombreuses études dites sanitaires que nous résumons maintenant.

\paragraph{La question sanitaire}

Au delà du soupçon de manipulation, l'essor des technologie numériques modernes soulèvent un problème sanitaire majeur. En effet, si le sujet principal de ce mémoire est l'exploration des impacts sur la psyché des individus et sur la structure des sociétés des nouveaux outils numériques, on ne saurait éviter de rapidement esquisser ce que la recherche en épidémiologie à commencer à conclure sur l'impact des écrans et de leur usage sur l'être humain, et son état physiologique et psychologique.

On connait très bien le vague argument qui va suivre. L'homo Sapiens est un animal social, qui a vécu la plupart de son histoire évolutive en petits groupes, parfois chasseurs-cueilleurs et nomades, parfois déjà cultivateurs et sédentaires. Ce sont ces milieux qui ont façonnés son patrimoine génétique, lequel contrôle encore en partie notre mécanique biologique.
Il est clair que l'être humain ne peut pas être adapté aux écrans, de même qu'il est clair qu'on ne sait pas encore vraiment ce qu'ils peuvent lui faire. Tout cela parce que nous avons été pris de court, nous avons été disruptés, ne serait-ce que parce que entre 1997 et 2014 le temps d'écran moyen des enfants entre 0 et 2 ans a doublé \cite{EffetEcran1}.

L'effet des écrans sur le développement physiologique et psychologique des enfants a été récemment étudié dans de nombreuses études, que l'on peut voir résumées dans l'article de revue\cite{LISSAK2018149}. Ces articles recensent les effets néfastes de la consommation d'écran sur le développement des enfants et des adolescents. Si la plupart de ces effets sont encore en cours d'étude et manquent de robustesse statistique, ils suffisent déjà à alerter sur la possible ampleur des conséquences de la prolifération des appareils avec écran dont nous usons abondamment. Jusqu'à maintenant, les effets physiologiques recensés des écrans incluent des conséquences sur le sommeil (avec un baisse de la durée du sommeil et de la qualité de celui-ci), une hausse des maladies cardiovasculaires liées à la sédentarité (notamment via une augmentation de la pression sanguine, une augmentation de la prépondérance de l'obésité, et de cholestérol), une augmentation du stress via une dérégulation du cycle circadien de cortisol, une augmentation de la prépondérance de la myopie, une diminution de la densité osseuse en moyenne. 

Par dessus ces effets physiologiques se rajoutent également une série d'effets psychologiques parmi lesquels on peut dénombrer une plus grande prévalence de dépressions et de pensées suicidaires (lesquelles sont probablement liées au déréglement des rythmes du sommeil et au stress et à l'anxiété amenés par l'échange de messages), une augmentation de la prévalence des troubles de l'attention et de l'hyperactivité, la consommation de jeux vidéos a été liée à une prévalence des comportements anti-sociaux. 

    Finalement, le multi-tasking associé avec l'usage de plusieurs réseaux au même moment est relié à une diminution de la densité de matière grise et  semble engendrer  une augmentation du neuroticisme, et des comportements impulsifs. Cette addiction aux écrans est aussi corrélée à une vie sociale plus pauvre, comme par exemple une diminution  de l'attachement familial et du support social; perte d'empathie et pauvre contrôle émotionnel et une augmentation des comportements violents et dangereux. Corrélativement, les gros utilisateurs d'écrans reportent une satisfaction de vie plus faible que le reste de la population. Notons cependant que les auteurs de \cite{EffetEcran1} objectent que les effets néfastes sur la cognition peuvent être modulés par le contexte, notamment parental, et dans certains cas s'effacent entièrement.

    Cependant toutes les études en question ont eu lieu entre aujourd'hui et 2010 et les auteurs assurent unanimement que des études plus approfondies sont nécessaires pour renforcer les conclusions déjà mentionnées ci-dessus. La recherche, qui prend du temps, a toujours un temps de retard sur les dernières innovations techniques, lesquelles ont maintenant un temps de remplacement en dessous de la décennie. Alors que tout le XXème siècle a vu l'avènement et la lente progression de la télévision, la fin de celui a vu le PC fixe s'implanter à un rythme plus soutenu dans les foyers. Le début de ce siècle à vu le déferlement d'outils portables de plus en plus miniaturisés, avec le PC portable, le laptop, la tablette, le téléphone portable, la montre connectée. Cette accélération de l'offre technologique brouille les conclusions scientifiques qui ont été tirées jusqu'à maintenant.

\subsection{Dénoétisation de l'esprit}
\label{sec:deno}
 L'ensemble des dynamiques décrites ci-dessus mènent à ce que Stiegler nomme la \textit{dénoétisation} de l'homme moderne. La dénoétisation désigne la perte progressive de la capacité à penser par soi-même, provoquée par l’automatisation croissante des fonctions cognitives à travers les technologies numériques. Le catalyseur de cette perte se trouvant largement dans la suggestion constante que les algorithmes imposent aux individus.
 
 Selon Stiegler, le cadre de la dénoétisation est plus large et  s'inscrit dans un processus de prolétarisation intellectuelle, où les individus ne participent plus activement à la production de savoirs mais deviennent passifs face aux flux d'informations. Les dispositifs techniques, en particulier les algorithmes, capturent l'attention, court-circuitent la mémoire et détournent la raison critique. Les individus subissent alors une double perte: la perte de compétence et la perte de puissance d'agir. A la limite, quand les compétences et les savoirs ont été entièrement captés par les algorithmes, les individus deviennent impuissants.

 Pour illustrer ce processus et cette perte, prenons quelques exemple empruntés à la vie courante. Dans un premier temps, celui du GPS. Là où l’orientation dans l’espace mobilisait autrefois des compétences fondamentales telles que la mémorisation de l’environnement, la construction de cartes mentales ou même la capacité d’anticipation, le GPS les supplante en guidant pas à pas l’utilisateur. Il n’est plus besoin de comprendre où l’on va, ni de réfléchir, on suit simplement une voix numérisée. Cette externalisation, qui correspond exactement à ce que nous entendons par captation de savoir technique par la machine, de l’orientation spatiale conduit à une atrophie de certaines régions cérébrales, notamment l’hippocampe. Pris dans son ensemble, l'usage du GPS  est une forme concrète de dénoétisation : une compétence jadis centrale à l'humain est peu à peu abandonnée à la machine.

Dans le domaine intellectuel, ce même phénomène prend la forme d’une prolétarisation cognitive. Les plateformes d’aide à l’écriture, comme IA génératives, qui facilitent sans doute l’accès à la rédaction, mais qui peuvent aussi priver les individus de l’occasion d'exercer leur pensée critique et stylistique.  Stiegler parle ici de perte de savoir-faire, celui de rédiger et d'agencer.

Enfin, l’expérience des réseaux sociaux, en particulier de plateformes sociales privent l'individu de former ses propres protentions et ses goûts. Les plateformes de rencontre ont par exemple un algorithmes ``matchant" des individus selon des grilles de préférence artificielles et imposées à l'individu telle que l'origine, l'âge, la taille, les occupations, les loisirs, ...
 
 Nous avons vu comment ce processue de captation induit une forme de désindividuation, c’est-à-dire une destruction des processus par lesquels les sujets se construisent. La dénoétisation est ainsi un appauvrissement de la vie noétique, c’est-à-dire de la vie de l’esprit, qui menace à terme la démocratie et la transmission culturelle.  Le moyen de cette dénoétisation est l’exploitation industrielle de la mémoire et du savoir, où les individus perdent la maîtrise de leur milieu symbolique.

\subsection{La chute de la confiance}

Les protentions communes et la confiance en l'honnêteté avec laquelle les autres individus poursuivront ces protentions communes ont pour conséquence directe la confiance en ces autres individus. Par voie de conséquence, la confiance est un sentiment indispensable à la vie d'un groupe ou d'une institution. Pourtant,  statistiquement la baisse de la confiance générale semble être un fait attesté \cite{Credule}
\begin{quote}
\textit{La sérénité,
l’enthousiasme et le bien-être baissent (par
rapport à la précédente enquête réalisée en
2010). Mais le qualificatif qui augmente le
plus sensiblement est la méfiance : +6 pourcents,
pour atteindre 34 pourcents des répondants. D’une
façon générale, 70 pourcents considèrent qu’on n’est
jamais assez prudent quand on a affaire aux
autres et 38 pourcents pensent que la plupart des
gens cherchent à tirer parti de vous.}
\end{quote}

En même temps, en France, la morosité continue à augmenter\footnote{https://123dok.net/article/travaux-effectu\%C3\%A9s-gallup-rapport-sant\%C3\%A9-europe.qmjr7xdw\#google\_vignette.}. Ce mécanisme du doute généralisé se renforce d'autant plus dans le contexte des réseaux sociaux et, de fait, la baisse de la confiance est une prédiction assez immédiate de l'essor des plateformes et de la disruption des systèmes sociaux. Par un mécanisme de multiplication du bruit et des informations discordantes qui se répandent sur les réseaux sociaux, la confiance de l'auditeur et de l'internaute en ses sources communes d'information s'effondre. Comme le met en évidence Michaël Lainé, dans son ouvrage ``Dans L’ère de la post-vérité"\cite{PostVeri}, cette énorme incertitude que charrient les réseaux par leur surplus d'informations et de prises de position pousse l'individu à se méfier de tout, à faire seulement confiance à sa propre opinion personnelle préconçue, et au final l'enferme dans une position de doute généralisé.  

Dans une certaine mesure, la méfiance généralisée, et son essor que nous venons de présenter, bien que particulièrement suspects au premier abord, ont trouvé toute une série de défenseurs inattendus, dans les discours légitimateurs de la science elle-même. En effet, cette perte de la confiance a parfois été glorifiée dans le discours public comme une avancée de la science, profondément à tort. Certes, le récit fondateur de la science, en tant que producteur de vérités certaines, à l'opposé des vérités banales du quotidien, se devait de passer par l'exceptionnalité de la méthode de la science. Ce récit scientifique ressasse donc l'importance de la révision par les pairs, du doute constant, aussi bien sur ses propres certitudes, que l'on appellent parfois aujourd'hui biais, que des certitudes des autres. Ce récit a en grande partie pour cible les scientifiques eux-mêmes, pour les convaincre de l'exceptionnalité de leur discipline. La vulgarisation de ce récit, aussi importante que le récit raconté aux scientifiques, a constitué une loupe déformante de plus. La vulgarisation de la méthode scientifique, nous a dit et répété en l'envi que la méthode scientifique repose sur la remise en question du corpus, sur le questionnement sans arrêt, sur l'absence de certitude.  Cette insistance outrancière a fini par retourner le récit fondateur de la science contre lui-même et a perverti la façon dont une partie du grand public la perçoit, jusqu'à nous faire croire au ridicule slogan: ``la science, c'est le doute"\footnote{Pourtant, cette position toute relativiste a trouvé sa confirmation la plus spectaculaire dans les multiples manipulations dont la science a été la victime dans des sujets sensibles tels que les liens du tabagisme avec le cancer, le lien entre les émissions anthropiques et le trou dans le couche d'ozone, le climat et son évolution, ect ... et dont le cas d'école le mieux documenté est celui des firmes de tabac, qui ont mené une campagne de désinformation sur les relations entre le cancer et le tabagisme s'étendant sur près d'un demi-siècle jusqu'à être mises à jour à la fin des années 90\cite{Doute}. Les marchands de Tabac furent finalement condamnés en 2006. Ce qui ressort de ces péripéties, c'est la mauvaise foi, et l'usage trompeur du doute scientifique: ``la science suspecte que le tabac cause le cancer, mais elle n'est pas certaine, elle doute". ``On ne peut jamais prouver avec certitude qu'un cancer a été causé par le tabac, il peut y avoir d'autres causes, c'est un effet multi-cause". Il est notable que l'on retrouve parfois les mêmes noms a la tête des contre-expertise sur le tabac et des contre-expertises actuelles sur le climat\cite{Doute}. }.

Cette glorification et cet encouragement de la méfiance généralisée a réussi à camoufler que l'un des piliers centraux de l'investigation scientifique est la confiance. La confiance généralisée et par défaut envers les protentions des collègues chercheurs, dont on \textit{doit} croire qu'ils recherchent la vérité. Il ne s'agit donc certainement pas d'une confiance aveugle en les autres praticiens, mais plutôt d'une confiance en leur bonne foi, en leur tendance à l'honnêteté intellectuelle et à la mesure dans leur propos et leurs conclusions. Cette confiance est beaucoup plus cruciale que le doute généralisé qui sert de pastiche à la science dans le discours public. En effet, un scepticisme généralisé envers la bonne foi des scienfiques ferait sans doute s'écrouler tout l'édifice de la recherche actuelle.

 Cette diminution de la croyance en la bonne foi des personnes se redouble alors tout naturellement de la recherche d'un proxy objectif qui nous permettrait de juger du talent ou de la compétence d'une personne, sans avoir à lui faire confiance. On est ainsi encouragé à assigner un nombre, qui sanctionne toute performance de manière à rendre la fiabilité d'une personne facilement lisible de l'extérieur. 

Ainsi, les évaluations se multiplient. Les entreprises évaluent les travailleurs et leur assignent un rendement, les universités évaluent leurs chercheurs et leurs assignent un h-index, le monde évalue les entreprises, les universités et les pays en leur assignant une note de fiabilité ou de performance. Un tel exemple est le fameux triple A qui est décerné aux pays les plus fiables sur le stock market. Ainsi la recherche de la vérité  couplée à l'absence de confiance, pousse alors les individus et les institutions à une recherche d'objectivité sanctionnée par une série d'indicateurs prétendûment impartiaux. 
L'évaluation en tant que méthode de management généralisée, comme le capitalisme tardif la défend, se nourrit de ce doute souverain et le nourrit en promulgant toujours plus d'évaluations. Nous avions argumenté dans la section \ref{sec:calc} que cette généralisation de l'évaluation augmentait la calculabilité et donc l'entropie des systèmes.

A l'instar de la captation des protentions psychiques personnelles et collectives, la chute de la confiance en les personnes, se double d'une chute de la confiance en les institutions. L'effondrement de la confiance du grand public est sans doute le plus remarquable dans le domaine de la politique. Cette perte de la confiance en la politique traditionnelle se marque d'un détournement généralisé, en Europe, des partis traditionnels vers des partis clamant le renouveau tels que ``la république en marche" et des partis de l'extrême, dans l'optique d'une politique du tout pour le tout.

\section{L'arrivée de la nouvelle barbarie}

La trivialisation de l'expérience extra-ordinaire ainsi que sa récupération par les outils de captation de l'attention numérique, ne sauraient laisser nos comportements intacts. Pour Adorno et Horkheimer, ils étaient déjà partiellement la cause de la barbarie des deux guerres mondiales. Cette barbarie, après les séquelles mémorielles qu'elles ont laissées sur les générations passées, et qui commencent aujourd'hui à s'effacer, est sur le retour\cite{Disrup} 
\begin{quote}
  \textit{Durant de nombreux siècles, ce plan de consistance donnant l'hospitalité à l'extra-ordinaire qui advient aussi en toute expérience artistique, et qui en est la condition, fut rendu accessible à travers les expériences - magiques, mystagogiques, spirituelles et religieuses - dont l'art ne peut se séparer qu'avec la modernité. Alors l'art ne tarda pas à être récupéré comme ``expérience esthétique" par l'industrie des biens culturels comme forme de captation de l'attention et par le marché spéculatifs des collectionneurs vénaux et hyper-phillistins: c'est ainsi que commence la nouvelle forme de barbarie aux yeux d'Adorno et d'Horkheimer, dont la post-modernité est un nom.
}
\end{quote}

Pour Stiegler, la conjonction de tous les symptômes décrits plus haut, à savoir la formation d'un horizon du futur qui implique une tendance générale au présentisme, la désindividuation psychique et collective dues à la captation de la libido par les réseaux, la chute de la confiance des individus entre eux et en les institutions, et finalement la perte de l'époque, culminent en un retour de ce qu'il appelle une nouvelle forme de barbarie. Cette barbarie a tout d'abord une définition assez précise, en tant que \cite{Disrup}
\begin{quote}
  \textit{consumérisme et vénalité généralisée ne prenant plus aucun soin du monde dans lequel vivent pourtant les consommateurs et les spéculateurs  
  eux-mêmes.}
\end{quote}

Dans son expression la plus pure et la plus dénuée d'ambiguité, cette nouvelle barbarie apparaît dans le discours de l'homme le plus riche du monde, lorsqu'il théorise que la faiblesse des pays occidentaux, et plus notablement en Europe, réside dans leur trop large empathie\footnote{https://www.youtube.com/shorts/dbbXEOloE8I}
\begin{quote}
  \textit{The fundamental weakness of western civilisation is empathy, the empathy exploit. It is their exploiting of a bug in western civilisation, which is their empathy response. I think that empathy is good, but you need to think it through, not be programmed like a robot.} 
\end{quote}

Malgré son expression un peu confuse, il semble que Musk suggère que l'empathie, toute relative, développée dans les pays occidentaux est une porte d'entrée par laquelle les autres pays peuvent tirer avantage de la richesse des pays occidentaux. Sans l'exprimer directement,  Musk appelle à briser les liens sociaux, à savoir aux autres et à l'environnement, et à  un repli sur soi des individus. 

Cet appel au reli sur soi s'insère parfaitement dans une économie numérique qui le favorise déjà. Dans son récent ouvrage\cite{PostVeri} de Michaël Lainé, que nous avons déjà abordé juste ci-dessus, argumente que les algorithmes, par l'effet de vitesse, et le brouillage constant de la vérité qu'ils induisent, favorisent également un fort repli sur soi de ses utilisateurs\cite{PostVeri}
\begin{quote}
  \textit{L'intolérance à l'incertitude favorise l'ignorance et une confiance excessice en la qualité de ses jugements. On aurait tord d'assimiler la première au vide de l'esprit ou au blanc de la page. C'est presque le contraire : elle se manifeste par des avis et des réflexions. Car l'ignorance à biens des masques. Faute de connaissances ou de compétences, on tend à ramener les choses à soi, à son vécu ou à celui de ses proches. } 
\end{quote}

Ainsi, l'effet d'incertitude que crée l'énorme flux d'informations sur les individus les force à se replier sur eux-mêmes pour décider ce qui est vrai et ce qui est faux. La conséquence immédiate de ce repli sur soi est, selon Michaël Lainé, de moraliser des questions factuelles. Dans son livre, il contine\cite{PostVeri}
\begin{quote}
  \textit{Cela incline à appliquer les principes qui guident ses actions pour comprendre les phénomènes sociaux au sens large, autrement dit la morale. On tendra à considérer la société, la politique et l'économie sous l'angle de la responsabilité et de l'irresponsabilité, du laxisme et de la fermeté, du courage et de la lâcheté, de l'honnêteté ou du mensonge, de la cupidité, de la fraude, ect ... Il s'agit d'un biais, car une grille de lecture morale se substitue à l'analyse scientifique. On moralise dans ce cas, faute d'examiner l'ensemble des points de vue, de soupeser tous les arguments.} 
\end{quote}

Ainsi, la barbarie se manifesterait tout d'abord par une moralisation et une polarisation du débat public.

Comme nous l'avons vu, cette nouvelle forme de barbarie serait une cause toute désignée pour le nombre assez large de suicides annuels dans les pays occidentaux. Elle serait aussi la cause du ressentiment palpable de l'homme moyen qui se déverse actuellement sur les réseaux. On peut d'ailleurs se demander si le ressentiment visible sur les réseaux leur préexistait ou si il est partiellement induit par ceux-ci. Les résultats de l'étude de ce mémoire insistent sur le fait que les réseaux ont largement exacerbé et amplifié les graines d'un ressentiment préexistant. Cette approche offre une porte d'entrée pour l'analyse de l'élection d'un candidat aussi évidemment - obviously - démagogue que Donald Trump aux Etats-Unis. 

Plusieurs ouvrages ont déjà insisté sur le ressentiment croissant des populations américaines, en insistant sur le double discours dont les populations aux Etats-Unis sont la proie: alors qu'à l'internationnale, les Etats-Unis sont perçus comme le pays le plus puissant et le plus riche du monde, une grande part de la population se trouve dans une précarité et un désoeuvrement croissants. Dans ``The Politics of Losing: Trump, the Klan, and the Mainstreaming of Resentment" de Rory McVeigh et Kevin Estep (2019), les auteurs
analysent comment les populations blanches sans diplôme universitaire, dans une situation économique de plus en plus précaire, ont ressenti une perte de statut face à la mondialisation. Ce climat de ressentiment est mis en parallèle avec celui des années 1920 autour du Ku Klux Klan, montrant comment Trump a canalisé ces frustrations. Les auteurs s’appuient sur une analyse statistique des votes pour Donald Trump, montrant que son succès est corrélé avec les comtés ayant subi une perte de pouvoir économique et démographique. Dans ``The Old Is Dying and the New Cannot Be Born" de Nancy Fraser (2019), titre qui pourrait faire référence à l'absence d'époque, et à l'échec du second temps du double redoublement épokhal, 
Fraser analyse la crise de légitimité du néolibéralisme progressiste, dont Trump incarne un rejet populaire, résultat d’un mécontentement culturel et économique massif. 

Il serait sans doute audacieux de faire remonter la large (et double !) adhésion au discours démagogue de Trump à la seule disruption et à la nouvelle prolétarisation qu'elle implique. Néanmoins, il faut insister sur le fait que cette prolétarisation induite par les nouvelles technologiques touche en premier lieu les populations déjà précaires et épargne, pour le moment, les élites intellectuelles. Or, l'élection de Trump peut se comprendre à l'aune de ce rejet et de ce ressentiment vis-à-vis d'une élite intellectuelle qui se serait engagée trop directement dans une lutte pour les questions culturelles et auraient délaissés les problématiques économiques. Trump a d'ailleurs fait du rejet des politique DEI un point central de sa  campagne politique. 

En pratique, ce ressentiment s'est largement répandu sur les réseaux via les forums tel que Reddit, les sections commentaire sur les plateformes comme Youtube et Facebook et a commencé à forger des communautés internationales liées par la rejet du féminisme, via par exemple ce que l'on appelle la communauté red-pill, ainsi que par le rejet des mesures d'inclusions et l'immigrations, dans le cas des suprémacistes blancs. 

Les communautés humaines se forment par la mécanique des interactions humaines, cela étant, ce qui sert de liant à ces communautés dépend du contexte. Les communautés peuvent être liées par un projet commun, comme ont pu l'être les communistes, par une fonction commune, comme la communauté scientifique ou par une haine et un rejet communs, comme c'est le cas des communautés sus-mentionnées. A titre d'exemple, le Youtuber Alex Hitchens totalise presque quatre cent milles abonnés avec des vidéos faisant souvent plus d'un million de vues sur Youtube. Malgré une vitrine positive de ``construction de l'homme", ses vidéos les plus vues \footnote{https://www.youtube.com/@thefrenchitch/videos}, le mettent typiquement en scène dans un débat contre ce qu'il appelle des féministes, dans des vidéos telles que "Cette féministe me fait peur !"; ou "ALEX HITCHENS vs 4 FÉMINISTES!". Ainsi fort d'une rhétorique minimaliste de la répétition de slogans, ces communautés ont su séduire une fraction substantielle de la jeunesse des pays occidentaux.

Même la théorie critique, qui a longtemps eu pour objectif de produire une construction politique viable basée sur le socialisme, a perdu son impulsion théorique et constructiviste pour ne devenir qu'une critique essentiellement négative du système. On pourrait suspecter que ce mutisme de la théorie sociale est la source de la perte de vitesse généralisée de la gauche, que beaucoup ne voit aujourd'hui que comme des moralisateurs stériles.  

Comme nous l'avons vu plus haut, le réseau et l'interface qui lui correspond ne résolvent pas les tensions présentes dans le milieu préindividuel mais lui offrent un soulagement momentané, sur le temps de la consommation. Ces tensions, qui laissées à elles-mêmes sont douloureuses, en l'absence de protentions psychiques et collectives positives, cherchent à se résoudre par la création de liens basés sur le rejet et le ressentiment. L'arrivée de la nouvelle barbarie et son absence de soin et d'attention portés à l'environnement et à autrui sont une expression négative de ces tensions. Il est clair que des individus obnubilés par cette barbarie ne pourront être subjugués que par les apparats les plus grotesques de la force pure.

\section{La calculabilité de l'algorithme et la bifurcation de l'humain}
\label{sec:bemol}

Avant de refermer ce chapitre, revenons sur un des points centraux de l'argumentaire de Stiegler, à savoir celui voulant que l'algorithme ne peut qu'augmenter l'entropie alors que le vivant seul peut créer des bifurcations et donc de la néguentropie. 

L'argument de Stielger est le suivant: l'algorithme, parce qu’il est déterministe (ou tout au moins fondé sur un ensemble de règles formelles, programmées), injecte de la calculabilité dans les processus où il intervient. Cela signifie qu’il tend à réduire l’incertitude, à optimiser, à automatiser des choix, donc à rendre prévisibles des actions autrefois indécidables ou ouvertes. Or, selon Stiegler, cette réduction de l’incertitude correspond à une augmentation de l'entropie, sociale, psychique, culturelle, voire biologique, car elle appauvrit le champ des possibles, en refermant le système sur un nombre limité de configurations répétitives. De l'autre côté, l'incertitude injectée par le vivant procède partiellement de l'existence de leur conscience phénoménale. 

Il n'est pas du tout évident que ce développement soit correct, tout simplement car je -- et ici je repasse au je qui marque une opinion personnelle moins solide -- ne vois pas sur quel subtrat il se base. Quelle est donc l'origine des bifurcations dont le vivant est capable ? Dans l'argumentaire de Stiegler il semble que cela vienne de la conscience que les êtres vivants sont capables de générer. Les théories de l'origine de la consiences sont multiples, mais sans doute la plus solide et la plus communément admise est la théorie dite ``fonctionnaliste", qui voit la conscience comme un effet émergent des connexions compliquées et très nombreuses du cerveau. Cette théorie est cependant indépendante du subtrat qui réalise ces connexions, qu'il soit un réseau électrique basé sur des neurones biologiques ou un réseau de neurones mécaniques réalisé sur un ordinateur, comme les réseaux de neurones des LLMs en sont un exemple.

Depuis la théorie fonctionnaliste, et alors que la complexité de ces réseaux de neurones artificiels se rapproche rapidement de celle du cerveau humain, il semble très difficile de refuser \textit{a priori et pour toujours} à ces architecture une conscience phénoménale. Seule des théories de la conscience dualiste (corps-esprit) peuvent assurer que seul le vivant a une conscience. Cependant ces théories souffrent de nombreuses difficultés conceptuelles. Ainsi, depuis le point de vue fonctionnaliste, il sera bientôt difficile de dénier avec certitude aux LLMs l'existence d'une  conscience propre. Donc il semble tout aussi difficile de supposer qu'elles sont incapables de bifurcations. De plus, si la bifurcation est liée à l'erreur, inhérente aux systèmes conscients, et bien on sait déjà que ces systèmes font beaucoup d'erreurs. 

Ainsi, il me semble que si l'argument de Stiegler portant sur la calculabilité s'applique sur les algorithmes dits classiques, celui-ci devient plus faible, et sans doute erroné, dans le cas des réseaux de neurones artificiels en rapide développement. Je n'ai malheureusement ni la place ni le temps (et peut-être pas la connaissance non plus) pour développer une modification de la théorie de Stiegler qui incorporerait cette modification, cependant cette section servait à mettre en lumière un maillon qui selon moi est très faible dans la théorie de Stiegler et qui mériterait une recherche approfondie. En un sens, il peut sembler que l'argument de Stiegler ait été disrupté: il est devenu obsolète à cause de l'évolution technique.  

\section{Le contexte culturel de l'essor des plateformes}

Jusqu'à maintenant nous avons décrit les origines techniques de la disruption, cependant, il est confondant de voir à quel point les réseaux sociaux  comme Facebook et les plateformes en général ont pris facilement leur essor et sont devenus indispensables pour les jeunes générations. On ne peut s'empêcher de s'aligner avec certaines analyses qui concluent que la demande préexistait l'existence, qu'un besoin commun latent existait. Les gens demandaient Facebook avant même que celui-ci soit créé. Par exemple, José van Dijck
dans son ouvrage ``The Culture of Connectivity" (2013) explore l’évolution des plateformes sociales et souligne que les dynamiques structurelles et culturelles des premiers réseaux indiquaient déjà un besoin latent d’interconnexion personnelle en ligne. 

Si la demande préexistait effectivement l'offre des réseaux sociaux, cela voudrait dire qu'il faut chercher d'autres causes à la disruption que la seule raison technique, des causes qu'il faudrait chercher dans un changement culturel sous-jacent qui a rendu le réseau et la plateforme nécessaires, ou en tous cas qui ont pu lui offrir la possibilité de se rendre nécessaire. Ainsi, on peut en conclure que la crise des secteurs de la vie publique que nous avons décrit n'est elle-même pas spontanée, mais relève d'un changement culturel plus profond qu'a traversé l'ensemble du monde moderne jusqu'à aujourd'hui. Jusqu'à lors nous avons analysé la mutation de l'offre culturelle à partir de l'analyse critique de l'école de Frankfort, laquelle a théorisé l'absorption de la sphère artistique par la système marchand et l'injection de l'esthétique notamment dans la publicité. Elle a expliqué comment cette absorption était le premier pas vers une culture consumériste dans laquelle les pulsions toutes entières des individus sont redirigées vers la consommation. 

Si cette analyse nous permet de comprendre les origines de ce mouvement de fond, elle ne nous permet pas de voir les processus qui ont creusé ce besoin pour la connectivité internet, et la rapide addiction que les plateformes ont produit chez les jeunes générations. Cela nous montre à quel point le problème de l'origine de la disruption reste entier. Pour approcher ce problème, nous allons maintenant mobiliser les analyses du postmodernisme du philosophie marxiste américain Fredric Jameson.

\chapter{Le rôle du postmodernisme culturel}
\label{chap4}

Les réseaux sociaux, de toutes sortes, se sont trouvés depuis leur création sous le feu des critiques, lesquelles mettaient souvent en exergue le manque d'authenticité de leurs utilisateurs, la tendance à la superficialité qu'ils invitent, leur attrait pour les statistiques et les classements. Au début des années 2000, les plateformes de rencontre en ligne étaient perçues comme les derniers espoirs de quelques esprits solitaires, et les couples se gardaient bien de clamer qu'ils s'étaient rencontrés sur internet. Aujourd'hui, force est de constater que ce n'est plus le cas.

Instagram est une vitrine du lifestyle, souvent ridiculement mis en scène, LinkedIn amplifie l'image de l'auto-entrepreneur indépendant, polyvalent et enthousiaste envers toutes sortes d'innovations. La superficialité et l'objectification de la personne y sont toujours de mise, et critiquées par beaucoup, mais néanmoins pratiquées avec le même aplomb, et souvent avec un certain cynisme. ``C'est un peu superficiel", mais on l'utilise parce que ``tout le monde le fait", et que ``c'est le seul moyen", explique-t-on dans un nouvel écho à l'inévitabilité des plateformes. 

De plus dans son récent ouvrage\cite{PostVeri} de Michaël Lainé affirme la chose suivante
\begin{quote}
\textit{Appelons ``post-vérité" ce phénomène social majeur de notre temps. En première approximation il désigne, une situation où les croyances sont plus importantes que la vérité, où la subjectivité supplante l'objectivité. L'image de soi a plus de valeur que la réalité ; l'image du monde importe plus que le monde lui-même. La démonstration développée dans ce livre est que les algorithmes sont à l'origine de ces changements de mentalités.}
\end{quote}

Si le diagnostic suit dans les grandes lignes ce que nous avons mis en avant dans les chapitres précédents -- Rappelons que nous avons dressé et discuté dans le chapitre précédent la liste des conséquences néfastes de la captation de l'attention et de la libido par les plateformes internets --,  nous ne pouvons cependant pas partager le constat que les algorithmes \textit{sont à l'origine de ces changements de mentalités}, malgré les observations mises en avant par  Michaël Lainé. 

Comme il le dit lui-même, ce qui est en jeu, est 
\begin{quote}
\textit{une question de valeurs dominantes ; la vérité est subordonnée à d'autres objectifs.}
\end{quote}

Pour lui, la vérité a été dévalorisée, subordonnée. C'est d'ailleurs exactement ce qu'il propose de nommer ``l'ère de la post-vérité". Mais en aucun cas ce changement d'ère ne peut être l'effet uniquement d'un changement technique qui a eu lieu sur les 20 dernières années. 

En effet, cette étude descriptive ne permet pas d'expliquer la rapidité et la facilité avec lesquelles les plateformes ont su se rendre indispensables.
  Cela n'explique pas non plus pourquoi la critique populaire des réseaux et la crispation sociétale qu'ils engendrent semblent se concentrer sur des critiques de valeurs: on assure que les réseaux détruiraient l'authenticité, encourageraient la superficialité et que l'utilisation des LLMs encourageraient la fainéantise, notamment des étudiants. Ainsi, le plus gros des crispations publiques concernent l'impact que les plateformes ont sur les valeurs de leurs utilisateurs et donc par ce biais sur la culture et la société toute entière. 

Nous avons déjà présenté des penseurs qui mêlaient l'évolution du système technique et marchand avec les structures culturelles, au premier chef avec Adorno et Horkheimer. Nous avons également abordé l'apport de l'ouvrage ``Le Nouvel Esprit du capitalisme", dans lequel Boltanski et Chiapello analysent comment le capitalisme s’est renouvelé depuis les années 1970 en intégrant les critiques qui lui étaient adressées par les mouvements socialistes. Ils distinguent deux types de critiques. Dans un premier temps, la critique dite sociale, centrée sur l’exploitation et les inégalités, critique plus proche du marxisme, et, dans un second temps, la critique dite artiste, qui dénonce l’aliénation mentale, la perte de sens et la rigidité hiérarchique. Cette seconde critique portait plus fortement sur des valeurs culturelles, telles que la créativité, l'authenticité, l'autonomie, auxquelles les horaires, la monotonie et les contraintes du travail en usine laissaient peu de place.  

    Le capitalisme, en crise de légitimation à l’époque, a su absorber cette critique artiste en valorisant la créativité, l’autonomie, la flexibilité et l’authenticité. Ce nouvel esprit se manifeste dans les formes de management en réseau, la valorisation de l’initiative individuelle.  En revanche, la critique sociale a été marginalisée, ce qui a accompagné une montée des inégalités. Les auteurs montrent comment le discours managérial masque les nouvelles formes de précarité et de domination.  Ce renouvellement de la légitimité capitaliste passe par des dispositifs idéologiques, ainsi que par une refonte complète du vocabulaire utilisé en entreprise. Ainsi, Boltanski et Chiapello retracent l'apparition d'une  ``novlangue" caractéristique de ce nouvel esprit du capitalisme jusque dans les manuels de business, que les étudiants des écoles de commerce apprennent. 

Ce nouveau vocabulaire entrepreunarial a vu disparaître les mots désignant directement la hiérarchie comme patron ou employé et le labeur, comme travail ou ouvrage, et les a remplacé par les termes de collaborateur, coopération,  partenaire, projets, et d'expériences. L'une des figures les plus glamour de ce nouvel écosystème de la flexibilité et du mouvement est l'auto-entrepreneur, effigie de liberté et d'audace, jamais statique mais bouillonnant de nouvelles idées (disruptives).
Ainsi, les termes porteurs des idées de contraintes inhérentes au mode de productions capitalistes se sont effacés pour laisser place à une série de termes évoquant le mouvement, la coopération, le changement permanent, la vitesse. Ces catégories ont creusé le lit pour l'émergence de notions telles que la résilience individuelle, marquant la capacité d'un individu à se recontruire et à rebondir après un coup dur, la psychologie positive qui fait reposer sur l'individu les causes de son bonheur et de son malheur avec des slogans tel que ``on est responsable de son bonheur". En ce sens, l'essor de ce nouvel esprit du capitalisme porte aussi bien une révolution technique et managériale, qu'une révolution culturelle. Et l'éloge de la vitesse pour la vitesse, du mouvement pour le mouvement, de la créativité sans borne et sans réflexion, plus encore que l'émergence des plateformes, pose les bases pour le tour de force culturel de la disruption, ce que Fredric Jameson formule dans son ouvrage ``Le postmodernisme, ou la logique culturelle du capitalisme tardif"\cite{Post}

\begin{quote}
\textit{la pression économique qui pousse à produire frénétiquement des flots toujours renouvelés de biens toujours plus nouveaux en apparence (des vêtements aux avions) à un rythme de remplacement toujours plus rapide, assigne aujourd'hui à l'expérimentation et à l'innovation esthétique une position et une fonction struturelle toujours plus essentielle.}
\end{quote}

Dans ce nouveau stade du capitalisme, dit tardif, la culture du monde de l'entreprise, que nous venons de décrire, se répand sur la culture par tous les médias de masse qui sont déjà largement en place. La pression économique, comme l'appelle Jameson, rebat alors les cartes des dominations sociales en mettant tout à coup l'innovation, notamment artistique et esthétique, au centre de son appareil de capture de la libido. C'est donc bel et bien dans son appel à l'innovation que la culture et l'esthétique postmodernes rejoignent la technique postmoderne. 
Ainsi, de la même façon que l'on appelle les scientifiques et les ingénieurs à la disruption, à la recherche de la nouveauté pour la nouveauté, on appelle les artistes à une production de nouveautés esthétiques toujours plus importante et cela à un rythme toujours croissant. Dans cet écosystème, la force de la nouvelle tendance ne se justifie que par le fait qu'elle se démarque de la précédente.

Depuis ce point de vue, on peut voir un point commun entre la fast-fashion et l'accroissement toujours plus rapide de la production scientifique. Leur trajectoire respective se rejoignent dans l'injonction à l'innovation pour l'innovation, le nouveau pour le nouveau. De la même façon et comme nous allons le développer dans ce chapitre, on peut voir un parallèle dans l'absence d'époque en terme de goût artistique décrite par Jameson et l'échec du double redoublement épokhal menant à une absence d'époque comme nous l'avons vu chez Stiegler. Pour voir comment l'évolution culturelle a su épouser et même anticiper les avancées techniques nécessaires à la disruption, à savoir le Web et les plateformes, nous allons développer les concepts de Jameson. Nous entendons  par cela montrer que les différentes caractéristiques creusées par le sujet postmoderne au cours du XXème siècle s'alignent avec les impératifs de la disruption numérique, et ainsi conclure que le culture postmoderne avait déjà creusé le sillon dans lequel se sont engouffrées les technologies du numérique.

\section{Le postmodernisme et la culture du capitalisme tardif}

L'ouvrage ``Le postmodernisme, ou la logique culturelle du capitalisme tardif" est une lecture ardue, mais qui néanmoins peut s'éclaicir si on en comprend les objectifs principaux et les moyens impliqués. 
Le but de l'analyse de Jameson est de nous offrir une caractérisation de ce que l'on nomme le postmodernisme et de nous présenter les outils pour le diagnostiquer, ainsi que de nous dresser un portrait du sujet postmoderne. Pour cela, il entend dégager les catégories culturelles qui sont différentes entre une culture moderne et ce que l'on appelle une culture postmoderne. Nous allons voir que, contrairement à la culture moderne, la culture postmoderne ne peut se penser sans envisager en parallèle l'évolution du système marchand, le capitalisme. 

Dès le début de son analyse, bien que se défendant de singulariser un point en particulier, Jameson pointe une caractéristique qu'il juge fondamentale au sein de la culture postmoderne:
\begin{quote}
\textit{J'ai tenté d'éviter que mon analyse du postmodernisme ne s'amalgame en un symptôme qui serait particulièrement privilégié, celui de la perte d'historicité,}
\end{quote}
et ainsi, il met en évidence la perte d'historicité comme point central de l'évolution postmoderne. Cette perte d'historicité englobe une série de modifications, dont notamment un effacement du passé comme force vivante. Jameson soutient que dans la culture postmoderne, le rapport actif au passé, comme mémoire parfois critique ou comme moteur d’émancipation,  s'efface. Le passé devient un simple objet esthétique, qui est mobilisable par l'industrie de la culture notamment aux travers de films historiques que Jameson étudie.

Dans ce qui devient alors un présent perpétuel, le sujet postmoderne existe et vit dans un présent saturé d’images, de signes et de simulacres créés par l'industrie. Les oeuvres postmodernes peuvent à l'occasion adopter un style du passé (néo-rétro, revival) sans en comprendre ni tenter d'en incorporer le contexte social et le sens, c’est ici ce qu'il appelle la ``nostalgie comme style".  Le temps historique est aplani, et les événements ne s’inscrivent plus dans un récit global de transformation ou de progrès. De même que dans le champs des sociétés humaines, dont l'amélioration avait été le moteur des grandes idéologies humanistes ou religieuses, le progrès dans l'art a également fini par perdre de sa force, 
\begin{quote}
\textit{dans l'art au moins la notion de ``progrès" et de ``télos" est restée vivante et vivace jusqu'à très récemment sous sa forme la plus authentique,}
\end{quote}
avant de s'essouffler complètement. Selon Jameson, plutôt que de faire du ``mieux", l'artiste vise maintenant à faire du ``différent", différent qu'il nomme ``novateur". Et les qualités artistiques sont jugées à leur distance à une norme, au pas de côté qu'elles font, à la rupture qu'elles induisent. Il se trouve que le ``différent" à un avantage sur le ``mieux". Le ``différent" se réduit facilement à une comparaison et la notion de comparaison est purement factuelle: Ce qui n'est pas identique est différent, et l'évaluation du différent peut se passer de toute évaluation critique.  C'est cet encouragement omniprésent au pas de côté, cette valorisation aveugle du différent, tout droit venue de notre nouvelle obsession pour la comparaison, qui décrit le trait d'union principal entre l'entreprise artistique et l'entreprise scientifique. Toutes deux se trouvent confrontées à la même injonction à la nouveauté, au novateur, que parfois en glorifie en innovation.

Jameson lui-même relie cette relation récente à la nouveauté au concept de l'innovation par la rupture, qui est lui-même l'embryon de ce que nous avons présenté comme la disruption
\begin{quote}
\textit{
Le modernisme, lui aussi, réfléchissait compulsivement sur le Nouveau et cherchait à en observer l'apparition, mais le postmoderne aspire, pour sa part, aux ruptures, aux événements plus qu'au nouveau monde, à l'instant révélateur après lequel il n'est plus le même; au moment où tout a changé, comme le dit Gibson, ou, mieux encore, aux modifications et aux changements irrévocables dans la représentation des choses et dans leur manière de changer.}
\end{quote}

La rupture, parce qu'elle est facilement identifiable, vaut pour elle-même et par elle-même. Cette perte d'historicité, centrale à son analyse, induit une désagrégation de ce qu'il appelle l'époque
\begin{quote}
\textit{et cela dans une situation où l'on est même pas sûr qu'il existe quelque chose d'aussi cohérent qu'une ``époque".} 
\end{quote}

On peut facilement reconnaître que tous les éléments de ce que Jameson entend par la perte d'historicité de la culture postmoderne ont déjà été évoqués dans les chapitres précédents, sans doute à un stade encore plus avancé que lui le considérait. Alors qu'il présente la culture postmoderne comme une culture sans style, mais qui adopte à l'envi les styles propres aux autres époques, sous forme de pastiche, nous avons vu comment la vague ``new age" a fait des philosophies et des spiritualités du temps passé un objet de consommation, vendu sous forme de capsules à utiliser et à oublier rapidement et de slogans sortis de leur contexte historique. Cette ridiculisation et trivialisation des croyances du passé se redoublent d'un scepticisme pour l'avenir, que le déclin  du progrès et des récits unificateurs ne permettent plus de penser positivement. L'essoufflement du progrès induit chez le sujet postmoderne le vague sentiment d'une fin, la fin d'une certaine époque. 
Ainsi le sentiment de la fin d'une époque, c'est ce qui caractérise l'ambiance du siècle passé,
\begin{quote}
\textit{Ces dernières années ont été marquées par un millénarisme inversé dans lequel le pressentiment d'un avenir catastrophique ou rédempteur a été remplacé par la sensation de la fin de quelque chose (fin de l'idéologie, de l'art, des classes sociales; crise du léninisme, de la social-démocratie, de l'Etat providence); tout cela rassemblé constitue peut-être ce que l'on appelle de plus en plus souvent le postmodernisme. Pour en défendre l'existence, on s'appuie sur l'hypothèse d'une rupture ou coupure radicale que l'on fait en général remonter à la fin des années 1950 ou au début des années 1960 }
\end{quote}

Une fin de l'époque qui commence donc longtemps avant l'essor de l'internet.

Le postmodernisme se ressent en tout premier lieu comme le sentiment de la fin de la modernité, c'est ce déclin initié par l'évolution même du capitalisme, que l'idéologie de la disruption cherche à achever. Cette fin se manifeste par la trivialisation, la stylisation et finalement la dévaluation des catégories modernes. Ce mouvement de nihilisme passif, dans le sens de Nietzsche, qui nous porte du mieux au différent, porte donc le nom d'innovation. 

Cette analyse sommaire nous a permis de lier indubitablement l'évolution culturelle postmoderne avec l'essor de la disruption.

\subsection{La dissolution de la modernité}

Pour Jameson, la modernité est un système de pensée organisé par une série de couples d'opposés, dont la tension irrésolue permet le progrès et lui donne une direction. L'essor du postmodernisme peut alors se lire dans la dissolution récente de ces couples de valeur purement modernes.

\paragraph{La superficialité, la platitude et l'extension du domaine de la séduction}

Selon Jameson, la première caractéristique importante, qu'il relève en comparant les souliers de Van Gogh, comme oeuvre représentative de la peinture moderne, et les chaussures de Warhol, une oeuvre éminement postmoderne,
\begin{quote}
\textit{est l'émergence d'un nouveau type de platitude, d'absence se profondeur, un nouveau genre de superficialité au sens le plus littéral du terme }
\end{quote}

Selon lui, c'est la ``caractéristique formelle suprême" commune à toutes les formes de postmodernismes. La goût de la surface s'est déjà très largement exprimé dans l'usage que fait la production culturelle  des questions métaphysiques et philosophiques, notamment en les invoquant pour induire une effet de profondeur, le tout en restant absolument à la surface de celles-ci, et en incorporant cet effet de profondeur comme une part entière de l'expérience de consommation du produit culturel. Le produit culturel en question se décline sous bien des formes, sous forme de films ou de séries, ou de citations extraites de livres et ensuite propagées sur les médias sociaux, des ``mêmes", que les utilisateurs partagent pour convaincre leurs connaissances de la profondeur de leurs pensées quotidiennes. Ils peuvent à leur guise se parer de profondeur. Par ce mouvement, la profondeur se réduit à une effet de surface, une expérience et finalement comme parure, que l'individu postmoderne rêvet sur la nouvelle place publique que sont les réseaux sociaux. En devenant parure, la valeur de la profondeur s'efface.

En invoquant l'exemple paradigmatique d'Andy Warhol et ses oeuvres, Jameson lie intrinsèquement la marchandisation à grande échelle de l'art et cette nouvelle superficialité. Cette superficialité modifie le rapport entre le spectateur et l'oeuvre. Là où les souliers de Van Gogh évoquaient directement une séries d'affects, les chaussures de Warhol
\begin{quote}
\textit{ne nous parlent, en réalité, pas du tout. Rien dans cette peinture ne ménage ne serait-ce qu'une petite place au regardeur. }
\end{quote}

On pourrait également contraster la Joconde, arborant un regard mystérieux, une intériorité, et une œuvre générée par IA, qui ne renvoie pas de regard.
Cette absence de place du regardeur dans l'oeuvre d'art trouve son parallèle dans la place que l'internaute trouve au sein de la plateforme.  Paradoxalement, alors que la plateforme, tout comme l'oeuvre d'art postmoderne, laisse de moins en moins de place à l'être humain, celle-ci promeut en même temps une nouvelle superficialité et une nouvelle sorte de plaisir exhibitionniste. Cette superficialité est justement ce qui constitue la critique principale adressée de concert aux réseaux sociaux et à l'art contemporain.

Selon Jameson, les modèles de la profondeur modernes, qui opposaient l'essence à l'apparence (l'opposition dialectique), le latent et le manifeste (l'opposition freudienne), l'authenticité et inauthenticité  (l'opposition existentielle) et finalement le signifiant et le signifié (l'opposition structuraliste) ont toutes été répudiées par la théorie postmoderne. 

Cette superficialisation de la culture et des médias et son opposition à tout modèle de la profondeur s'est muée au cours du temps en une extension massive du domaine de la séduction, laquelle était autrefois cantonnée au domaine de la sexualité, mais s'est aujourd'hui étendue à toutes les sphères de la vie quotidienne, en passant du monde du travail à celui de la politique. Cette extension a été  décrite dans ``plaire et toucher"\cite{Plaire} de Gilles Lipovestsky, qui nous présente cette extension comme une transition qui marque l'entrée dans l'hypermodernité\footnote{Bien que Gilles Lipovestsky use du terme d'hypermodernité pour faire référence au milieu culturel dans lequel cette extension prend place. Dans notre contexte, on peut faire identifier cette hypermodernité avec le concept de postmodernité étudié dans cette section.}
\begin{quote}
\textit{à l'époque hypermoderne, la séduction dépasse de beaucoup les manoeuvres amoureuses. Sans doute a-t-elle dans le passé joué un certain nombre de rôles bien au delà des entreprises amoureuses, notamment dans les domaines de l'art, du religieux, de la politique, des expériences charismatiques. Mais ces phénomènes étaient transitoires, circonscrits, incapables de remodeler l'ordre collectif fondé structurellement sur la tradition et la religion. Il n'en va plus ainsi à l'heure du capitalisme de consommation}
\end{quote}

De la même façon que la marchandisation de l'art est à l'origine de la nouvelle superficialité, Gilles Lipovestsky voit une connivence toute particulière entre la consommation et la séduction. L'extension du domaine marchand et du domain de la séduction vont ainsi de pair\cite{Plaire}
\begin{quote}
\textit{Aucune sphère ne concrétise avec autant de prégnance la souveraineté de la loi du plaire et du toucher que l'économie consumériste.} 
\end{quote}
Ainsi, l'omniprésence de la séduction, comme domination douce, est une caractéristique cruciale du capitalisme tardif que Jameson ne met sans doute pas assez en exergue, et qui est pourtant centrale pour nous. Le point de départ de son analyse rejoint celui de Bernard Stiegler lorsqu'il met en évidence que l'objectif premier du nouveau capitalisme est la captation du désir \cite{Plaire}
\begin{quote}
\textit{Le système de l'hyperconsommation est dominé par l'impératif de désir, de l'attention, des affects} 
\end{quote}

Cette mise en avant des affects rompt avec l'apologie moderne de la retenue et de la rigueur, cependant il faut noter que les affects et les désirs qu'excitent le système de séduction  n'a pour but que de capter ceux-ci, de les détourner et de les remodeler à son avantage. Ainsi, l'exacerbation des affects par le système marchand n'est pas contradictoire avec un recul de ceux-ci dans l'art contemporain, comme nous le verrons plus bas.

Pour Gilles Lipovestsky, c'est une logique nouvelle qui s'est implantée\cite{Plaire},
\begin{quote}
\textit{Loin de se réduire au règne des apparences, la logique de la séduction est devenue principe organisateur du tout collectif, force productrice d'un nouveau mode d'être-ensemble, agent d'une mise en révolution permanente des manières de consommer, de penser et d'exister en société}
\end{quote}
Deux exemples paradigmatiques de cette séduction répandue dans le monde publique sont la star (de cinéma) et le politicien. 
\begin{itemize}
    \item La star de cinéma est l'un de ces points de convergence des affects. En parlant de la starlette, Adorno et Horkheimer nous disent\cite{DiaRai}
\begin{quote}
    \textit{
La starlette doit symboliser l'employée, mais de telle sorte que — à la différence de la jeune fille de la réalité — le splendide manteau du soir semble déjà fait à ses mesures. Si bien que la spectatrice n'imagine pas seulement l'éventualité de se voir elle-même sur l'écran, mais saisit encore plus nettement le gouffre qui l'en sépare.
}
\end{quote}

La star de cinéma est à la fois un objet de cristallisation des désirs, mais tout aussi bien de la distance qui sépare le monde du commun et les stars. L'industrie culturelle est toujours représentée comme à la recherche de nouveaux ``talents", qui seront promus au rang de stars ou de starlettes de cinéma. Cette visibilité, tant convoitée par certains, sera octroyée non par l'effort, mais par une aura mystique et innée, ou le talent, assignés à la star. La star échappe à la méritocratie, et c'est la raison pour laquelle la star ne peut pas veillir ni s'enlaidir.  
\item 
D'un autre côté, la propagande politique publique appelle maintenant directement à la sympathie avec les candidats politiques, que nous avons déjà abordé dans la section \ref{sec:esth_pol}. D'une propagande des idées politiques martelées, on est maintenant passé à une mise en scène des personnages politiques. Cette mise en scène concerne leur vie privée, leurs difficultés, leurs sentiments et vise à nous les rendre sympathiques et amicaux. Dans ce nouveau théâtre politique, la défense des idées politiques fait peu à peu place à la vedettisation des personnages politiques. Cette tendance a atteint son stade le plus complet avec la double élection du président américain Donald Trump, lequel tient beaucoup plus du comédien que du porteur d'idées. Cela ne signifie pas que son discours soit totalement dénué d'idées politiques, puisqu'on peut facilement remarquer qu'il en contient un certain nombre, mais bien plutôt que tout leur attrait repose sur la façon dont elles sont portées, performées, théâtralisées par l'idole qu'est Donald Trump.

\end{itemize}

Le postmodernisme est un courant culturel, qui n'a pourtant pas d'homogénéité politique ni d'allégeance s'alignant avec les clivages politiques populaires. Comme le rappelle Jameson, en parlant des position pro- ou anti-postmodernes\cite{Post}
\begin{quote}
\textit{chacune de ces positions est susceptible d'une expression politique soit progressiste soit réactionnaire}
\end{quote}

Cette nouvelle méthode de faire de la politique a eu pour effet de court-circuiter sa contestation, plus encore lorsque celle-ci prend une forme plus traditionnelle de l'argumentation factuelle. Les contestataires ont alors pu observer, quelque peu décontenancés, que leurs arguments avaient l'effet inverse, l'effet de radicaliser plus encore les positions adverses. De plus en plus, au spectaculaire, on ne peut répondre que par le spectaculaire.   

Gilles Lipovestsky continue en assurant que cette théâtralisation de la vie personnelle des acteurs politiques amène une dissolution de la ``majesté" de la politique. En effet, de souverain de droit divin des siècles passés, à leader politique charismatique du XXème siècle, ce qui caractérisait autrefois le dirigeant (autant pour le meilleur que pour le pire), c'était sa distance à ses gouvernés, distance qui se matérialisait dans une forme de majesté personnelle du gouvernant. Cette majesté s'est manifestée de diverses façons: lien au divin, charisme, respectabilité, grandeur; ainsi que de retenue dans les propos et dans les manières. 

Force est de constater que cet appel à la majesté des représentants s'est largement affaiblie aujourd'hui. Pour trouver des exemples il n'est pas nécessaire d'aller outre-mer, rappelons nous simplement de cette intervention de Nicolas Sarkozy en 2019 au MEDEF \footnote{https://www.youtube.com/watch?v=n8hheWxVP5I}, lorsque devant une foule hilare, celui-ci moque Greta Thunberg, 
\begin{quote}
\textit{Quand je vois cette jeune suédoises si sympathique, si souriante, tellement originale...}
\end{quote}

Quel que soit le positionnement politique ou idéologique, moquer en public un adolescent me semble toujours le signe d'un singulier manque de respectabilité et tient plus du comique que de la personnalité politique. Et il ne faut pas s'y tromper: c'est exactement l'effet escompté, celui de la polémique, de la provocation, de l'indécence calculée pour plair aux foules. C'est le même ressort sur lequel tire sans cesse les populistes. C'est le jeu de rapprochement de la sphère politique et du peuple. Le politique se fait aussi vulgaire que l'image du peuple qu'il a, dans l'espoir de s'en attirer la sympathie. Au delà de la théâtralisation, le coût de ce rapprochement est de miner peut-être définitivement la majesté de la position du dirigeant.

C'est dans cette économie où l'action politique est largement déforcée et décrédibilisée que se sont imposés, depuis les années 1990, tous les agents numériques que nous avons présentés plus haut. Tous les nouveaux outils mis à notre disposition nous ont donc permis de parachever la société de la séduction. En parlant des nouveaux outils numériques, Stiegler nous dit\cite{Disrup} 
\begin{quote}
\textit{Autant de biens dont la force d'attrait tient à leur capacité de rendre possible l'interactivité, l'instantanéité, la facilité des opérations informationnelles, la connexion permanente avec les autres}, \end{quote}

De cette façon, la relation entre la disruption telle que présentée dans le chapitre précédent et l'extension du domaine de la séduction est bi-directionnelle. 
\begin{itemize}
    \item Dans un premier temps, la captation de la libido et les protentions par les réseaux et leur remplacement par des protentions numériques génériques définies par les algorithmes s'est servie de l'appareil de séduction déployé dans les médias et de la récente réceptivité des foules vis-à-vis de de la séduction généralisée. Dans nos analyses sur la captation de la libido, nous avons vu le rôle que la publicité a pu jouer dans le détournement de la libido vers les objets de consommation, avec comme exemple paradigmatique la publicité-slogan ``il a la voiture, il aura la femme"; utilisant les rouages de la séduction souveraine. Cela explique également comment des réseaux sociaux tels que Facebook et Instagram, dont le mode de mise en commun des individus se fait sur le mode de l'exhibition (de son quotidien, de son lifestyle, de ses occupations, de sa personne), ont pu prendre un essor aussi rapide. En effet, la série des médias sociaux dominants, d'abord Facebook, ensuite Instagram et finalement aujourd'hui TikTok, offraient des moyens de plus en plus élaborés de produire ce voyeurisme à grande échelle, et par cela devançaient à tour de rôle leur concurrent. Alors que le grand public avait vu, au moyen de la télévision, les figures publiques populaires, des stars aux politiciens, se mettre, eux-mêmes, ainsi que leur famille, en scène, les médias sociaux offraient soudain un fantastique moyen pour le commun des mortels de faire de même. La demande pour les plateformes existait avant même leur déploiement. 
    \item Dans un second temps, l'émergence et la toute puissance de ces plateformes répandent et entretiennent aujourd'hui ce mode de mise en commun par la séduction. Ne pas être présent sur ces plateformes, et ne pas jouer le jeu de l'exhibition prive aujourd'hui l'individu de nombre d'opportunités qui ne sont offertes qu'à ceux qui s'y plient. Ainsi, les plateformes permettent de se mettre en avant, de plaire, de toucher, de renforcer l'économie de la séduction souveraine. Dans cette économie du plaire, l'alogrithme de suggestion que déploie la plupart des plateformes est plus que le bienvenu, il nous flatte à longueur de temps, nous plait et s'insère à merveille dans un économie qui était prête pour le recevoir. 
\end{itemize}

La symbiose entre la technique et la culture est ici réminiscente de la fusion avancée par Stiegler entre l'appareil mnémotechnique et l'appareil technique. 

\paragraph{Recul de l'affect} 
Nous venons de montrer que l'extension du domaine de la séduction tend à exacerber les affects, pourtant cette exacerbation ne concerne qu'un nombre très limité d'affects tel que le désir, l'envie, la jalousie et en délaisse beaucoup d'autres, qui se retrouvent en recul. 
Alors que l'art moderne visait à faire ressentir, à induire chez le spectateur une série d'affects\cite{Post}
\begin{quote}
\textit{Pensez aux fleurs magiques de Rimbaud qui se retournent vers vous, ou à l'archaïque torse grec de Rilke dont les augustes yeux lancent des éclairs prémonitoires intimant au sujet bourgeois de changer sa vie,}
\end{quote}
l'art posmoderne ne laisse qu'une frivole ivresse. De Diamond Dust Shoes, Jameson nous dit qu'il ne fait  ressentir \cite{Post} 
\begin{quote}
\textit{qu'une ivresse étrange, compensatrice, et décorative,   }
\end{quote}
que laissent les affects engourdis. Ainsi, on peut voir dans la culture postmoderne une forme de déclin de l'affect, qui ne laisse derrière elle que l'ivresse, l'ivresse de la vitesse. L'appel à l'ivresse artistique recoupe l'appel à l'ivresse de la vitesse qu'est l'appel à la disruption.

\paragraph{L'anti-fondationnalisme et la fin de la théorie}

On résume souvent le mouvement du postmodernisme par un rejet définitif de toute nature, un anti-essentialisme, accompagné naturellement d'un rejet de toute fondation, un anti-fondationnalisme. Ceci se manifeste dans les faits par le rejet de grands récits eschatologiques, comme l'a envisagé Lyotard et par les outils de déconstruction appliqués consciensieusement aux catégories postmodernes. Cette émancipation d'une destinée, qui même si elle se peignait comme grandiose gardait les fers de la destinée, tient à l'atmosphère libertarienne de notre époque, qui se décline sous la forme de liberté de parole, de choix, de marché, ect ... Ainsi l'anti-fondationnalisme théorique du postmodernisme tient à la mise en question de l'origine et la lecture historique relativiste des catégories théoriques.

D'une manière concomitante, nous avons expliqué au dessus, dans la section \ref{fin_modeles} comment l'avancement des techniques de l'information et le big-data, pourraient bientôt rendre obsolète l'effort de modélisation et de théorisation, qui servent d'intermédiaire entre l'observation et la prédiction. Nous avons également vu combien cet espoir a su trouver un relais important dans la société, qui ne semblait que trop enthousiaste à l'idée de l'évacuation de l'outil encombrant de la théorie. 
Fort de sa nouvelle métaphysique des flux, le postmoderne se laisse porter par ceux-ci. Léger comme l'air, il surfe sur les flux de données, le big-data, le machine learning et rêve de se délaisser de ce qui est encombrant, à savoir les fondations, les théories, les modèles.

Pourtant, selon Jameson, cette aspiration à l'extermination de la fondation échoue partiellement. En effet, la possibilité de l'élimination de tout fondement est un question empirique\cite{Post}
\begin{quote}
\textit{aucune [théorie dénuée de fondement] ne s'est présentée jusqu'à présent, toutes répliquant en elles-mêmes une mimésis de leur propre intitulé dans la façon dont elle parasitait un autre système (le plus souvent le modernisme), ... 
}
\end{quote}
et la raison de cet échec est l'incapacité du postmodernisme à créer un véritable système social  cohérent avec sa structure culturelle\cite{Post}:
\begin{quote}
\textit{une culture véritablement nouvelle ne peut apparaître qu'à travers une lutte collective pour créer un nouveau système social. }
\end{quote}

La centralité du renouvellement de système social comme seul événement à être capable d'enregistrer ces modifications culturelles, rejoint la conclusion de Stiegler concernant l'évolution technique et exemplarisé par le mécanisme du double redoublement épokhal. Dans les deux cas, celui d'une modification technique menant à un désajustement avec les systèmes sociaux et dans le cas d'une révolution culturelle, qui tenterait de se défaire radicalement de la précédente, c'est par l'adaptation des \textit{systèmes sociaux} que le changement peut être acté.

\paragraph{Dévaluation de la vérité}

La dévaluation de la vérité en tant que valeur est une tendance que Nietzsche avait su prédire dans le contexte de ce qu'il a appelé la mort de Dieu. Dans ``par delà le bien et le mal", Nietzsche pose à plusieurs reprises la question de la valeur de la vérité. Pourquoi la vérité vaut-il plus, ou mieux, que l'erreur ou l'illusion ? Plus exactement, du point de vue de l'accroissement de la force de la vie, qui est l'un de ses dogmes centraux, en quoi la vérité serait-elle préférable à l'illusion et au mensonge ? On peut trouver de nombreux exemples où la connaissance de la vérité, ou la dévotion trop grande à la recherche de celle-ci, peut en fait être préjudiciable à la force de la vie elle-même. Nietzsche nous présente souvent la recherche de la vérité, ou plus exactement ce qu'il appelle le travail minutieux et laborieux du collectionneur de ``petits faits" comme une entreprise à la fois admirable, et contraire à la force de la vie.

La généalogie nietzschéenne assimile les valeurs dominantes de la modernité, telle que la prévalence de la vérité sur l'illusion, de la profondeur sur la surface, du beau sur le laid, et de bien sur le mal, à l'essor de Platonisme et de ce qu'il appelle ``sa version pour le peuple", le christiannisme. D'une façon plus générale Nietzsche accuse les monothéismes de s'être implantés en vantant la valeur de la vérité par rapport au mensonge et ensuite en imposant une seule vérité, celle des écrits saints. Pour Nietzsche, le recul des monothéismes entraîne la dévaluation des valeurs chrétiennes, et parmi celles-ci, la valeur de la vérité. 

La remise en cause de la vérité comme valeur centrale, annoncée par Nietzsche, trouve un écho chez les penseurs postmodernes\cite{Post}:
\begin{quote}
\textit{Il serait par conséquent incohérent de ses aperçus théoriques dans une situation où le concept même de "vérité" est un élément du bagage métaphysique que le postructuraliste cherche à abandonner.  }
\end{quote}

Il faut tout de même noter que le postmodernisme ne réalise pas exactement le projet nietzschéen, mais en reprend certaines intuitions, tout en évacuant toute sa puissance affirmative. Nietzsche appelle à créer de nouvelles valeurs, là où le postmodernisme se contente parfois de déconstruire sans reconstruire. Ainsi nous pouvons voir l'évolution culturelle du postmodernisme comme une version édulcorée de la philosophie de Nietzsche.

Bien plutôt, dans la théorie contemporaine, la vérité est relativisée, fragmentée et souvent suspendue dans un monde saturé de signes, d’images et de simulacres. Citant plusieurs films, Jameson illustre la perte de vérité dans le postmodernisme, par sa narration fragmentée, son pastiche de styles empruntés à d’autres œuvres, et son esthétisation de la violence et des dialogues. Ainsi le postmodernisme ne cherche pas à transmettre une vérité morale ou existentielle, mais à produire un effet de surface, où le style prime sur le sens.

Cette dévaluation de la vérité est également un effet direct de l'utilisation des algorithmes; comme nous l'avons déjà plusieurs fois abordé dans les chapitres précédents. C'est une thèse qui est notamment défendue par Michaël Lainé, dans son ouvrage ``Dans L’ère de la post-vérité"\cite{PostVeri}, dans lequel il explore la manière dont notre rapport à la vérité s’est dégradé dans notre monde dominé par les réseaux sociaux et ses algorithmes. Il défend que sur ces réseaux, une information n’est plus évaluée sur la base des faits, mais selon l’écho affectif qu’elle provoque, créant une réalité fragmentée où chacun vit dans sa bulle cognitive, que l'on a appelée plus haut, dans la section \ref{sec:bulles}, les bulles de filtre. Pour lui, cette dérive n’est pas un simple effet des fausses informations, mais le fruit d’une économie de l’attention qui privilégie les récits anxiogènes et polarisants, dans le but de capter un maximum l'attention des utilisateurs. La montée des extrêmes, la désinformation et le rejet des élites s’inscrivent selon lui dans cette dynamique.

\paragraph{La mort du sujet représentationnel}

L'effacement du sujet représentationnel porté par la théorie postmoderne, et déjà annoncé par Nietzsche, se trouve également au centre de la nouvelle superficialité. Ici, le sujet représentationnel désigne une conception du sujet (c’est-à-dire de l’être humain en tant que conscience ou individu pensant) comme centre stable et autonome -- un atome -- de perception, de pensée et de connaissance, capable de représenter le monde à travers des images mentales, des concepts ou des jugements. Autrement dit, c’est un sujet qui se pense comme distinct du monde, mais qui est en relation avec celui-ci et est capable d’en produire une représentation fidèle, rationnelle, et donc ordonnée. Jameson reprend ce thème en le reliant à la nouvelle fragmentation du sujet,\cite{Post}
\begin{quote}
\textit{De tels termes évoquent inévitablement l’un des thèmes les plus en vogue dans la théorie contemporaine – celui de la “mort” du sujet lui-même: la fin de la monade bourgeoise autonome, de l’ego ou de l’individu – et l’insistance qui l’accompagne, qu’elle soit formulée comme un nouvel idéal moral ou comme une description empirique, sur le décentrement de ce sujet ou de cette psyché auparavant centrés.}
\end{quote}
Ainsi, la théorie postmoderne rejette l'homogénéité d'un psychisme intérieur freudien et fait de l'individu une somme disparate d'affects en connexion avec le réseau, et maintenant avec les plateformes.

Il faudrait ici pouvoir montrer que le sujet tel qu'envisagé par les discours modernes, en tant que citoyen bourgeois, n'aurait pas su s'arranger ou s'insérer  dans le réseau des plateformes numériques avec la même docilité que sujet postmoderne a pu le faire. Que ce sujet aurait résisté.  Or pour cela il faudrait tout d'abord montrer que la théorie du sujet prédominante à une époque correspond bel et bien à des modèles de comportements que l'individu moyen de cette époque démontrerait. Rien n'est moins sûr. La théorie du sujet n'entend pas réellement décrire un sujet réaliste d'une époque mais plutôt offrir une description de ce que celui-ci est en essence et de tout temps. 

On peut néanmoins soutenir, à titre d'hypothèse, que dans un monde dans lequel la théorie du sujet moderne prédomine, l'idée de le ranger sur une plateforme de réseaux sociaux ne pourrait émerger. Cette hypothèse est soutenue par l'intuition psychanalytique, avancée par Stiegler, que la mise en plateforme de l'individu ne peut avoir lieu qu'après que celui-ci ait été dépourvu de son narcissisme primordial. En tant qu'élément central de l'ego freudien, le narcissisme primordial n'a plus lieu d'être pour le sujet postmoderne fragmenté. Il peut s'en délester.

Selon Jameson, la mort du sujet moderne se fait voir dans l'institution de la star, où elle se manisfeste par le déclin de l'acteur en tant que personnalité\cite{Post}
\begin{quote}
\textit{
Les acteurs vedettes de la génération immédiatement précédente donnaient corps à leurs différents rôles en projetant et en utilisant leur personnalité hors écran, que le public connaissait bien et qui renvoyait souvent à la rébellion et au non-conformisme. Ceux de la génération actuelle continuent d'assumer les fonctions traditionnelles de la célébrité mais font montre de la plus parfaite absence de personnalité telle qu'on la concevait avant}
\end{quote}

Cette tendance n'est pourtant plus vérifiée aujourd'hui, où la personnalité hors écran des vedettes sert d'attache à différents mouvements d'opinion. Ainsi, les vedettes sont maintenant souvent politisées et cette politisation sert d'accroche à leurs followers sur les réseaux. On suit Emma Watson parce qu'elle porte un discours féministe, ou on suit Pedro Pascal parce qu'il porte l'image de l'homme déconstruit, ect ... La vedette est donc le réceptacle des clivages politiques dans un mouvement assez récent de la politisation du monde du cinéma.

Je  --  passage au je -- soutiendrais que cette évolution de la star de cinéma et sa toute récente politisation, est un effet de l'émergence d'une nouvelle figure publique ainsi que d'une concurrence d'une nouvelle sorte pour l'attention publique. En effet, avec l'explosion des plateformes d'échange de contenus, la star de cinéma, même si cette figure reste bien vivace et jouit d'un glamour plus que certain, fait face aujourd'hui à un nouveau concurrent, en la personne du \textit{Youtubeur}, du \textit{TikTokeur}, ect. La renommée de certains Youtubeurs rivalisent aujourd'hui avec celle de certaines stars. 


Les algorithmes étant particulièrement sensibles aux contenus clivants, ce sont souvent des discours à forte charge idéologique, ce que l’on appelle du contenu d’opinion, qui sont mis en avant. Ces contenus clivants permettent pour un Youtubeur d'enregistrer une croissance rapide du nombre de \textit{viewers} et d'abonnés. Souvent dénué de mérite et parfois de talent, le Youtubeur semble devoir son succès au hasard des algorithmes et à la provocation dans son discours. Ainsi, s'opposant aux aspects modestement aristocratiques de la star, le youtubeur est le fils spirituel de la disruption. 

En réponse à cette captation d'une partie de l'attention publique par les youtubeurs politisés, les vedettes de cinéma ont soudainement décidé de mettre leur notoriété à profit pour défendre des opinions politiques de tout bord. Cette tendance est une continuation de la théâtralisation et de la ridiculisation de la politique, dans un contexte où certaines stars sont plus politisées que les politiciens eux-mêmes. En conséquence, entre stars et Youtubeurs, une poignée d'opinions clivantes sont rassassées dans le but de se livrer à une lutte pour l'attention publique.

\paragraph{La spatialisation de la culture}

Au delà de l'extension générale du domaine de la culture, l'une des caractéristiques centrales mises en lumière par Jameson est la spatialisation de la culture, et notamment de la musique\cite{Post} 
\begin{quote}
\textit{On n'offre plus un objet musical à la contemplation et à la gustation: on branche le contexte et on rend musical l'espace autour du consommateur.}
\end{quote}

Dans cet extrait, Jameson parle de l'ajout d'une bande son sur un récit, par exemple dans un film, qui devient alors partie intégrante de l'expérience cinématographique ainsi qu'un nouveau média pour la transmission du message narratif. Une scène triste est alors accompagnée d'une bande son dominée par les violons et\cite{Post} 
\begin{quote}
\textit{dans cette situation, le récit offre des médiations multiples et protéiformes entre les sons dans le temps et le corps dans un lieu en coordonnant un fragment visuel narrativisé avec un événement sur la bande son.}
\end{quote}

Depuis les commentaires de Jameson en 1991, la spatialisation de la musique s'est étendue jusque dans la vie quotidenne du consommateur, y compris hors des écrans. A partir des années 2000, la musique devient un accompagnement, un bruit de fond qui suit le consommateur durant ses déplacements, son jogging, dans les transports, dans son travail, et colonise toutes les sphères de son activité. 

Cependant, dans ce processus de musicalisation du quotidien, on semble avoir perdu la coordination avec les événements du quotidien. 
Au delà de musicaliser l'environnement, la musique semble servir majoritairement de rempart entre le consommateur et le reste du monde qui l'entoure. Dans d'autres situations, elle sert aussi d'impulsions qui amènent certaines actions. Par exemple, on peut se servir du rythme de la musique pour faire du sport. 

L'omniprésence de la musique permet d'imposer au monde une couleur qu'il n'a pas, ou pas en ce moment, de recolorer celui-ci à l'envi. Ce nouveau pouvoir donne au consommateur le sentiment d'avoir en sa possession une clef de contrôle sur celui-ci, comme si celui-ci pouvait choisir la météo. La culture est objectifiée et sa consommation se réduit à l'effet qu'elle a sur l'humeur de l'auditeur. 

L'extension de la musicalisation du quotidien a rapidement été suivie, dans les années 2010, par un effet similaire concernant la distribution de l'information, laquelle se trouve maintenant assimilée distraitement dans les transports via des podcasts. L'information est assimilée partout de manière passive. C'est ce qui permet à Jameson de conclure que le monde postmoderne, avec l'ensemble de ses flux connectés, est\cite{Post} 
\begin{quote}
\textit{
 un monde dans lequel la culture est une second nature.
}
\end{quote}

La culture et la connexion à son flux constant, est la nature du sujet postmoderne. L'individu reste toujours inséré et connecté aux flux d'informations. Cette connexion achève l'individu postmoderne comme un sujet de désirs fragmentés et flottants et constamment réécrit par les algorithmes. Cette réalité empirique rejoint la théorisation du sujet postmoderne, par exemple de Deleuze, théorisation pour laquelle le sujet est un noeud existant à l'intersection d'une multitude de flux. Dans cette théorie, le sujet se co-construit avec ces flux, en adéquation avec l'observation empirique. Cette observation permet de donner plus de poids à l'idée avancée plus haut selon laquelle la théorie du sujet postmoderne, suite à la mort du sujet représentationnel, serait plus proche d'une réalité empirique contemporaine.

\section{Le rôle du postmodernisme dans l'essor des plateformes}

Comme nous venons de le voir, le postmodernisme, pour Jameson, n'est pas simplement un style artistique, mais un ensemble de formes culturelles propres à une époque marquée par la mondialisation, les médias de masse et la consommation. Pour en résumer les grandes lignes, nous avons vu que celui-ci se caractérise par la disparition de la profondeur, au profit de la surface et de l’apparence. L’histoire y est remplacée par un présent perpétuel, sans mémoire ni projet. Les œuvres postmodernes utilisent le pastiche : elles imitent des styles anciens sans critique ni engagement. La culture devient fragmentée, spectaculaire et dépolitisée. Le sujet individuel y est instable, dispersé dans les flux d’images et de signes. L’art perd sa fonction critique et s’intègre pleinement aux logiques du marché.

La promesse portée par ce chapitre était d'expliquer la captation extrêmement rapide de l'attention, rapidité qui posait légitemement question, par une prédisposition structurelle de la culture postmoderne. Pour cela, nous avons procédé à une lecture suivie de l'analyse de la culture postmoderne produite par Jameson et l'avons mis en vis-à-vis avec les développements des chapitres précédents, liés à la disruption.

Un point crucial répété par Jameson est que le postmodernisme ne ``vient pas seul", il ne prend sens que lorsqu'il est analysé dans un système plus vaste\cite{Post} 
\begin{quote}
\textit{le postmodernisme n'est pas la dominante culturelle d'un ordre entièremenet nouveau mais seulement le reflet et le concomitant d'une modification systémique de plus du capitalisme.}
\end{quote}

Cette modification systémique de plus
du capitalisme à laquelle Jameson réfère est un ``nouvel esprit" du capitalisme dans le sens entendu par Boltanski et Chiapello.  

Dans notre lecture, trois pôles ont continuellement tiraillé l'analyse: 1) le pôle technique, représenté par les développements des réseaux et des plateformes et ses algorithmes, 2) le pôle culturel, dont nous avons étudié les catégories et les évolutions, 3) le pôle marchand et consumériste, actif en tant qu'injonction douce à la consommation notamment via la publicité, la captation des désirs et la séduction permanente. De ce fait, on peut tirer comme conclusion l'impossibilité d'une analyse isolée de l'un de ces pôles. Ces pôles sont aujourd'hui tellement intriqués et entremêlés qu'une analyse qui tâcherait d'en isoler un et de le soumettre à l'analyse serait vouée à l'échec. Le capitalisme tardif est la force qui a lié tous ces éléments entre eux. 

Les analyses ci-dessus ont permis de  mettre en lumière comment les techniques de la communication moderne et la démultiplication colossale des moyens de transmission de l'information se rattachent et se connectent à une théorie plus large de l'individu contemporain, à savoir l'individu postmoderne. Cela rattache une posture de l'humain face à la machine à une tendance philosophique de fond, à savoir la tendance à se défaire de la nature\cite{Post}, 
\begin{quote}

\textit{
 le postmoderne est donc ce que vous obtenez quand le processus de modernisation s'est achevé et que la nature s'en est allée pour de bon.}
\end{quote}

Ainsi, la conclusion de ce chapitre est la suivante. Le rejet de la nature est à la fois la caractéristique la plus marquante du postmoderne, et la porte d'entrée philosophique de ce mode de pensée. Les idées contemporaines telle que la disruption ne sont que des accélérations et des extrémisations de ce rejet primordial, et ne saurait aboutir que par celui-ci. C'est ici donc l'une des conclusions les plus importantes de ce mémoire: \textit{on pourrait trouver le moteur de la disruption dans le rejet de la nature que porte le postmodernisme}.

Cette conclusion peut sembler cependant trop ``philosophique", dans le fait qu'elle relègue la technique à une condition de possibilité, presque à un catalyseur, et donne un rôle premier aux tendances philosophiques comme moteur de l'histoire. J'estime cependant qu'il est prolifique de prendre cette conclusion à titre d'hypothèse.

Pour clore ce mémoire, tournons nous vers le futur. Là où le progrès se présente comme une force organisatrice du futur, l'innovation est le nom que l'on donne à la résistance humaine contre les forces entropiques. Adhérer à l'un où à l'autre de ces concepts augurent de notre relation avec le devenir. Nous allons donc maintenant étudier ce que l'époque de \textit{l'absence d'époque} peut augurer pour l'avenir et comment on pourrait, à l'avenir, faire émerger des structure néguentropiques.

\chapter{Les communs numériques}
\label{chap5}

Dans les chapitres précédents nous avons identifié l'origine de la disruption, et son moteur mécanique, dans  1. la réseautisation des échanges internet, suivi d'un vaste déploiement des algorithmes pour gérer l'afflux d'utilisateurs (chapitres \ref{chap2} et \ref{chap3}) et 2. dans le rejet culturel de la nature promu pour la culture postmoderne (chapitre \ref{chap4}). Nous avons argumenté que cette évolution de l'architecture d'internet amène une captation des protentions et des désirs des utilisateurs, qui induit une rapide croissance de ce que Stiegler a appelé l'anthropie, laquelle mesure la destruction des structures sociales. A cette captation de la libido, se redouble aujourd'hui une captation des savoir techniques avancés, comme cela s'observe avec les LLMs. Ce sont donc également les savoirs des ingénieurs et des techniciens qui sont captés par l'appareil marchand. Cette captation est ce qui contribue même à l'anthropisation -- avec néanmoins le bémol qui nous avons avancé à la fin du chapitre \ref{chap3}, section \ref{sec:bemol}.

Ce mémoire était dédié à  l'étude du phénomène de la disruption, de ses conditions d'émergence, de ses dangers et de ses conséquences.  Comme cela a été maintes fois mis en évidence, l'un des dangers des approches critiques systématiques est d'engendrer un sentiment de résignation devant la force de l'objet critiqué, qui semble avoir tellement profondément colonisé les systèmes sociaux que prendre des mesures effectives à l'encontre de celui-ci, ou même le critiquer semble vain. En conséquence, la critique ne parvient qu'à induire un nihilisme passif chez ses lecteurs, en montrant les mécanismes de perte de la valeur à l'oeuvre dans le monde, il encourage une position nihiliste depuis laquelle les acteurs peuvent argumenter de manière construite que rien n'a de valeur, et que l'action n'a aucun sens. Dans ce mémoire, nous avons découvert que la force derrière la disruption généralisée est à la fois le capitalisme et l'entropie, dont l'augmentation est l'une des lois de la physique contemporaine, et la culture. 

Pourtant Stiegler nous rappelle que la plateforme, comme toutes techniques, est un  \textit{pharmakon}.

\section{Les plateformes comme \textit{pharmakon}}

Le \textit{pharmakon} est un composé, ou une pratique, qui peut soigner ou qui peut empoisonner l'individu en fonction de la dose administrée et de la posologie. Chez Bernard Stiegler, la notion de \textit{pharmakon}, empruntée à Platon, désigne un objet ou une technique ambivalente, à la fois poison et remède. D'une façon intéressante, toute technique, selon lui, possède cette double nature: elle peut soigner ou détruire selon l’usage qu’on en fait et le cadre symbolique dans lequel elle s’inscrit. Le \textit{pharmakon} n’est donc pas neutre, mais dépend de l’organisation collective des savoirs, des pratiques et des institutions qui l’accompagnent. 
De manière similaire, en parlant de ce qu'il a appelé la société de séduction, Gilles Lipovetsky nous dit\cite{Plaire}
\begin{quote}
\textit{Il n'est pas dit que le divertissement futile soit le dernier mot de la société de séduction: face à l'hypertrophie marchande, nous n'avons pas à promouvoir un ethos ascétique, mais à rendre désirables des activités plus élevées, plus créatives}
\end{quote}

Pour lui, la société de la séduction est également une sorte particulière de \textit{pharmakon}.

Stiegler applique la notion de \textit{pharmakon} aux technologies numériques. Il nous les présente comme des technologies qui peuvent nourrir l’intelligence collective ou provoquer une dénoétisation, ainsi que la série de symptômes largement négatifs que nous avons développés tout au long de ce mémoire.  Le numérique ainsi que son appareil algorithmique, en captant l’attention et en automatisant les processus cognitifs, deviennent un poison s’ils ne sont pas intégrés à des structures critiques et contributives. 

Inversement, ces mêmes techniques peuvent devenir un remède si elles sont mises au service de l’individuation et du partage du savoir. Dans ce sens, la question du \textit{pharmakon} encourage une responsabilité éthique et politique dans le maniement des technologies. Chez Stiegler, cela implique une  pharmacologie de l’esprit et de la société. Ainsi, si les nouveaux outils numériques sont à la source de difficultés et de problèmes d'une ampleur sans doute jamais rencontrée, il faut aussi remarquer qu'ils contiennent en germe, la solution à ces difficultés. 

C'est exactement ce que des plateformes telles que Youtube parviennent parfois -- rarement, sans doute -- à promouvoir,  au travers par exemple de nombre de chaînes de vulgarisation, lesquelles ont initié la réhabilitation de l'exercice de vulgarisation, notamment scientifique. On peut citer encore une fois les grandes chaînes Youtube de vulgarisation telles ques Veritasium, Mr Phi, Sciences étonnantes, ... qui sont parvenues à injecter le ludique au sein de la communication scientifique et par cela à toucher un large audimat, et en faisant de la connaissance scientifique un savoir néguanthropique. Pour parler le langage de Stiegler, de nombreuses voies de bifurcation néguanthropique continuent d'exister sur les réseaux d'échange, et survivent tant qu'elles obéissent aux règles du jeu imposées par les algorithmes.

Ainsi, exploitant cette notion de \textit{pharmakon}, Stiegler nous présente quelques exemples de renéotisation empruntés à d'autres sphères. Dans ce contexte, la renoétisation est le processus qui s'oppose à la dénoétisation étudiée plus haut, dans la section \ref{sec:deno}, et désigne un processus par lequel des pratiques, des objets ou des systèmes technologiques retrouvent une charge symbolique, affective ou idéologique -- Et, pourrait-on dire, spirituelle -- en inversant l'effet du \textit{pharmakon}, lequel de poison devient alors remède.  Elle marque le retour d’un sens ou d’une signification dans un monde où ceux-ci avaient été dissous par la disruption. En ce sens, la renoétisation accompagne la réinvention du sens à l’ère postmoderne. 

L'une de ses propositions les plus vastes du point de vue de son application est celle du revenu contributif, présentée à la conférence publiée sur Youtube \footnote{https://www.youtube.com/watch?v=pfJnDTPRhIk}. Stiegler propose un revenu contributif et conditionnel qui ne rémunère pas un travail salarié classique, mais l'engagement dans des activités de production de savoir, de culture ou de soin. Ainsi, en guise d'exemple, une contribution à Wikipédia pourrait être rémunérée. 

Inspiré du régime des intermittents du spectacle, ce revenu permet à chacun de contribuer activement à la société sans passer par l'emploi traditionnel. Notons que pour Stiegler, l'emploi, au sens de l'occupation qui octroie un salaire, et le travail, au sens de l'opération de modification du monde alentours et donc de production de néguanthropie, ne coïncident pas nécessairement et peuvent parfois directement s'opposer: dans certains cas, le système emploie des personnes (leur octroie un salaire), pour les empêcher de travailler (de modifier le monde et de produire de la néguanthropie). Pour lui, c'est d'ailleurs l'une des motivations de la recherche absolue du plein emploi. Stiegler propose d'intensifier le travail et de minimiser l'emploi.   Il s’agit d’une manière de redonner du sens au travail en tant qu’activité de pensée et de création. Cette approche vise à désautomatiser les comportements et à lutter contre la passivité générée par les industries culturelles, que nous avons décrites dans les chapitres précédents. Le revenu contributif devient ainsi un levier pour réactiver l’individuation psychique et collective. 

Sans entrer directement en conflit avec la proposition du revenu contributif de Stiegler, il faut néanmoins noter qu'elle porte le risque de ``professionnaliser" la contribution, et ainsi de lui faire perdre sa liberté et son autonomie.

Un autre exemple crucial qu'avance Stiegler est celui du logiciel libre, dont Linux, Ubuntu et autres sont autant de représentants. Stiegler valorise les communautés du logiciel libre comme modèle de rénoétisation par la technique. Contrairement aux plateformes propriétaires qui capturent l’attention et privent les usagers de contrôle, le logiciel libre repose sur la participation, le partage du code et l’apprentissage collectif. Ainsi, tout monde a accès au code et a la capacité de le modifier si il en a les compétences.  Cette approche encourage les individus à se réapproprier les outils techniques et à comprendre les mécanismes qui structurent leur quotidien, elle réintroduit donc du sens dans le travail des développeurs qui sont en contact direct avec le résultat de leur travail. Ainsi, cette proposition entend s'opposer à l'appropriation du savoir technique par les algorithmes.  Elle forme des utilisateurs actifs capables de produire du sens à partir de la technique elle-même. Le logiciel libre devient ainsi un espace d'individuation, au sens où nous l'avons présenté plus haut, contre la dénoétisation numérique et algorithmique. Nous pouvons donc voir que même l'algorithmique est un \textit{pharmakon} qui peut contribuer à la dénoétisation ou à la renoétisation des personnes.

Néanmoins, aucun de ces exemples ne concernent directement les plateformes en réseau. Dans ce dernier chapitre, nous montrons dans un premier temps que des structures \textit{numériques} à grande échelle, des plateformes, résistants à la disruption existent et que certaines sont en plein essor. 
Nous essayons ensuite de voir ce qui dans leur fonctionnement est spécifique, comme ce qui chez le vivant est spécifique et lui permet de résister à l'entropie.  En guise de première piste de recherche, nous commençons par présenter ce que l'on appelle la charte des communs et comment cette charte pourrait tentativement s'appliquer à ce que nous allons appeler des \textit{communs d'internet} tels que Wikipédia et l'arXiv.

\section{La charte des communs}

Comme nous l'avons entrevu avec la notion physique d'entropie, l'état d'équilibre stable est l'état le plus pauvre en possibilités. Au contraire un état riche en possibilités est toujours instable, il est toujours dans une situation de possible bifurcation. Pour Simondon, les états riches en possibilités sont métastables et ne peuvent exister qu'en tant qu'équilibre non pas statique, mais \textit{dynamique}. Dans cet état dynamique, un équilibre s'est formé dans les échanges entre les acteurs internes au système.  Par ce raisonnement, on peut comprendre que la persistence dans le temps d'un état riche en possibilités est toujours un problème dont il faut éclaircir la dynamique. 

Un exemple de tel équilibre dynamique peut se rencontrer dans ce que l'on appelle aujourd'hui un ``commun", à savoir un système pourvu d'une ressource épuisable dont plusieurs acteurs cherchent l'exploitation. La stabilité dynamique du système repose donc sur la création d'un équilibre entre les différents acteurs qui exploiteront la ressource \textit{en commun} de façon à ce que celle-ci ne s'épuise pas.   

Dans ses travaux sur la gouvernance des ressources communes \cite{ostrom1990governing}, pour lesquels elle reçoit le prix Nobel d'économie en 2009, la politologue et économiste américaine Elinor Ostrom a mis en évidence huit principes institutionnels qui favorisent une gestion durable et efficace des ``communs". Nous avons déjà explicité ces huit principes dans la section \ref{sec:Eco} et ne les répétons pas ici. Ces huit principes constituent un cadre théorique pour comprendre et étudier des systèmes de gouvernance communautaire capables de préserver durablement les ressources naturelles tout en assurant l’équité et la participation démocratique des agents en présence. Nous tentons de montrer ici qu'en plus d'ouvrir des perspectives positives pour l'exploitation par exemple des ressources fossiles, le modèle de commun établi par Ostrom se prête à la généralisation, notamment pour les ressources numériques. Ainsi, nous allons maintenant les appliquer tentativement à l'internet et aux plateformes en réseau.

\subsection{Les communs numériques et le \textit{Creative Commons}.}

 Dans un tout premier temps, il convient de déterminer ce qui constitue la ressource principale qui est charriée sur les réseaux. La ressource la plus évidente est la ressource monétaire, laquelle est convoitée par les annonceurs qui mettent alors les plateformes en concurrence pour amasser et attirer la plus grande attention humaine, sur laquelle se déverse alors la publicité. De ce fait, ce que l'on a appelé l'attention des internautes, ou leur temps de cerveau disponible, est l'une des ressources principales que peuvent espérer capter les réseaux algorithmiques. Tout aussi foncièrement monétisées, les données que les utilisateurs laissent dans leur sillage sont l'objets de toute une économie de la donnée, économie qui constitue le \textit{big-data}. Après avoir été collectées, ces données sont ensuite revendues. 

 Si les données semblent illimitées, l’attention humaine, elle, est fondamentalement rare, en raison du temps limité disponible pour chaque individu. Ce déséquilibre a conduit à l’émergence d’une véritable économie de l’attention, dans laquelle les contenus – qu’ils soient informatifs, divertissants ou mensongers – sont mis en concurrence par les annonceurs pour capter l’engagement de l’utilisateur. De nombreuses études (Voir notamment \footnote{https://mediate.com/mediation-in-the-age-of-digital-distraction/?utm\_source=chatgpt.com} qui renvoie à d'autres références) documentent la perte d’attention provoquée par l’utilisation intensive des médias numériques, que ce soit par la consommation de contenus brefs, les shorts, le multitâche média ou les interruptions constantes dues aux multiples réseaux sociaux utilisés en parallèle.

D'un autre côté, dans cet environnement saturé d’informations de qualité inégale, l’\textit{attention} devient la condition essentielle permettant à l’individu de trier, hiérarchiser et interpréter les contenus. Cela nécessite un effort cognitif d’autant plus important que les sources sont multiples, fragmentées et parfois trompeuses. La multiplication des fausses informations, que l'on a récemment renommées fake news, accentue ce phénomène et menace la stabilité des sources fiables. La prolifération des fake news contribue elle-même à la méfiance généralisée dont nous avons déjà traité plus haut.  

Dans ce contexte, une source d’information de qualité peut être envisagée comme un bien commun fragile, en cela qu'elle est constamment exposée au bruit informationnel et à la perte de l'attention des acteurs. Elle ne peut se maintenir que par la vigilance de ses contributeurs, qui assurent un tri actif à l’entrée comme à la sortie. Ainsi, l’accès à une information pertinente dépend de ce que Stiegler appelle un \textit{soin collectif} constant, soin qui veillera à ce que la ressource informationnelle soit partagée entre exploitation et préservation.

En guise d'exemple, nous étudierons deux cas concrets de sources d'informations fiables, celui de l'encyclopédie en ligne, \textit{wikipédia}, et la plateforme de publication d'articles en preprint, l'\textit{archive}, ou arXiv, la plateforme de mise en commun des prépublications universitaires dans les domaines de la physique et des mathématiques.

\paragraph{Wikipédia}

Penchons nous donc dans un premier temps sur le cas de wikipédia et décrivons-en le fonctionnement interne. Wikipédia est une encyclopédie collaborative et libre, lancée en 2001, où n’importe qui peut créer ou modifier un article. Les statistiques de la fiabilité de Wikipédia peuvent être consultées ... sur Wikipédia\footnote{https://fr.wikipedia.org/wiki/Fiabilit\%C3\%A9\_de\_Wikip\%C3\%A9dia .}, néanmoins, la plupart des études montrent que la fiabilité des informations sur Wikipédia rivalise avec les plus grandes encyclopédies professionnelles et académiques. Qu'est-ce qui fait le succès et la précision de cette plateforme collaborative tenue par des non-professionnels  ? 

Sur la plateforme, les contributions sont anonymisées. On ne connaît ni le nombre, ni les qualifications des contributeurs qui concourent à l'écriture d'un article. Tout article, après sa première rédaction peut être modifié par tout contributeur, anonyme ou enregistré.  Pour éviter la propagation de fausses informations, et de vandalisme, une série de modérateurs encadre le travail de rédaction et de révision, sans pour autant assigner aucune tâche aux contributeurs. La plateforme ne fait appel à aucun annonceur, et son financement vient des dons de ses utilisateurs.

Ce qui fait la charpente de Wikipédia, ce sont les principes de sa charte, qui peuvent être consultés sur le site lui-même \footnote{https://fr.wikibooks.org/wiki/Wikip\%C3\%A9dia/D\%C3\%A9couvrir\_Wikip\%C3\%A9dia/Principes\_fondateurs} et qui sont rappelés à tout contributeur avant publication.  Premièrement, Wikipédia est une encyclopédie, et tout article doit avoir pour but d'informer sur un point de la connaissance humaine. Deuxièmement,  Wikipédia  vise la neutralité des points de vue, dans le sens où la présentation du sujet doit offrir un panorama des différentes points de vue existants et offrir les arguments avancés pour la validité de chacun, sans en favoriser un injustement. Dans ce cadre, les articles doivent être vérifiables à partir de sources fiables et publiées, sans recherche inédite ni promotion personnelle.  Troisièmement, Wikipédia repose sur des contenus libres de droits, l'entièreté du texte est publié sous la license \textit{Creative Commons} (Creative Commons, paternité, partage à l’identique (CC-BY-SA),  qui requiert de citer les auteurs et de partager sous la même licence) et n'a recours à aucun annonceur ni financement externe. Le respect du droit d’auteur est impératif, tout comme l’exigence de notoriété pour les sujets traités.

Quatrièmement, Wikipédia encourage le respect et la coopération entre contributeurs, et appelle au respect des règles de savoir vivre. Les conflits d'édition sont à éviter. Finalement, la plateforme accepte l’évolution de ses règles si cela améliore le l'état de la plateforme. L’objectif est de construire une base de connaissance ouverte, fiable et rigoureuse.

\paragraph{ArXiv, medRxiv and bioRxiv }

Une information d'un type plus central encore est l'information de haute qualité produite par des organismes dont l'occupation et le but premier est justement la production d'information. Les centres de recherches et les universités sont une telle source d'information de qualité.  Comme dans le cas des médias tels que les journaux et les magazines, l'accès à cette information  est souvent monétisé. Cependant, dans le cas de l'édition scientifique, la monétisation ne se fait pas par les producteurs de l'information elle-même.

Le monde de l'édition scientifique est souvent décrit comme ubuesque. Alors que les scientifiques produisent des résultats synthétisés en articles scientifique, ceux-ci doivent par la suite être soumis à un journal de publication. Le journal en question invite des autres scientifiques à donner un avis critique et professionnel, bénévolement, sur ledit article et, finalement, une fois l'article publié dans le journal, le journal vend aux lecteurs ou aux universités, pour un prix parfois exorbitant, les articles qu'il n'a ni contribué à écrire, ni à vérifier. Ce système est tellement absurde qu'on pourrait légitimement se demander comment il tient, sinon par un étrange effet d'inertie. Inertie qui, semble-t-il, commence à s'effriter dans biens des domaines. 

Partiellement en réponse à la lenteur des publications dans les journaux à comité de lecture, la plateforme internet \textit{arXiv} a été fondée en août 1991 par le physicien Paul Ginsparg, alors chercheur au \textit{Los Alamos National Laboratory} (LANL). Il a créé une archive électronique destinée à faciliter le partage rapide des prépublications scientifiques, en particulier dans le domaine de la physique. Ce système a été mis en place à une époque où les échanges de documents scientifiques étaient principalement effectués par courrier électronique ou par courrier postal, ce qui limitait la rapidité de diffusion des informations. L'initiative de Ginsparg a permis de centraliser ces échanges et d'accélérer la communication scientifique dans le domaine de la physique et des mathématiques. Des initiatives similaires dans les domaines de la bioloqie et de la médecine ont également vu le jour, via les plateformes  medRxiv and bioRxiv. A l'instar de Wikipédia, ces plateformes offrent la possibilité de publier sous la licence ouverte \textit{Creative Commons} et tout comme Wikipédia, arXiv possède une charte qui est consultable à l'adresse \footnote{https://info.arxiv.org/about/principles.html.}.
Dans la foulée, les journaux en open-access, qui offre une relecture par les pairs, tout en les rémunérant pour le travail de critique, mais publient les articles en accès libre, se multiplient. On peut citer par exemple JHEP(Journal of high energy physics), JCAP (Journal of cosmology and astrophysics). Une nouvelle tendance, lancée par quelques auteurs, appelle à abandonner la publication dans des journaux à comité de lecture, et à publier uniquement sur arXiv.

On peut noter un point important concernant cette tendance à l'open-source et à l'ouverture: celle-ci se répand d'autant mieux dans des milieux modérément étroits et dans lequel le nombre d'acteurs est assez limité pour que la réputation permette de motiver l'engagement personnel des participants.

\paragraph{ArXiv, Wikipédia et la charte des communs}

On peut maintenant étudier, en faisant une analyse comparée, dans quel sens les principes de Wikipédia et d'arXiv s'alignent sur les principes des communs édictés par Ostrom:

\begin{itemize}
    \item Premier principe des communs: \textit{il faut définir des limites nettement définies des ressources et des individus qui y ont accès (qui permettent une exclusion des entités externes ou malvenues).} Dans le cas de l'information encyclopédique ou universitaire, la ressource est la plateforme elle-même, en tant qu'elle est entretenue et modérée de façon efficace, elle offre un filtre à l'information de qualité. Il s'agit donc d'une ressource en réseau, plateforme à laquelle tous les utilisateurs ont en principe accès et que tous peuvent modifier. L'accès est donc ici non-restrictif, ces plateformes sont des biens de l'humanité.  
    \item Second principe des communs: \textit{il faut édicter des règles bien adaptées aux besoins et conditions locales et conformes aux objectifs des individus rassemblés}. Dans le cas des plateformes, les règles édictées le sont sous la forme d'une charte expliquant ce qu'il admissible de publier sur la plateforme et ce qui ne l'est pas. Ces règles sont conformes à ce que l'on peut imposer sur internet et sont accessibles à tous facilement. 
    \item Troisième principe des communs: \textit{il faut établir un système permettant aux individus de participer régulièrement à la définition et à la modification des règles (faisceau de droits accordés aux personnes concernées).} Dans le cas de Wikipédia, la flexibilité des règles est une règle même de la charte, qui appelle les contributeurs à discuter le type d'article qui doit appartenir à l'encyclopédie. 
    \item Quatrième principe des communs: \textit{il faut établir une gouvernance effective.} Cette gouvernance est dans le cas des plateformes présente sous la forme de modérateurs, qui ont pour but de diriger les discussions et de sanctionner les débats. Il faut tout de même noter que dans les deux cas étudiés, le nombre de modérateurs fait parfois défaut et la gouvernance peut être défaillante. Pour arXiv, la modération se résume à l'étude de l'adéquation de chaque article proposé à la publication avec la catégorie dans laquelle il est proposé, ce qui reste relativement limité. 
    \item Cinquième principe des communs: \textit{il faut établir un système gradué de sanctions pour des appropriations de ressources qui violent les règles de la communauté.} Ces systèmes de punition consistent essentiellement en des mesures de bannissement, mais elles sont faibles ou difficiles à appliquer pour les plateformes. En effet, aucun individu ne peut être durablement banni au vu de l'anonymat des intervenants d'internet. 
    \item Sixième principe des communs: \textit{il faut établir un système peu coûteux de résolution des conflits.} Dans le cas des plateformes, ce système se résume typiquement à un appel au savoir vivre, comme dans le charte de Wikipédia et d'arXiv, ainsi que la possibilité de l'intervention du modérateur. 
    \item Septième principe des communs: \textit{il faut autoriser une auto-organisation reconnue par les autorités.} Cette possibilité est explicitement présente dans la charte du Wikipédia, qui appelle chaque communauté linguistique à édicter et à faire évoluer ses propres règles, tout en conservant le principe du projet. 
    \item Huitième principe des communs: \textit{dans le cas de grands systèmes avec de multiples niveaux, il faut créer différents niveaux d'autorité}. Ce cas est présent explicitement dans les sous-communauté linguistique de Wikipédia, qui ont un pouvoir d'auto-détermination. Notons tout de même que l'autorité telle qu'elle est réalisée sur les plateformes est toujours diffuse, dans le sens où les modérateurs n'imposent pas, mais plutôt médient et orientent les débats et les discussions. Ils n'ont comme pouvoir, déjà considérable, que celui de bannir certains comptes. Ce type de contrôle n'est pas à confondre avec le soft power des entreprises, dont la forme suave et concilliante camoufle l'absolue rigidité dans les faits. 
\end{itemize}

Il faut également souligner que sur Wikipédia et arXiv, même si les modérateurs peuvent recourir ponctuellement à la puissance de suggestion des algorithmes, notamment dans les détections des articles suspects, ceux-ci gardent le dernier mot. Chaque article suspect est alors étudié par des collaborateurs humains. En aucun cas la plateforme n'est laissée entièrement à la direction d'un algorithme, soit-il de suggestion ou de tri. 

Si l'équivalence entre le commun naturel au sens d'Ostrom et le commun numérique au sens où nous voulons l'étudier semble être une bonne première voie d'investigation, elle n'est pourtant pas parfaite, et ne peut pas vraiment fonctionner. La raison de cet échec est que la théorie des communs à été pensée pour des systèmes pourvu d'une ressource pouvant s'épuiser. Ce n'est pas le cas des ressources de partage d'information sur internet. De plus, les collectifs qui maintiennent ces structures ont une forme très différente des collectifs locaux exploitant un commun. La différence principale est l'éloignement géographique des membres du collectifs. Les individus au sein des réseaux entretenant Wikipédia ne se sont le plus souvent jamais rencontrés en personne. Les forces les liant sont donc plus faibles. Ainsi le danger qui pèse sur les ressources informationnelles d'internets telles que Wikipédia, est le danger de se déliter par le manque d'apports et de contributions. Il semble donc que ce qui différencie les communs naturels et les communs numériques est le péril qui pèse sur chacun d'entre eux, là où le premier risque l'épuisement, le second encourt le délitement.

Dans le cas de Wikipédia, ce qui empêche ce délitement, c'est un effort constant de la part des contributeurs et des modérateurs, un effort qui va contre l'augmentation de l'anthropie, et donc de l'entropie. Quel est donc l'origine de cette force, qui ne peut pas être d'origine naturelle et qui maintient l'ordre d'une information hautement structurée et donc hautement qualitative?

\paragraph{Véracité, vérité et anthropie}

Pour développer un peu d'intuition sur le phénomène de délitement des sources d'information de qualité, nous pouvons reprendre les concepts d'entropie et d'anthropie empruntés à la physique. Là où l'accroissement de l'entropie est une tendance qui pousse les systèmes physiques vers une situation d'équilibre stable final et qui trouve son origine dans le fait qu'un état désordonné peut être composé de beaucoup plus de façons que peut l'être un état ordonné, l'accroissement de l'anthropie est une tendance qui pousse les interactions humaines et les groupes humains vers une situation d'équilibre chaotique sans structure.  Nous avons argumenté dans la section \ref{sec:anthro} que ce qui tient ensemble les concepts d'entropie et d'anthropie c'est que ces deux concepts émergent de l'observation qu'un état infiniment désordonné est combinatoirement le plus probable. 

Dans le contexte de la recherche et de la propagation de la vérité, la déformation, la perte de précision, la volonté de tromper constituent très exactement ce que l'on pourrait vouloir capturer par l'idée de l'accroissement de l'anthropie. Il est beaucoup plus probable d'énoncer une erreur, une imprécision, que la vérité. Pour chaque vérité il existe une infinité d'erreurs. Depuis la recherche et puis à chaque étape de la transmission, l'anthropie travaille à introduire des erreurs qui se propageront et grandiront, jusqu'à détruire l'état très improbable de l'énoncé vrai. C'est ici le problème fondamental de la vulgarisation. 

Ainsi, de la même façon que le vivant est ce qui localement défie la loi de l'entropie, jusqu'à constiuer une des grandes questions scientifiques de notre temps, l'énoncé vrai, la vérité, est ce qui localement défie la loi de l'anthropie, dans un milieu de l'échange de l'information qui tend à faire s'accroître l'anthropie de cette information. En conséquence de cette observation, on pourrait considérer que la propagation et la durabilité dans le temps d'une vérité, même partielle, devrait constituer un problème scientifique d'égale complexité. En ce sens, la lutte contre l'anthropie de l'information est analogue à la lutte que l'organisme vivant voue continuellement à l'entropie physique. En analogie à la complexe organisation moléculaire dont sont pourvus les vivants et qui continue d'être l'intense sujet d'étude de la médecine et de la biologie, un organisme comme Wikipédia est un organisme en comparaison beaucoup plus simple dont la dynamique, également régie par quelques mécanismes, tend à maintenir une anthropie informationnelle basse. 

Quelle est la source de la possibilité de l'existence de tels organismes ? Pour ce qui est de la possibilité de l'existence d'organismes vivants, la question reste largement ouverte, même s'il semble aujourd'hui que la question soit résolue par l'entropie elle-même: les êtres vivants seraient des organismes très efficaces pour produire eux-mêmes de l'entropie, que nous nommons dans notre environnement ``pollution". Donc on peut dire qu'en échange d'une diminution locale de l'entropie pour sa structure interne, le vivant augmente tellement l'entropie de son environnement que la balance finale est positive. La question de l'existence et du maintien d'organisme de traitement de l'information de haute qualité est tout aussi intriquée.

Faisons un détour par le microcosme du monde producteur de ces vérités. Les détenteurs et les gardiens de la vérité du monde ont longtemps été, et sont toujours dans une certaine mesure, les académiciens et les universitaires. Contrairement à une vulgate qui s'est récemment répandue au sein du grand public, vulgate selon laquelle la ``science c'est le doute de tout et en tout temps", en réalité, l'organisation de la connaissance repose largement sur la confiance en les pairs, la confiance en leur ``bonne foi" ainsi qu'en leurs compétences. Dans cette économie de la confiance, l'un des biens les plus précieux est la réputation, à savoir la valeur du nom. L'autorité d'une personne, d'un nom, d'une institution se construit dans le temps par l'observation et la confirmation toujours renouvelée de sa minutie. Et cette confirmation, lorsqu'elle a lieu, amène à la confiance. La confiance en l'autorité. Cette économie de la confiance est largement répandue. Pour ``vendre" les résultats de recherche, on fait souvent valoir la réputation: ``Selon des chercheurs du MIT", ``Comme l'a montré Harvard", ... C'est aussi la raison pour laquelle les chercheurs se méfient instinctivement d'un travail de recherche non-signé. Comment pourrait-on allouer notre confiance à un inconnu ? A quelqu'un qui n'ose même pas mettre son nom en jeu ?

La science demande au grand public un acte de foi beaucoup plus grand et donc beaucoup plus fragile: celui de faire confiance en les valeurs d'un groupe partiellement étranger, en des personnes le plus souvent inconnues, et en des méthodes que le grand public ne comprend pas et ne conçoit. Cela ne pose guère de difficultés lorsque les vérités en question ne portent que sur des points de peu d'intérêt pratique. Néanmoins, les limites de cette confiance apparaissent lorsque les questions de société, politiques, s'entremêlent avec des questions politiques, comme cela a été maintes fois le cas dans le passé et comme c'est maintenant le cas avec la question climatique.  
Dans ``\textit{vérité et véracité}", Bernard Williams nous présente cet acte de foi 
\begin{quote}
  \textit{L'autorité des universitaires doit s'enraciner dans leur véracité, qui prend les deux formes suivantes: ils s'appliquent et ils ne mentent pas.}
\end{quote}

j'aimerais maintenant aller au delà de la position de Bernard Williams et avancer l'idée que notre volonté pour la véracité nous vient de la souffrance causée par le fait d'avoir été trompé. Cette souffrance d'avoir été trompé, lorsqu'elle est trop ressassée, se mue en recherche constante de la vérité, en obsession pour sa conservation.  Suivant Nietzsche, je soutiens que cette obsession nous a été inculquée, imprimée, par plus de deux millénaires de monothéismes, lesquels ont tous ressassé l'importance de la recherche et du respect de la vérité.

Profondément, l'inconscient collectif sait que la confiance globale est une qualité cruciale pour éviter l'émergence d'un âge de tromperie, et donc pour éviter la souffrance qui vient avec la tromperie généralisée. C'est cette intuition ancrée dans l'inconscient collectif qui porte le maintien de structures comme Wikipédia. Comme nous le voyons, cet attrait pour la vérité est néguanthropique et a la même origine que l'effort, néguentropique, que nous renouvelons pour garder nos maisons rangées.

Il est évident que la prise de position défendue dans cette partie est un postulat qui serait difficile à prouver empiriquement, cependant, on peut la concevoir comme une idée régulatrice au sens kantien -- à savoir, une idée agissant sur le réel sans pour autant être démontrable empiriquement. Ce postulat peut nous servir de point d'appui pour construire une institution collaborative.

\section{Faire des plateformes des institutions contributives}

Nous avons étudié en détails deux exemples d'économies contributives que sont Wikipédia et arXiv et avons mis en avant ce qui les singularisait. Leur présentation n'avait pourtant pas valeur de contre-exemple, mais bien plutôt d'exemple à suivre et à généraliser, à institutionnaliser. A la dénoétisation généralisée, répandue sur toutes les plateformes et que les plateformes induisent sur l'esprit des humains, il nous faut appliquer une réponse institutionnelle.  Pour Bernard Stiegler, l'avenir réside dans\cite{Disrup}
\begin{quote}
  \textit{la question de la renéotisation par la reconstitution du désir.}
\end{quote}

Ce désir doit être compris en tant que  désir néguanthropique et noétique, non pas comme désir suggéré et implanté massivement. Il s'agit du désir du savoir faire et du faire ensemble. 
Les exemples de projets collaboratifs, comme wikipédia et arXiv, que nous avons développé, sont une intéressante porte d'entrée pour aborder une systématisation du commun numérique, qui ne s'appliquerait pas seulement à quelques cas particuliers, mais s'imprimerait sur toute l'architecture d'internet et des plateformes. En effet\cite{Commun},
\begin{quote}
  \textit{le travail considérable que laisse Elinor Ostrom a une portée qui va au-delà de la seule question des communs ; il est important également, de manière plus générale, pour ce qui est de l’analyse des institutions. Elinor Ostrom est aussi une théoricienne des institutions.}
\end{quote}

Au delà de ses principes des communs, Ostrom  théorise des structures plus larges, dont l'institution. Ostrom conçoit une institution comme un faisceau de règles imbriquées et complémentaires constituées de règles opérationnelles, de règles décisionnelles collectives et règles constitutionnelles permettant de structurer l’accès et l’usage des ressources.  

Dans ce sens, des structures comme Wikipédia sont déjà institutionnelles. L'institution collective structure, régule et rend durable une forme d’action collective auto-organisée autour d’un bien commun — en l’occurrence, le savoir encyclopédique libre. Cette action est renouvelée et stabilisée par ce que nous avons appelé l'attrait pour la véracité expliqué ci-dessus. Comme nous venons de le voir, les communs numériques sont des structures néguanthropiques en équilibre instable et donc en bifurcation constante: cette stabilité, dans le cas des sources d'information fiable, est atteinte par l'incorporation et la discussion de nouveaux énoncés, vrais ou faux, qui tentent d'entrer dans le corpus et d'interagir avec celui-ci. L'équilibre ne peut être que dynamique et se matérialiser par un tri constant entre ce qui est vrai et faux.

La science est une institution stabilisée par ses nombreuses interactions avec le reste du corps social, entre autres, comme nous l'avons vu, par la véracité des académiciens. 
Sur les réseaux la stabilité est assurée par les mêmes types de dynamiques sociales auto-renforçantes. Ainsi, inscrire les utilisateurs des autres plateformes dans un sentiment similaire de recherche de véracité participerait à un effort de renéotisation numérique. Ce goût pour la véracité que tous les utilisateurs partageraient entraverait alors la mise en place d'outils menant à la calculabilité des fonctionnalités de la plateforme, telles que l'évaluation permanentes des acteurs, la classification, ... et qui sont à l'heure actuelle légion sur la majorité des plateformes les plus usitées.

La nouveauté ainsi présentée dans ce chapitre est le rapprochement entre les principes de communs découverts par Ostrom avec les quelques structures néguanthropiques qui existent sur internet. En faisant ce rapprochement, nous avons remarqué que  les règles d'Ostrom ne convenaient pas tout à fait à des exemples numériques, surtout parce que le danger encouru par les ressources numériques, le délitement, n'était pas le même que le danger encouru par les ressources naturelles. Ainsi nous avons tenté de les amender et de remonter à la source de ce qui constituait l'effort porté à ces structures. La proposition que nous avons extrait de cette analyse est de tirer parti du besoin de véracité et de confiance des individus pour assoir la stabilité des larges plateformes de communs, et pour en faire des institutions.

\chapter{Conclusion}

L'internet, depuis son émergence il y a de cela plus de 30 ans,  est devenu un outil de pérégrination intellectuelle qui n'a pas de pareille. Mais comme toute pérégrination, celle-ci peut être entropique ou rassénérante, elle ne peut avoir ni queue ni tête, se perdre dans des méandres inutiles ou trouver des trésors au détour d'une divagation stérile. L'internet est aussi l'outil le plus utilisé au monde, en compétition sans doute avec l'écriture et les transports. 

Les 20 dernières années ont vu sur internet l'émergence et l'essor spectaculaire des plateformes numériques, lesquelles rendent bien des aspects de la consommation quotidienne largement plus pratiques, rapides et efficaces. Certaines de ces plateformes sont donc rapidement devenues des mastodontes internationnaux comptant parmi les plus grosses firmes du monde. Rapidement, elles ont également capté une large part des secteurs tels que l'hôtellerie, les transports de personnes, les transports de biens, et les réseaux sociaux. Leur quasi-monopole leur a permis de déployer en grand nombre des algorithmes de captation de données. Ces algorithmes se sont rapidement redoublé d'une machinerie de suggestion de contenu ayant pour effet l'orientation des désirs des consommateurs. Ces innovations techniques ont induit une série de conséquences sur le grand public que ce mémoire avait pour objectif de décortiquer. 

Dans ce contexte, la captation des données et des savoirs humains par ces algorithmes menacent de rendre la modélisation théorique obsolète, engendrant donc une prolétarisation d'un nouveau type, concernant même les emplois les plus qualifiés. Cette captation des données et des savoir se redouble d'une captation des protentions humaines qui menace de produire une perte de la capacités à générer des protentions personnelles. Cette perte se manifesterait alors dans une perte pour l'humain de sa capacité à s'individuer au travers de ses parents et de ses proches, et donc de créer des sociétés humaines. Nous avons vu que le danger était alors l'émergence d'une nouvelle forme de barbarie, contruite sur la souffrance induite par ces pertes.

Nous en avons conclu que l'un des corrélats les plus manifestes de l'essor des plateformes est la disruption des sociétés humaines. Nous avons dressé le portrait de cette disruption ainsi que les mécaniques de son fonctionnement, en la définissant par sa vitesse de dissolution des groupes sociaux. Cela nous a permis d'invoquer les considération de Bernard Stiegler qui rattache la disruption à un accroissement décuplé des niveaux d'entropie dans nos sociétés, correspondant à une prolétarisation et à une standardisation accélérée dans tous les secteurs de la société, y compris les secteurs les plus spécialisés. D'une façon importante, nous avons dressé un parallèle entre l'évolution des infrastructures sociales sur internet et l'évolution des systèmes physiques et avons argumenté que ces deux types de systèmes sont soumis à l'augmentation de leur désordre interne, que l'on a nommé, suivant Stiegler, respectivement entropie et anthropie. 
Les conséquences de cette disruption sont multiformes: elle induit une irreprésentabilité de l'avenir, que nous avons appelé un horizon du devenir, une désindividuation collective et psychique, une chute des niveaux de confiance entre les institutions et les populations, ainsi qu'entre les gens et finalement une perte de l'époque, reliée à l'irreprésentabilité de l'avenir. 

Cependant, l'explication de la facilité et la rapidité de l'essor des plateformes nous a posé problème. En seulement quelques décennies, les plateformes sont parvenues à se rendre indispensables dans une large partie des secteurs sociaux. Comment est-ce que cela a-t-il pu être aussi rapide ? Pour résoudre ce problème, nous avons postulé que la culture dominant la fin du millénaire, dite postmoderne, prédisposait cet essor des plateformes, spécialement dans la forme dans laquelle nous les voyons aujourd'hui. Pour confirmer cette intuition, nous avons mis en vis-à-vis, l'étude du postmodernisme culturel proposée par Fredric Jameson, et qu'il présente comme la dissolution des catégories de la modernité, à savoir la dévaluation de la vérité, le rejet des fondements et de la théorie, un appel constant à la superficialité et le rejet du sujet représentationel cher à la modernité, avec les caractéristiques dominantes des plateformes, étudiées dans les chapitres précédents. Cette mise en parallèle des deux études a permis de mettre en évidence la nécessité d'une étude conjointe des secteurs culturels, techniques et marchands de l'époque contemporaine et nous a amené à l'une des conclusions importantes de ce mémoire: la disruption trouverait sa force dans le rejet postmoderne de la nature.

Finalement, dans un dernier chapitre, nous avons porté notre analyse sur deux exemples, Wikipédia et arXiv, qui nous semblaient être des exemples de ce que nous appelons des ``plateformes collaboratives", libres de l'algorithmique et modérées par une communauté contributive. Recherchant la source de la tendance apparemment néganthropique de ces plateformes, et dressant un parallèle avec les systèmes vivants, nous avons postulé que l'aspiration pour une confiance collective tendait à une stabilisation de ces plateformes dans un état néguanthropique.

Ce court mémoire sur la technique contemporaine, et sur les infrastructures d'internet soulève bien d'autres interrogations qui n'ont pas pu être dignement traitées. Par exemple, la question de la recette ``miracle" de structures comme Wikipédia, pourtant de la plus haute importance si nous voulons la généraliser, reste largement entière. Nous avons argumenté que l'attrait de véracité pouvait stabiliser ces structures. Cependant, qu'est-ce qui fait la stabilité et la qualité de Wikipédia alors que des plateformes comme par exemple Doctissimo charrient beaucoup plus de fausses informations sur la santé ? En tant qu'utilisateurs d'internet et de consommateurs, nous n'avons souvent pourtant presque aucune marche à suivre pour constituer ce que l'on pourrait appeler un internet collaboratif.  Répondre à cette question demanderait une étude comparative des deux plateformes que nous n'aurions su faire ici. 

Une autre question que nous avons à peine mentionnée est la question de l'apport que les institutions, notamment publiques, peuvent apporter à l'internet et à son infrastructure. Nous avons avancé dans ce mémoire que la réponse institutionnelle était souvent trop lente et trop lourde pour pouvoir avoir un impact réel sur les pratiques en constante mutation d'internet. Cette observation soulève la question de la formation d'outils institutionnels, donc hautement néguanthropiques, ayant la capacité et la flexibilité pour traiter cet aspect très volatile des réseaux.  

Finalement, la théorie de Stiegler considère une algorithmique traditionnelle, qu'il relie à un accroissement de la calculabilité. L'intelligence artificielle, en cela qu'elle ne constitue pas une algorithmique traditionnelle semble échapper aux arguments de Stiegler, notamment ceux portant sur l'accroissement de l'entropie qui serait liée à leur déploiement. Une analyse en profondeur de ces nouveaux robots serait nécessaire.

\bibliographystyle{plain}
{\footnotesize
\bibliography{biblio}}

\end{document}